\documentclass[mscthesis]{usiinfthesis}
\usepackage{mdframed}
\usepackage{lipsum}
\usepackage{listings}
\usepackage{wrapfig}
\usepackage[caption=false]{subfig}
\usepackage{float}
\usepackage{tabularx}
\usepackage{amsmath}
\usepackage{algorithmicx}
\usepackage{algpseudocode}
\usepackage{verbatim}
\usepackage{array}
\usepackage{multirow}
\usepackage{csvsimple}
\usepackage{rotating}
\usepackage{float}
\usepackage{multicol}
\usepackage{pdfpages}
\usepackage{xtab}
\usepackage{nameref}
\usepackage{caption}
\usepackage{color}
\usepackage{xcolor}
\usepackage{longtable}

\mdfdefinestyle{myframestyle}{backgroundcolor=black!5}

\DeclareCaptionFont{white}{\color{white}}
\DeclareCaptionFormat{listing}{\colorbox{gray}{\parbox{\textwidth}{#1#2#3}}}
\newcommand{\BigO}[1]{\ensuremath{\operatorname{O}\bigl(#1\bigr)}}
\lstdefinelanguage{algebra}
{morekeywords={import,sort,constructors,observers,transformers,axioms,if,
else,end},
sensitive=false,
morecomment=[l]{//s},
}

\restylefloat{figure}
\everymath{\displaystyle}

\title{Study and evaluation of an Irregular Graph Algorithm on Multicore and GPU Processor Architectures} 
\bigskip
\subtitle{\includegraphics[scale=0.2]{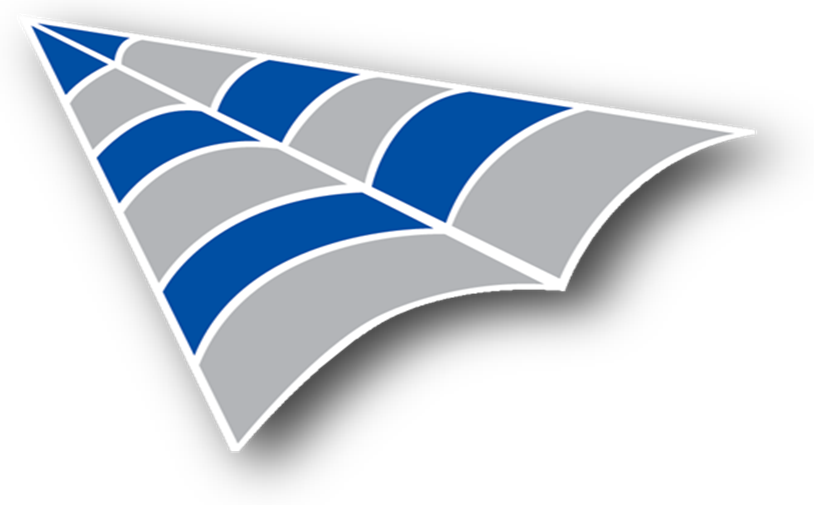}} 
 
\author{Varun Nagpal} 
\Day{14}
\Month{January}
\Year{2013}
\place{Lugano}

\begin{committee}
    \advisor{Prof.}{Mariagiovanna Sami}{}{} 
    \coadvisor{Dr.}{Leandro Fiorin}{}{} 
\end{committee}

\dedication{I dedicate this work to my late grandfather Shri. Vidya Sagar Nagpal ( 15th June, 1921 to 31st Aug, 2012 ). For me, he was the highest source of discipline, determination and undying enthusiasm. I miss him dearly} 
\openepigraph{\emph{Karmanye Vadhikaraste, Ma phaleshou kada chana, Ma Karma Phala Hetur Bhurma tey Sangostva Akarmani} - Do your duty without being desirous of the results. The results are not in your hands, so you shall not think of them. Your attachment should not be to the results or to the non-performance of action}{Sankha Yoga - chapter-2, verse 47, Gita}

\begin{document}

\maketitle 

\frontmatter 

\begin{abstract}
To improve performance of an application over the years, software industry assumed that the application would automatically run faster on new upcoming processors. This assumption mainly relied on ability of the microprocessor industry to extract more ILP(Instruction level parallelism) in single-threaded programs through technology improvements and processor architecture innovations ( higher clock rates, greater transistor density due to transistor scaling, deeper pipelines etc.). Consequently, software programmers have traditionally focused on writing correct and efficient sequential programs and have rarely needed to understand hardware or processor details. 
\\
\\
In the last few years however, this reliance of the software industry on the processor industry to automatically extract and scale application performance with new generations processors has hit a wall due to processor industry moving towards Chip Multiprocessors. This sudden change in trend was motivated mainly by factors of technology limitations( exploding power, ILP flattening ), costs and changing market trends ( increased popularity of portable devices ). As a result, most upcoming processors in desktop, server and even embedded platforms these days are Multicore or Chip Multiprocessors(CMP). Technology scaling continues to date thanks to Moore's law, although now in the form of more processor cores per CMP. Consequently, the focus of programming is slowly moving towards multi-threading or parallel computing. Writing scalable applications with increasing processor cores cannot be done without knowing architectural details of the underlying Chip Multiprocessor.
\\
\\
One area of applications which poses significant challenges of performance scalability on CMP's are \textit{Irregular applications}. One main reason for this is that irregular applications have very little computation and unpredictable memory access patterns making them memory-bound in contrast to compute-bound applications. Since the gap between processor \& memory performance continues to exist, difficulty to hide and decrease this gap is one of the important factors which results in poor performance acceleration of these applications on chip multiprocessors. 
\\
\\
The goal of this thesis is to overcome many such challenges posed during performance acceleration of an irregular graph algorithm called \textit{Triad Census}. We started from a state of the art background in Social Network Analysis, Parallel graph algorithms and \emph{Triad Census} graph algorithm. We accelerated the Triad Census algorithm on two significantly different Chip Multiprocessors: Dual-socket Intel Xeon Multicore (8 hardware threads/socket) and 240-processor core NVIDIA Tesla C1060 GPGPU(128 hardware threads/core). Intel Multicores are designed mainly for general purpose and irregular applications, while GPGPU's are designed for data parallel applications with predictable memory access.  We then looked at techniques for efficiently implementing graph data structures and optimizations targeted for Irregular applications(memory and control flow). During implementation and experiments we looked at the problem from different points of view: from the algorithmic perspective, code optimization perspective, parallelization perspective and processor-aware perspective. We then experimented with different algorithmic, parallelization, load balancing and architecture-aware approaches to accelerate this algorithm. The experimental results obtained for Multithreaded Triad Census implementation on our Intel Multicore Xeon system shows performance speedups (w.r.t baseline sequential) of maximum 56x , average 33x and minimum 8.3x for real world graph data sets. On NVIDIA Tesla C1060 GPGPU, we were able to match almost equally the Multicore results - 58.4x maximum, 32.8x average and 4.2x minimum speedups w.r.t baseline sequential. In terms of raw performance, for the graph data set called Patents network, our results on Intel Xeon Multicore(16 hw threads) were 1.27x times faster than previous results on Cray XMT(16 hw threads) while results achieved on GPGPU were comparatively slower(0.72x). To the best of our knowledge, this algorithm has only been accelerated on supercomputer class computer named Cray XMT and no work exists that demonstrates performance evaluation and comparison of this algorithm on relatively lower-cost Multicore and GPGPU based platforms.
\end{abstract}


\begin{acknowledgements}
\emph{No man is an island} - \textit{John Donne}. This work wouldn't have been possible without encouragement and support of so many people. Firstly I wish to extend my sincere thanks to my advisor Prof. Mariagiovanna Sami and co-advisor Dr. Leandro Fiorin for allowing me to work on this thesis topic and also for their support and guidance throughout this work. Special thanks to Dr. Oreste Villa at PNNL, US for proposing the thesis topic and for replying patiently to my emails. I would also like to thank Umberto Bondi, Janine, Elisa and other members of the ALaRI staff for their support and encouragement throughout the master studies. Special thanks to Prof. Antonio Carzaniga for allowing me to attend his lectures on algorithms which helped me immensely in many ways during this work. I would also like to thank Prof. Vladimir Batagelj at University of Ljubljana for very patiently answering my queries. Special thanks to Dorian Krause at ICS for arranging access to NVIDIA Tesla C1060 GPGPU and patiently replying to emails.
\\
\\
Learning Embedded Systems Design at ALaRI has been an enormous learning experience for me which I could have hardly imagined before coming here. In many ways, my motivation and enthusiasm for Computer Architecture has been shaped by the inspiration and exposure to lectures and lab sessions from ALaRI faculty and so, I feel that I also owe sincere acknowledgement to them.
\\
\\
Furthermore, my studies and time spent in Switzerland wouldn't have been worthwhile without the love and support of all friends I made during this time at ALaRI - Darayus, Anubhav, Chetan, Basundhra, Jane, Tanish, Ameya, Amrit, Gaurang, Naresh, Sanathan Krishna, Kanishk, Sudhanshu, Prasanth, Avinash, Nisha, Ekaterina, Varun, Vasudha, Ankur. Lastly, I am at loss of words to acknowledge how much I owe to my parents and family for their constant encouragement, and support throughout my academic career - to my grandfather for being source for undying enthusiasm, to my father and mother for believing in me and for their unconditional care and love.
\end{acknowledgements}

\tableofcontents 
\listoffigures
\listoftables

\mainmatter

\chapter{Introduction}
\label{chap:Introduction}
The diminishing returns of single threaded performance(Instruction Level Parallelism or ILP) derived from uniprocessor architectural and micro-architectural techniques as well as the growing chip power dissipation have in recent years led the general purpose and embedded computing industry to shift the processor design from single-threaded processors to Multithreaded Chip Multiprocessors(CMP) called \textbf{Multicore or Manycore Processor}\cite{SutterLarus2005, Geer2005}. In such CMP's, multiple processor cores are integrated on a single chip die and each processor core provides one or more hardware threads that can run one or more software threads either in interleaved concurrency or true concurrency. As a result, it is possible to get work done faster on such CMP's by distributing work onto the multiple cores.

As a side effect, this shift in processor design from \emph{performance-driven computing} to \emph{performance per watt computing} has lead software development to move towards concurrency and has opened up new areas of research in parallel programming, automatic parallelization frameworks, compilers etc. In comparison, the challenges for efficient programming of Embedded Multicore processors(\textit{also called Multiprocessor SoC( MPSoC )}) are somewhat greater mainly due to the additional constraints posed in embedded systems such as hard real-time requirements, ultra low power demand for battery-driven devices, reliability, security, heterogeneous architectures( multiple functions on a SoC) and lack of standard tools(compilers, debuggers etc.).

Further, these CMP based Multicore and Manycore GPGPU(General Purpose GPU) platforms are now easily accessible and affordable. As such platforms become more ubiquitous and powerful due to continued CMOS scaling, many applications which have been limited to High Performance Computing domain will slowly make their way into the mainstream consumer applications on desktops, servers and portable embedded devices. 
\\
\\
One application area which requires a large amount of parallel processing power is \emph{Graph Algorithms}. Many real world data can be most naturally modelled as a \emph{graph} as it provides intuitive abstraction for representing objects and relationships between them. For example, the entire web on the internet can be represented as a graph with vertices representing web-pages and edges representing the hyperlinks connecting those web-pages. Internet search engines often use such graph abstractions and algorithms for processing search queries. Social networking sites use graph abstractions to maintain their social networks. Understanding the properties and dynamics of such social networks is important for example to target marketing. EDA tools used for designing VLSI circuits use graph abstraction for representing net-lists(circuit connectivity description) where vertices represent circuit elements(ALU, Bus, Memories etc.) and edges represent nets (interconnections between circuit elements). These graph abstracted net-lists can then be explored and optimized for various figures of merit (power, performance, area, testability, fault-tolerance, floor-planning etc.). Biological networks represented as graphs are used to study the spread and origin of communicable diseases. Transportation companies use graph algorithms to find the shortest possible routes for cargo planes and ships. Fault tolerance studies use bi-connected component graph algorithm to understand robustness of the system when one or more parts of system fail.

As a result, there exist many graph algorithms \cite{Mehta.Sahni2004, Atallah.Blanton2010} that process real world data-sets. Examples of some graph algorithms are: Breadth-first search, Depth-first search, Minimum spanning tree(Prim's, Kruskal's and Boruvka's algorithm), Single source shortest path (Dijikstra's, Bellman-ford algorithm), All pairs shortest path, Bi-connected components(Tarjan-Vishkin algorithm). Studies and uses of such parallel graph algorithms have focused on computational science and engineering applications such as mesh structures, unstructured grids, linear/PDE solvers, N-body methods, VLSI layout, compiler design, system modelling, event-driven simulation, wireless communication, distributed networks, social network analysis, data mining, bio-informatics and security. A graph can be dense or sparse depending upon the number of edges. For a graph with \emph{n-vertices} and \emph{m-edges}, the graph is

\begin{center}
Sparse Graph, iff  m=\BigO{f(n)}\\
Dense Graph, iff  m=\BigO{n^2}
\end{center}

Graph algorithms for \emph{dense graphs} lend themselves to efficient parallel realization and speedups in both theory and implementation on HPC platforms. On the other hand, graph algorithms for \textbf{sparse graphs} have been traditionally known to be difficult to accelerate on CMP's and to date this continues to be a challenging exercise. One of the inherent problems with sparse graph algorithms is that theoretically achievable sequential and parallel performance cannot be easily realized in practice. There are many reasons for this:

\begin{itemize}
\item The gap between parallel models(like PRAM) assumed by the graph algorithm and current parallel computer architectures that masks machine dependent details(Cache memory hierarchy, Communication costs, Distributed Vs Shared memory architectures) crucial for obtaining good speedups. Bridging this gap requires novel algorithm and data structure design techniques which are aware/oblivious of the target processor architecture.

\item Some graph algorithms are combinatorial in nature and thus require an immense amount of computation and communication. Other graph algorithms are more memory-intensive and have very little computation such that they spend most of the time either in memory-traversal or communication(through data sharing or message passing). Irregular sparse graphs algorithms belong to this latter category and are characterized by poor data locality and irregular memory access patterns which results in poor cache performance.

\item Few organized techniques exists which can evenly and efficiently partition an irregular sparse graph problem instance onto available processors. The problem of effectively load-balancing such workloads across PE's(or hw threads) of a Chip Multiprocessor is another challenging aspect.
\end{itemize}

HPC class Supercomputers are still the typical machines used for efficient acceleration of many such graph algorithms. With growing trend for Multicore desktop and embedded platforms, it is very likely that in near future many graph algorithm based applications will be ported to these platforms. The challenges for fast execution of  graph algorithms on these platforms are somewhat higher because these platforms still cannot match Supercomputing capabilities of HPC domain. And therefore, obtaining good execution speedups for such irregular sparse graph algorithms on these emerging Multicore CPU's and Manycore GPGPU's is an important problem. 

\section{Thesis Objectives}
\label{sec:Objective}	
\emph{Main objective} of this thesis work is to study and accelerate a Supercomputing-class irregular sparse graph algorithm called \emph{Triad census} \cite{Wasserman1975, Moody1998a, Batagelj2001}. This algorithm was previously accelerated on Cray XMT Supercomputer \cite{Chin2009}. The Multicore processor architectures used in our work for accelerating this algorithm includes Dual-socket Intel Xeon NUMA processor(16 hardware threads) and 240-core NVIDIA Tesla C1060 GPGPU processor(128 hardware threads per core). The other objectives of this thesis work are as follows:
\begin{itemize}
\item Study relevant work on state of the art in various parallel graph algorithms to understand the issues involved in creating efficient parallel realizations. 
\item Study state of the art in the Multicore and Manycore processor architectures to understand how its various architectural features can be effectively exploited to improve application performance.
\item Improve memory complexity of targeted graph algorithm by exploring architecture aware and architecture oblivious techniques.
\item Explore efficient algorithm and data structure design techniques - \textit{both processor architecture independent as well as dependent} - to improve performance
\item Explore efficient data partitioning and load balancing strategies which are fast, scalable, having low memory overheads and possibly amenable to parallelization.
\item Finally, get good overall speedup(strong/weak scaling) for the Triad Census algorithm on targeted Multicore and GPGPU architectures.
\end{itemize}

\section{Contribution of thesis}
\label{introcontrib}
The contributions of the thesis are the following:

\begin{enumerate}
\item \emph{Main contribution of thesis}: Compared to previous results on the Cray XMT(16 hw threads), for a graph data set called Patents Network, our results(raw performance) on Intel Xeon Multicore(16 hw threads) were slightly better (1.27x times faster) while the resulting performance achieved on GPGPU was comparatively slower(0.72x). To the best of our knowledge, this algorithm has only been accelerated on a supercomputer class machine named Cray XMT \cite{Chin2009} and no work exists that demonstrates performance evaluation and comparison of this algorithm for relatively low-cost commodity processors such as Intel Xeon Multicore and NVIDIA Tesla C1060 GPGPU.

\item \emph{Performance results on our platforms}: On our Intel Xeon Multicore platform, with respect to baseline performance results for the Sequential Triad Census algorithm implementation, the Multithreaded Triad Census implementation resulted in \textbf{maximum 56x, average 33x and minimum 8.3x performance speedups}. In comparison, the CUDA Multithreaded implementation of Triad Census algorithm on our NVIDIA Tesla C1060 GPGPU platform resulted in nearly competitive performance results - \textbf{maximum 58.4x, average 32.8x and minimum 4.2x speedups}.

\item \emph{Graph data structure}: On Intel Xeon Multicore processor, we improved performance(from 1.3x upto 6x faster) by replacing linked list based implementation of adjacency lists with cache blocked linked lists. On GPGPU, we evaluated algorithm performance for graph represented using CRS(Compressed Row Sparse) format.

\item \emph{Control flow optimizations}: We identified two branch optimizations and used pre-computation to improve performance upto 1.23x.

\item \emph{Graph data sets}: Our work used only real world data graph sets(Amazon, Google, Slashdot, Patents etc.) in comparison to some synthetic data sets used in previous work. Evaluation of algorithm on real world graph networks increases reliability of our speedup results.

\item \emph{Thread coupling}: Previous work shared a single census array among threads which required synchronizing concurrent updates to it using fast atomics. In our work, we obtained better results(2x relative speedup on GPGPU) by creating per thread local census array(\textit{decoupled threads}), which are later on merged to get the final census array.

\item \emph{Graph partitioning and Load balancing}: Previous work used distributed task queues approach for graph partitioning and load balancing. In our work on Multicores, we used this same graph partitioning strategy but improved the load balancing. Also, on GPGPU's we demonstrated use of centralized task queue based approach for graph partitioning.

\end{enumerate}

\section{Outline of the thesis}
\label{sec:Outline}	
This chapter was an introduction to the graph processing on CMP's, thesis objectives and contributions. Following is the outline of the remaining parts of thesis:

\paragraph{Chapter~\ref{chap:background}}
This chapter discusses necessary background to set the tone of the thesis. It first discusses Social network analysis(SNA) and the targeted algorithm - \textit{Triad Census}. Then some related topics are discussed namely \textit{graph data structures},  \textit{irregularity} and techniques for efficiently implementing adjacency lists based graph representations. The chapter also discusses optimizations to tackle memory and control flow irregularity. Thereafter, the chapter presents details of the Multicore and GPGPU platforms on which the Triad Census algorithm is implemented, optimized and accelerated. Last part of chapter discusses important differences between Multicore and GPGPU processors.

\paragraph{Chapter~\ref{chap:relatedwork}}
This chapter starts with related work on various Parallel Graph Algorithm implementation and evaluation on CMP's along with some discussion on state of the art graph data structures used in such implementations. Then we compare in detail our work with previous relevant work (\textit{Acceleration of triad census algorithm on Cray XMT Supercomputer}) and present specific contributions of our work .

\paragraph{Chapter~\ref{chap:multicoreval}}
Presents evaluation of Triad Census algorithm on Intel Multicores. It first discusses the adopted methodology for sequential and parallel implementation of Triad Census. Then it discusses details of implementation, optimization and acceleration results for sequential and multithreaded Triad Census algorithm.

\paragraph{Chapter~\ref{chap:gpgpueval}}
Presents evaluation of Triad Census algorithm on NVIDIA Tesla GPGPU platform. It describes the details of implementation, profiling, optimization and results of CUDA based Multithreaded implementation of Triad Census algorithm.

\paragraph{Chapter~\ref{chap:conclusion}}
This chapter discusses the results, conclusion and prospective future directions.

\paragraph{Appendix~\ref{app:concpar}}Discusses definitions of frequently used terminology in Concurrency and Parallelism.

\paragraph{Appendix~\ref{app:procarchmicro}}Discusses definitions of frequently used terminology in Processor Architectures and Micro-architectures.

\paragraph{Appendix~\ref{app:algograph}}Discusses background topics in Graph theory related to small world networks.

\paragraph{Appendix~\ref{app:cudamodel}}Discusses details of NVIDIA's CUDA Programming model used for programming NVIDIA based GPGPU's.

\paragraph{Appendix~\ref{app:comprof}}Presents definitions of different profiling counters for NVIDIA profiler tool called \emph{computeprof}.

\paragraph{Glossary}This section describes acronyms.

\chapter{Background}
\label{chap:background}
In this chapter, we discuss necessary background topics to set the tone of the thesis. We first discuss \textit{Social network analysis (SNA)} along with the irregular sparse graph algorithm that we have chosen to accelerate - \textit{Subquadratic Triad Census}. After this, we give overview of basic graph data structures and briefly describe problem of \textit{Irregularity}. Then we present techniques that can be used for efficient implementation of adjacency lists based graph representations. Next, the chapter presents details of the Multicore and GPGPU platforms on which the Triad Census algorithm is implemented, optimized and accelerated. Last part of chapter discusses important differences between Multicore and GPGPU processors. The topics discussed in this chapter cover necessary background to understand discussions in later chapters.

\section{Social Network Analysis(SNA)}
\label{sec:sna}
\emph{Social network analysis or SNA} deals with analysis of social network structures. It focuses on social entities and ties or relationship between them. Relationship could be friendship, beliefs, common interest, economic interactions etc. Examples of social structures are communities, market, societies, and countries. In SNA vocabulary, any entity involved in a social network is called an \emph{Actor}. An actor can refer to a person, organization, nation etc. basically any entity that can be conceptualized as an object with behaviour and which can interact with other objects.
\\
\\
SNA involves conceptualizing, modelling, representing, visualizing such social structures and then detecting and interpreting patterns of social ties among the actors. It also deals with measurement of characteristics or metrics of social networks. Examples of such metrics are centrality (which actors are best connected to others, or have the most influence), betweenness centrality of a vertex v(the number of shortest paths between all pairs of vertices in the network going through v), connectivity(how actors are related to one another), cohesion(how strong is the relationship between connected group of actors), transitivity(Is enemy of your friend is also your enemy?), clustering(are two neighbours of a vertex are also neighbours themselves), prestige, trust etc.

\begin{figure}[!ht]
	\subfloat[][]
	{
	    \label{fig:sna-ex-1}
	    \includegraphics[angle=90, scale=1.5]{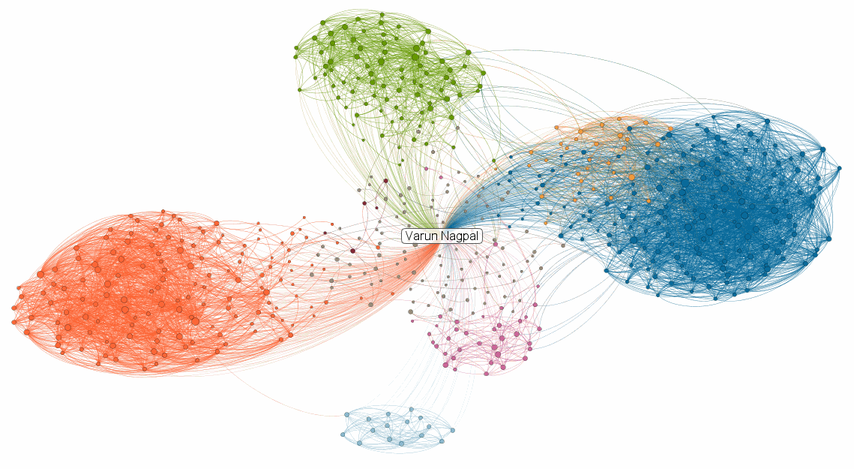}
	 }	
	\subfloat[][]
	{
	    \label{fig:sna-ex-2}
	    \includegraphics[scale=0.4]{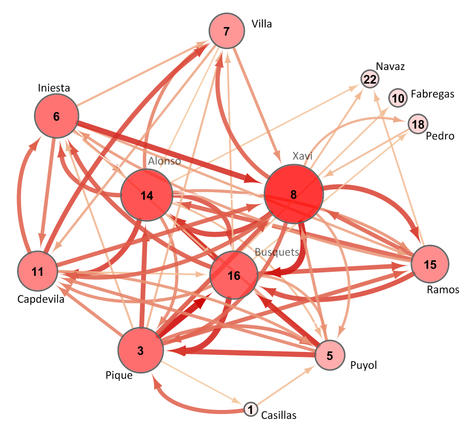}
	}
	\\
	\subfloat[][]
	{
	    \label{fig:sna-ex-3}
	    \includegraphics[scale=0.5]{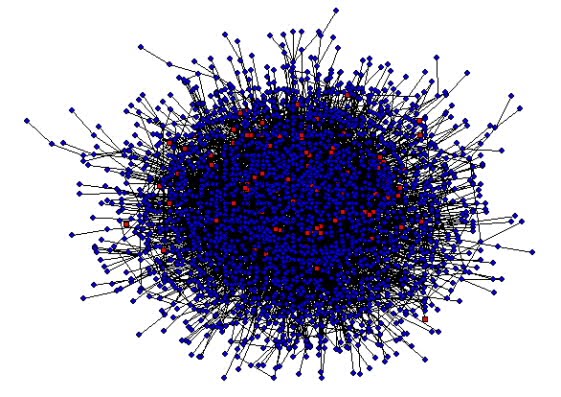}
	}
	\subfloat[][]
	 {
	    \label{fig:sna-ex-4}
	    \includegraphics[scale=1]{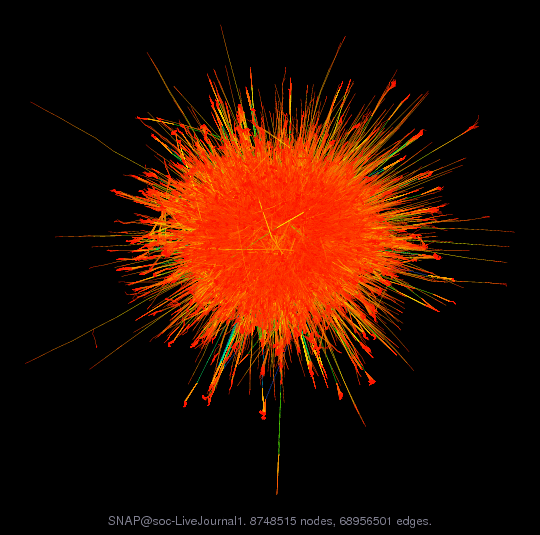}
	 }

	\caption[]
	{ Examples of various social networks:
	    \subref{fig:sna-ex-1} My linkedin professional network
	    \subref{fig:sna-ex-2} Soccer pass network
	    \subref{fig:sna-ex-3} Human protein network
	    \subref{fig:sna-ex-4} Livejournal social network
	    \label{fig:sna-examples}
	}
\end{figure}

Social network analysis has also been linked with biology, communication studies, organizational studies, internet social networks, sociology, geography, finance and terrorism studies (Figure~\ref{fig:sna-examples}). As a result new forms of network analysis have emerged. For example SNA is used in epidemiology to help understand how patterns of human contact aid or inhibit the spread of diseases in a population such as HIV\protect\footnote{\href{http://www.respondentdrivensampling.org/reports/sociological_focus.pdf}{http://www.respondentdrivensampling.org/reports/sociological\_focus.pdf}} and obesity.

Graph abstractions from graph theory provide an intuitive way of representing such social networks and applying various algorithms to capture social network metrics. Actors are represented as \emph{vertices} and relations between them are represented as \emph{edges} or \emph{arcs}. To make networks out of these graph representations, all real world information associated with actors and ties are associated with the vertices and edges of graph. For example consider a social graph of a soccer team\footnote{http://scientometrics.wordpress.com/} during a match (Figure~\ref{fig:sna-examples}), with vertices representing soccer players and arcs representing a pass from one player to another. By associating information such as player names with vertices and number of passes made between two players with thickness of the arc, one can make a social network out of it. This network can be analysed for various metrics like which players made or received the most passes, percentage of passes made or received by defenders, which sequence of passes had a higher probability of scoring a goal, etc. Many tools  are available for analysis and visualization of such social networks namely Pajek, GUESS, ORA, Cytoscape, NetworkX, SNAP and igraph.

\section{Triad Census Algorithm}
\label{sec:triadalgorithm}
\emph{Triad census} based analysis are statistical methods that study and examine characteristics of social networks by studying properties of local 3-vertex-subgraphs present in a graph. The description of \emph{triad} and \emph{dyad} census in SNA was first studied and proposed by \cite*{Holland1970}. They introduced the concept of \emph{dyad and triad census} - number of 2-subgraphs and 3-subgraphs in a graph and also generic \emph{k-subgraph} census. \cite*{Wasserman1975} studied triad census for various random graph distributions. 
\\
\begin{figure}
	\centering
	\includegraphics[scale=0.8]{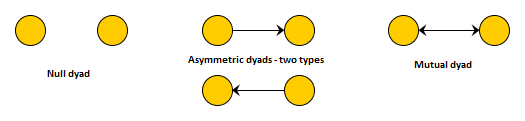}
	\caption{Dyad types}\label{fig:dyadtypes}
\end{figure}
\\
A \emph{triad} is defined as a digraph with 3 vertices and zero or more arcs connecting them. Similar to triad, a \emph{dyad} is a pair of vertices which have zero or more arcs between them. There can be four types of dyads(Figure ~\ref{fig:dyadtypes}): 1 - \emph{null dyad}, 2 - \emph{asymmetric dyads} and 1 - \emph{mutual dyad}. Only null dyad is a \emph{disconnected dyad}, other dyad types are \emph{connected dyads}. One can think of a triad to be made up of three dyads and therefore 64-possible triads(4*4*4, as each pair of vertices can be one of four possible dyads) can be formed from a \emph{strict digraph}(no loops) of order three. Further, these 64 triads can be merged into 16-unique \emph{isomorphic triads}. All triads can be classified into three triad classes namely \emph{null, dyadic and connected triads}
\\
\begin{figure}
    \centering
    \includegraphics[scale=0.6]{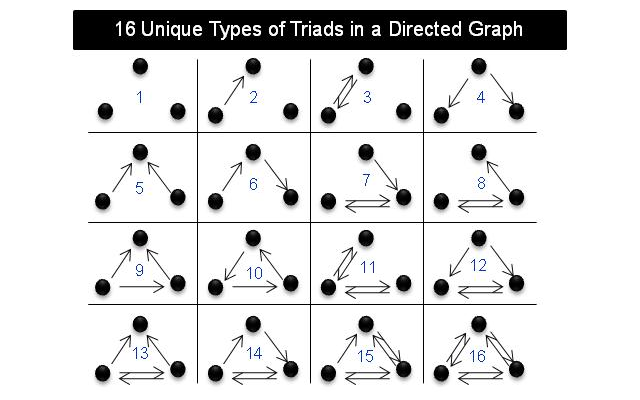}
    \caption{Types of Triads}\label{fig:triadtypes}
\end{figure}

Figure~\ref{fig:triadtypes} shows the isomorphic types of triads numbered 1 to 16 and classification of these triads. Established convention of naming the 16 triad types uses a 3-digit code suffixed with an optional triad property: \(\mathsf{MAN<P>}\), where \(\mathsf{M}\) is number of \emph{mutual dyads}, \(\mathsf{A}\) is number of \emph{asymmetric dyads}, \(\mathsf{N}\) is the number of \emph{null dyads}, \(\mathsf{P}\) is an optional triad property. 
\\
\\
Each of the digits \(\mathsf{M, A, N}\) can range from 0 to 3. \(\mathsf{P}\) can take values U-up, C-cyclic, D-down, and T-transitive. So triad types 1-16 are classified and named as follows:

\begin{itemize}
\item \emph{Null triads}: 1- 003 
\item \emph{Dyadic triads}:  2 - 012, 3 - 102
\item \emph{Connected or complete triads}: 4 - 021D, 5 - 021U, 6 - 021C, 7- 111D, 8 - 111U, 9 - 030T, 10 - 030C, 11 - 201, 12 - 120D, 13 - 120U, 14 - 120C, 15 - 210, 16 - 300
\end{itemize}

Only null and dyadic triads are disconnected, all other triads are connected. \emph{Triad census algorithm} is a combinatorial problem that aims to count number of each of these 16 unique isomorphic or 64 triads in a given graph i.e. census of triads. By finding out frequencies of these 16 or 64 triad types, different characteristics associated with social network can be examined such as transitivity, cyclicity, mutuality (reciprocity) etc. Triadic methods have been used for studies in psychology, network evolution, business networks, neuroscience and security.
\\
\\
For a digraph with \(n-vertices\) there can be \(T = {^n\text{C}_3}\) triads. Naive algorithm for calculating triad census will take too much time - \BigO{n^3} time complexity. Faster algorithms for calculating triad census have been proposed. \cite*{Moody1998} proposed an adjacency-matrix based algorithm with quadratic complexity \BigO{n^2}. \cite*{Batagelj2001}  proposed algorithm with sub-quadratic time complexity \BigO{m\Delta} where \(m = \BigO{n.k(n)}\) is the number of arcs of the graph and \(\Delta\) is maximum degree of the graph.
\\
\\
We now introduce some notations and then explain the working of Subquadratic triad census algorithm. \(\{ \}\) denotes set notation, \(( )\) denotes ordered pair notation. Consider a strict digraph (no loops) \(G = [V, E]\) where \(V = \{1, 2... n\}\) is set of vertices and \(E \subseteq V x V \) the set of arcs (directed) i.e. \(E = \{(u, v), \, \exists u,v \in V \}\). Order of graph \(n = |V|\), size of graph \(m = | E |\). N is an array of neighbour lists indexed on vertex such that \(\forall u \in V\),  N(u) is the set of open neighbourhood of vertex u i.e. \(N(u) = \{ v \in V | \{u, v\} \in E \}\). Note that N(u) contains all vertices v adjacent to u in undirected sense. 
\\
\\
The Subquadratic triad census algorithm(Figure~\ref{fig:triadcensus}) works by computing count of three triad types separately. It first counts all dyadic triads($T_{2}$) and connected triads($T_{3}$). Then it counts number of null triads $T_{1} = T - T_{1} - T_{2}$. Counting of the triads is basically based upon dyads in the digraph. The algorithm starts with initializing Census array to zero(lines 1-3). It then runs a for loop for each vertex of graph(lines 5-23). Within this loop, dyadic($T_{2}$) and connected triads($T_{3}$) are counted. 

\begin{figure}[t!]
\fbox{\begin{minipage}{13 cm}
\begin{flushleft}
\textbf{Function}: Subquadratic Triad Census Algorithm\\
\textbf{Input}: Graph \(G = [V, E]\) where V is set of vertices, E is array of edge lists and N is array of neighbour lists\\
\textbf{Output}: Census array with frequencies of isomorphic triadic types\\
\end{flushleft}
    \begin{algorithmic}[1]
	\For{$i = 1 \to 16$} 
	    \State Census[$i] \gets 0$
	\EndFor
	\\
	\For{ $u \in V$} 
		\For{ $v \in N[u]$}
			\If {$u$ \textless $v$}
			    \State $S \gets N(u) \cup N(v)  \setminus \{u,v\}$
				\If { $IsEdge(u,v) \wedge IsEdge(v,u)$ }
				    \State $TriadType\gets 3$
				\Else
	                \State $TriadType\gets 2$
				\EndIf
				\State $Census[TriadType] \gets Census[TriadType] + n - |S| - 2$
				\For{ $w \in S$ }
					\If{$v$ \textless $w \vee ( w$ \textless $v \wedge u$ \textless $w \wedge \neg IsNeighbour(u,w) )$}
		                \State $TriadType \gets TriadCode(u,v,w, )$
		                \State $Census[TriadType] \gets Census[TriadType] + 1$
		            \EndIf
				\EndFor
			\EndIf	
		\EndFor
	\EndFor
	\\
	\State $sum \gets 0$
	\For{$i = 2 \to 16$} 
	    \State $sum \gets sum + Census[i]$
	\EndFor
	\State $Census[1] \gets n*(n-1)*(n-2)/6 - sum$
	\end{algorithmic}
	\caption{Subquadratic Triad Census Algorithm}\label{fig:triadcensus}
\end{minipage}}
\end{figure}

\begin{itemize}
\item \textbf{Counting number of dyadic triads($T_{2}$)}: a connected dyad (u, v) with u < v (line 7), contributes number of dyadic triads equal to all vertices of the graph except u and v themselves and except all vertices which are in open neighbourhood of the connected dyad  i.e. \(n-|{u,v}|-|S|\) where \(S = N(u) \cup N(v) \setminus {u,v}\) is the set of vertices in open neighbourhood of connected dyad (u,v)(see lines 8,14). Depending upon if the connected dyad (u, v) is asymmetric or mutual, all dyadic triads contributed by it are either of triad type 2-012 or 3-102 (line 9 to 13). The condition u < v (line 7) is necessary for canonical selection of a dyad. By canonical it means every non-null triad contributed by the connected dyad must be counted only once during different iterations of the algorithm. For example the connected dyad (1,2) is canonical but (2,1) is not. This is so because after counting the triads contributed by dyad (1,2), processing dyad (2,1) would lead to repetition of triad counting.

\item \textbf{Counting number of connected triads($T_{3}$)}: Each connected dyad (u,v) where u<v, also contributes number of connected triads equal to vertices in its open neighbourhood set S(line 15-20). Again, all non-canonical selections must be eliminated. Three vertices forming a connected triad can be selected in six ways: (u,v,w), (u,w,v), (v,u,w), (v,w,u), (w,u,v) and (w,v,u). Line 7 already already made last four of these selections non-canonical by enforcing u<v condition. Line 16 determines which of the two remaining selections (u,v,w) and (u,w,v) is canonical. 

\begin{itemize}
\item Selection (u,v,w) with u<v< w  is always canonical. 
\item However if u < w < v, then this selection can be non-canonical only if u and w are neighbours because then the triad (u,v,w) was already counted in previous iteration of algorithm when dyad (u,w) was processed. 
Line 16 enforces condition to check if w forms part of a canonical triad: 

\begin{verbatim}
    v < w or ( w < v and u < w and !IsNeighbour(u,w) )
\end{verbatim}

\end{itemize}

$IsNeighbour(u,v)$ function checks if vertices u and v are neighbours in undirected sense. Line 17 finds isomorphic triad types using $TriadCode(u,v,w)$ function ( Figure~\ref{fig:TriadCode} ). This function first calculates triad code value between 0-63 corresponding to 64 non-isomorphic triad types. It returns this value + 1 if main algorithm calculates non-isomorphic triad census. Otherwise a mapping table $TriadTable$ maps the 64 non-isomorphic triad types to 16 isomorphic types and function then returns a value between 1 and 16. $IsEdge(u,v)$ function checks if a directed edge exists from u to v. Returns 1 if true, 0 otherwise. Finally the Census corresponding to triad type is incremented to count the connected triad (line 18).

\begin{figure}
\fbox{\begin{minipage}{13 cm}
\begin{flushleft}
\textbf{Function}: TriadCode\\
\textbf{Input}: vertices u, v, w\\
\textbf{Output}: isomorphic or non-isomorphic triad code
\end{flushleft}
\begin{algorithmic}[1]
\State $TriadCodeVal = IsEdge(u,v)$
\State $TriadCodeVal += 2 * IsEdge(v,u)$ 
\State $TriadCodeVal += 4 * IsEdge(u,w)$
\State $TriadCodeVal += 8 * IsEdge(w,u)$
\State $TriadCodeVal += 16 * IsEdge(v,w)$
\State $TriadCodeVal += 32 * IsEdge(w,v)$
\State $return ( IsIsomorphic ? TriadTable[TriadCodeVal]:TriadCodeVal ) + 1$
\end{algorithmic}
	\caption{Triad Code Algorithm}\label{fig:TriadCode}
\end{minipage}}
\end{figure}

\item \textbf{Counting null triads($T_{3}$)}: Lines 24-27 stores sum of $T_{2}$ and $T_{3}$ in variable sum and subtracts this sum (line 28) from total count of all triads of the graph(line 28) to get count of null triads $T_{1}$
\end{itemize}

Time complexity of the of the above Subquadratic triad census algorithm is \BigO{m\bigtriangleup}. Exact complexity depends upon the sparsity of graph data set and the maximum degree of the graph. For small world networks with power law degree distribution that are sparse i.e. m is \BigO{n.k(n)} where $k(n)<< n$, and have small maximum degree i.e. $\bigtriangleup << n$, the algorithm complexity is linear - \BigO{n}. If maximum degree of graph is high i.e. $\bigtriangleup \backsim n$, then algorithm complexity is quadratic - \BigO{n^2}. For complete graphs, the time complexity becomes cubic - \BigO{n^3}

\section{Graph Data Structures}
\label{sec:graphds}
\begin{figure}	
	\includegraphics[scale=0.6]{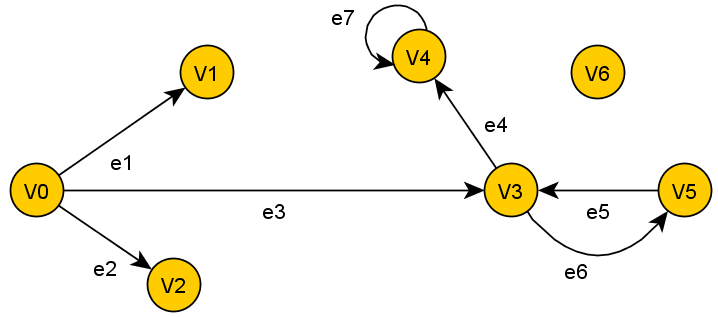}
	\caption{Directed graph}\label{fig:digraph}
\end{figure}

A \emph{graph data structure} stores graph data and allows for traversal and manipulation of the graph. In graph algorithmic theory, there are mainly two representations that are used for representing graphs: \emph{Adjacency Matrix} and \emph{Adjacency Lists}. Figure~\ref{fig:digraph} shows an example directed graph

\emph{Adjacency Matrix}: For a graph \(G = [V, E]\) with $n$ vertices and $m$ edges(undirected graphs) or arcs(directed graphs), an adjacency matrix uses an $n x n$ matrix to represent list of edges or arcs. Each location $[i,j]$ of the matrix stores a 1 or edge weight $w_{ij}$ if there is an edge(or arc) between(from) vertex $v_{i}$ and(to) $v_{j}$ otherwise a 0 is stored. This graph representation(Table~\ref{tab:adjmatrix}) is good for dense graphs and small order graphs (dense/sparse) but uses lot of space \BigO{n^2}. Its nice property is that elements of the data structure are accessed in contiguous fashion and so is cache and prefetch friendly. For sparse graphs, adjacency matrix wastes lot of space as there will be lot of zeros and instead adjacency list is preferred which uses space of order \BigO{|V|+|E|}.

\begin{figure}
		\centering
		\includegraphics[scale=0.6]{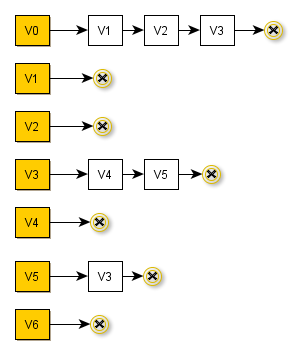}
		\caption{Adjacency list representation of a digraph}\label{fig:adjlist}
\end{figure}

\emph{Adjacency List} representation( Figure~\ref{fig:adjlist} ) of a graph \(G = [V, E]\) is a one dimensional array Adj of size $N = |V|$. For a vertex $v$ of graph $G$, $Adj[v]$ stores pointer to list of vertices adjacent with $v$. For weighted and digraphs, each adjacent list has an additional field to hold weight for that edge. 

\begin{table}
  \centering
  \caption{Adjacency matrix representation of a graph}
  \begin{tabular}{ *{8}{c}}
     & V0 & V1 & V2 & V3 & V4 & V5 & V6\\
  V0 & 0  & 1  & 1  & 1  & 0  & 0  & 0\\		     
  V1 & 0  & 0  & 0  & 0  & 0  & 0  & 0\\
  V2 & 0  & 0  & 0  & 0  & 0  & 0  & 0\\
  V3 & 0  & 0  & 0  & 0  & 1  & 1  & 0\\
  V4 & 0  & 0  & 0  & 0  & 1  & 0  & 0\\
  V5 & 0  & 0  & 0  & 1  & 0  & 0  & 0\\
  V6 & 0  & 0  & 0  & 0  & 0  & 0  & 0\\
  \end{tabular}
  \label{tab:adjmatrix}   
\end{table}

\section{Irregularity}
\label{sec:irregularity}
Many algorithmic problems in computer science exhibit a property called \emph{Irregularity} \cite{Yelick93datastructures} which makes it hard to obtain their efficient sequential and parallel implementations of algorithms. Irregularity is closely linked to the problem which the algorithm solves and also to the way it is represented. Irregularity stems itself in many ways:

\begin{itemize}
\item \emph{Irregularity in flow control}: This irregularity results from presence of many flow control statements inside the program such as if-else, GOTO, switch, while/for loops, function calls etc. This is mostly the property of the algorithm itself and not necessarily of its representation/implementation. Different processors and machines will handle and give different performance for given control flow instructions. For example general purpose processors(like Intel x86 or x64) are typically designed to perform well for control-flow intensive programs. On the other hand, hardware platforms like GPGPU's, FPGA's or DSP's typically do not perform well for control flow intensive programs as these platforms are simply not designed for them and are instead designed to target low-latency, high computational throughput applications.

\item \emph{Irregularity in data structure}: Typically every program uses some sort of data structures to store data associated with algorithm so as to allow primitive data operations - create, insert, delete, update or traverse. Broadly speaking, a data structure can be array based(data static or known priori), pointer chasing based (for unknown/dynamic data) or a combination of array / pointer chasing. Depending upon the processor and memory architecture, data structure may be stored physically different than logically imagined by data structure designer. Due to the fact that there always has been a performance gap between processor and memory, the actual achieved performance may differ from the logically expected performance(usually based on some machine model like PRAM). Other factors which influence this kind of irregularity are data partition technique, data mapping/layout, data traversal order, data dependency, operating system internals. This kind of irregularity thus results from the data structure and algorithm representation/implementation and is not necessarily a property of the algorithm or its associated data structures. 

\item \emph{Irregularity in communication}: Many programs need data structures to be partitioned across memory architecture so that data can be processed by many similar threads as in data parallel applications or by different threads as in task parallel applications or a combination thereof. Most of the time, application threads have to communicate with each other or share some data. Threads may also have to synchronize with other threads at some point. This thread communication is typically handled either using shared memory(as in tightly coupled multiprocessors) or through explicit message passing(as in loosely coupled multiprocessors). Depending upon the processor coupling, communication architecture (shared/distributed address-space/memory) and the data partition technique employed, the communication cost can can take significant amount of the overall application execution time. Irregularity in communication is the problem in matching theoretical optimal communication cost with actually achieved communication cost on hardware platform.
\end{itemize}

\section{Efficiently Implementing Irregular Graph Algorithms}
\label{sec:opttechirrg}
We now present techniques to address performance limiting issues. As the algorithm targeted for acceleration - \textit{Triad Census} - deals with irregular sparse graph data, so we first focus on techniques to design efficient adjacency list graph representations. In the remaining part of the section, we describe techniques to address the irregularity problem with focus on memory and control-flow optimizations.

\subsection{Efficiently implementing adjacency lists}
\label{subsec:effadjlst}
Performance of any graph algorithm depends a lot on the design of graph data structure. Which graph data structure is good for the graph algorithm depends upon many factors like graph size, dynamic/static graph algorithm, graph density, hardware resources, algorithm specific requirements (data partitionability, fast insertions, deletions, traversal etc.). The challenges lies in creating compact space efficient and cache friendly representations of such sparse graphs. 
\\
\\
In particular, because we are dealing with irregular sparse graphs, it is important to have efficient design of adjacency list structure. Main problem with adjacency list representation is that finding whether an edge $(u, v)$ is present in graph requires traversal of adjacency list of $Adj[u]$. As lists are normally implemented as linked lists, they suffer from problem of  cache unfriendly design due to poor spatial locality.  In worst case, a cache miss occurs for every node access - \BigO{n}. There are few approaches to overcome some of these problems:

\begin{enumerate}
\item \emph{Cache blocked or unrolled linked lists} \protect\footnote{\href{http://www.wikipedia.com}{http://www.wikipedia.com}}: This technique replaces adjacency linked lists of graph data structure with blocked or unrolled linked lists. An \emph{unrolled or cache blocked linked list} is a linked list of fixed size arrays( Figure~\ref{fig:unrolledlist} ). Each node of an unrolled list consists of

\begin{enumerate}
\item A linear array of fixed size(blocking) called \emph{Element array}
\item A variable that stores number of valid elements in the element array. 
\end{enumerate}

Compared to normal linked list, unrolled linked lists have two major advantages. Firstly the memory overheads(pointers, alignment, meta-data etc.) are lower as a blocked node amortizes the memory overheads. Secondly, because of better spatial locality of arrays, overall cache misses that occur during traversal also reduce. Cache misses in unrolled linked list is close to optimal cache misses in arrays i.e. \BigO{n/B}. An additional benefit that comes with improved spatial locality is the increase in accuracy of compiler and hardware assisted prefetching.

\begin{verbatim}
 If
      n = Elements inserted
      B = Cache line size in terms of number of elements
      N = blocking size in terms of number of elements
\end{verbatim}

Cache misses on traversing the list of $n$ elements in
\begin{itemize}
\item Array - \BigO{n/B}
\item Linked list - \BigO{n}
\item Unrolled linked list - \BigO{(n/N+1)(N/B)} [Note traversing elements in each node require \BigO{N/B} cache misses. Further assumption is all nodes are full]
\end{itemize}

Unrolled list can be easily tuned to the specific processor architecture by choosing size of the element array according to the cache line size. One disadvantage of unrolled lists is that in some cases, overall memory consumption may increase. For example if there are many graph vertices with low degrees, then element arrays of such vertices will be partially filled leading to memory wastage. Even then an unrolled list can perform better than a normal linked list. 
\\
\\
A blocked or unrolled linked list thus combines spatial locality benefits of arrays with dynamic capabilities (quick insertion, re-sizing) of a linked list and therefore should be used whenever feasible.

\begin{figure}
	\centering
	\includegraphics[scale=0.5]{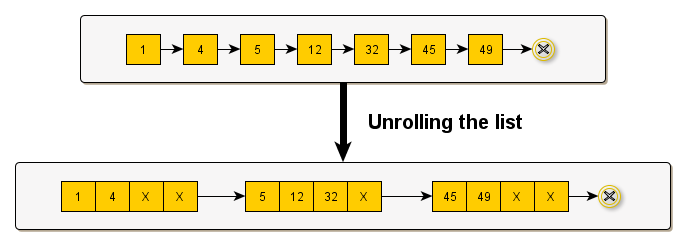}
	\caption{Normal linked list and Unrolled(blocked) list}\label{fig:unrolledlist}
\end{figure}

\item \emph{Dynamic Array}: Another data structure which combines spatial locality benefits of linear array with dynamic data structure capabilities is a \emph{Dynamic array}. A dynamic array is an array which can be resized at runtime. Other advantages of dynamic arrays over linked lists include compactness, faster indexing(constant time), constant time(amortized) insertion/deletion at end, faster linear time iteration and faster insertion/deletion at beginning of array(due to better spatial locality). Standard Template Library(STL) Vector container is a popular implementation of a dynamic array. This container is guaranteed  by C++ standard to be contiguous \protect\footnote{\href{http://herbsutter.com/2008/04/07/cringe-not-vectors-are-guaranteed-to-be-contiguous/}{http://herbsutter.com/2008/04/07/cringe-not-vectors-are-guaranteed-to-be-contiguous/}}

\item \emph{Adjacency array representation}\cite{Mehlhorn.Sanders2008}: This representation concatenates adjacency arrays of all vertices in a single array called \emph{Edge array} $E$. An additional array $V$ stores the starting index positions of the individual adjacency arrays in $E$, i.e., for any vertex $v$, $V[v]$ is the index in $E$ of the first edge out of $v$. 

\begin{figure}
    \centering
    \includegraphics[scale=0.6]{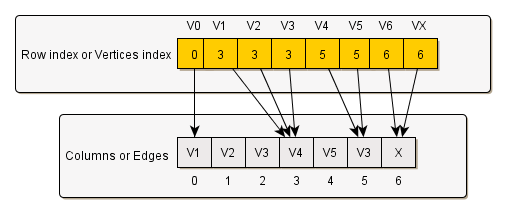}
    \caption{Adjacency array representation of a digraph. Top array $V$ is the row index array of length $Size(G) + 1$. Lower array $E$ is the column or edge array of length $Order(G)$}\label{fig:adjarray}
\end{figure}

The edges out of any vertex v are then easily accessible as $E[V[v]] \ldots E[V[v + 1]−1]$; Normally a dummy entry $V[n + 1]$ with $V[n + 1] = m + 1$ is added to ensure that this also holds true for vertex $n$. Figure~\ref{fig:adjarray} shows an example. The memory consumption for storing a directed graph using adjacency arrays is $n + m + Theta(1)$ words.
\\
In addition, adjacency array representation( Figure~\ref{fig:adjarray} ) is analogous to CRS(Compressed Row Storage) format used for representing sparse matrices
\protect\footnote{\href{http://software.intel.com/sites/products/documentation/hpc/mkl/mklman/GUID-9FCEB1C4-670D-4738-81D2-F378013412B0.htm}{Intel MKL CRS Format}}. In CRS terminology, array $V$ is called \emph{Row index array} and array $E$ is called \emph{Column Array}. \emph{Value array} of CRS format normally would represent weights or any other edge property and is thus optional. Advantage of this graph data structure is higher spatial locality and compactness.

\end{enumerate}

\subsection{Optimizing irregularity}
\label{subsec:optimitech}
In this section, we will present taxonomy of techniques that can be employed to address irregularity issues. What makes it difficult to create generalized methodologies to address irregularity is that multiple abstraction layers of software and hardware hide  many software/hardware details which could be crucial for obtaining good performance. As a result, during program execution many non-deterministic events(cache misses, page faults, context switching, thread interleaving/migration, resource sharing, branch mispredictions, bandwidth/latency variance, etc.) occur which simply cannot be controlled even when they can be monitored. Also, as different irregularity types could be interdependent, optimizing a particular irregularity type may result in a detrimental effect for other type of irregularity 

Therefore, in order to achieve optimal performance, the programmer must make careful trade-off's, do thorough analysis of the problem, understand the machine architecture, map the problem in best possible way to the machine architecture and finally do many experiments to get optimal results.

Irregularity issues can be mitigated both at software level and processor hardware level. At software level, irregularity can be handled from application programming layer to compiler layer. Some of the software based techniques are generic and some are specific to processor architecture(at compiler level). Hardware based Sophisticated micro-architecture circuit blocks in modern day Multicore and Manycore processors try to address irregularity. We will now explain important hardware and software techniques to address irregularity for control flow and memory optimization

\subsubsection{Control flow optimizations}
\label{subsubsec:cntrloptim}
All control flow programming constructs on compilation generate branch instructions. Instruction stream containing high percentage of branch instructions cause control hazards(irregular control flow), resulting in processor pipeline stalls and thus can deteriorate performance. For irregular algorithms, many branches tend to be data dependent and thus branch predictors in Multicore processors have hard time in correctly predicting these branches. On GPGPU processors, presence of unpredictable data dependent branches can significantly lower the performance if different threads tend to follow different branch paths ( called as \emph{warp divergence} ). 

In other words, processors work best when control flow is sequential. Consequently, optimizing branches should not be ignored as it can yield significant performance improvements. Some of the techniques to optimize branches are: 

\begin{itemize}
\item \textbf{Eliminate branches}: The simplest way to optimize branches is to try to eliminate unnecessary branches. Most of the time, this requires analysis at algorithmic and code level. This is the simplest and least effort branch optimization which can reduce irregular control flow and yield good performance improvements. 
\item \textbf{Loop unrolling} reduces number of loop iterations and consequently reduces the branch penalties. Loops with static or predictable iteration bounds can be easily unrolled either by the compiler or explicitly by programmer. For loops whose iteration range at compile time is not predictable or unknown, the programmer can explicitly unroll the loops with some programming effort.
\item \textbf{Function inlining} is another branch optimization technique which is also normally done by the compiler but can also be controlled programmatically. When a function is \emph{inlined}, all calls to this function in the entire program are replaced with the actual function code by the compiler. For functions with small code size and which are called very frequently, function inlining makes sense and can yield good performance improvements. Note that compilers may ignore the inlining of a function.
\item \textbf{Predication} is a branch optimization technique which is normally employed by compilers. When a branch is predicated by the compiler, all instructions in different branch paths are predicated with the branch condition. All the predicated instructions are executed by the processor irrespective of the branch condition, however only those instructions for which the predicate condition is true are committed. Compiler based predication thus requires support at ISA level of the processor. Although compilers are able to optimize many branches with predictable paths but data dependent branches pose significant challenges for compiler. For such data dependent branches, predication can be implemented programmatically.
\item \textbf{Modifying algorithm}: Many a times, formulating alternative algorithmic solution to the problem alters flow control and thus there could be possibility of realizing algorithmic solutions with better control flow.
\end{itemize}

At micro-architecture level, many processors employ sophisticated techniques to efficiently execute branch instructions. These includes \emph{branch prediction(instruction level speculative execution)} and \emph{thread level speculative execution}. Branch prediction involves predicting the most likely control-flow path of a branch instruction and then speculatively fetching-executing instructions from the predicted branch path. TLS or thread level speculation is analogous to branch prediction in the sense that it operates at thread level instead of instruction level.

\subsubsection{Memory focussed optimizations}
\label{subsubsec:memoptim}
Irregular computations and data structures suffer from locality problems making them memory bound. The underlying cause for this problem - \textit{processor-memory performance gap} - can be mitigated by mainly two approaches. First, by \emph{hiding} this gap(\textit{memory latency hiding} by keeping processor busy while it is awaiting a slow memory transaction to finish. Second approach is to \emph{reduce} this gap by making sure data is held close to the processor whenever possible. Following are some state of the art hardware and software techniques that implement these approaches
\\
\\
\textit{\textbf{Hardware architecture level techniques}}: Sophisticated hardware architecture design techniques try to address memory irregularity problem in software independent manner. Traditional hardware techniques used for hiding and reducing memory latency are as follows:

\begin{itemize}
\item \emph{Hardware Prefetching} also called Data level speculation
\item \emph{Cache memory hierarchy} (see next sub section for more details)
\item \emph{TLB (Translation look-aside buffers)}
\item \emph{Out-of-order execution}
\end{itemize}

Another hardware approach to hide memory latency is through \emph{Hardware Multi-threading}. Hardware Multi-threading hides memory latency by exploiting high amount of thread level parallelism. The key to hardware multithreading is to keep the processor busy with work by scheduling a ready thread while current thread is waiting for a memory transaction to complete. Following is a taxonomy of different hardware multithreading techniques :

\begin{itemize}
\item \emph{Block or Coarse grained Multithreading}: Scheduling another thread on processor pipeline when present thread stalls on high latency operation(like cache miss). At a given clock cycle, a single pipeline stage contains(issued or to be executed) instructions from single thread only.
\emph{Cooperative or fine grained Multithreading}: \item Scheduling many threads in processor pipeline by interleaving their execution in pipeline stages. At a given clock cycle, a single pipeline stage contains(issued or to be executed) instructions from single thread only.
\item SMT(Simultaneous Multithreading) architectures like Intel Hyperthreading combine Superscalar processing with Hardware Multi-threading where instructions from multiple independent threads can be simultaneously issued and executed to/in a given pipeline stage
\item \emph{Thread level speculation} is another technique for hiding memory latency through speculative execution of hardware threads.
\item \emph{Chip Multithreading(CMT)}
\end{itemize}

For more information on Hardware Multithreading and other processor related concepts, please see the  \textbf{Appendix~\ref{app:procarchmicro} on \nameref{app:procarchmicro}}.
\\
\\
\textit{\textbf{Software level techniques}}: Software approach to address irregular memory access problems is by using a mix of algorithmic, data structure design and compiler techniques. Some of these techniques are aware of the underlying processor architecture and some of them are oblivious to it. Before looking at the techniques, we will first give overview of processor memory hierarchy which will set tone for understanding the techniques.

Most modern state of the art processor architectures use some form of \emph{Multilevel Memory Hierarchy} where each level act as a cache for next. A typical hierarchy consists of registers, level 1 cache, level 2 cache, level 3 cache, main memory(DRAM), and disk. The underlying importance of this hierarchy is that as the memory level gets farther from the CPU, it becomes larger in size and slower(high latency). Typically registers and level 1 caches are the fastest when compared to other memory levels. The difference in memory access time increases dramatically from level caches to DRAM.

An instruction which needs to access data from a memory address, can execute only if the required data is present or brought in closest cache level. Instruction thus waits for the data if its in farther memory levels. A consequence of this is that in order to obtain high algorithmic performance, it is important to make sure the accessed data is held by closest memory level. This problem of making sure that accessed data always resides in closest level (atleast in some level cache) - i.e. memory access latency is low - is non-trivial to solve, depends upon many factors and requires careful algorithmic and architecture specific design optimization's.

As already explained before, many a times pointer-chasing based dynamic data structures need to be used in algorithms. Examples of such data structures are  lists, sets, trees, graphs, unstructured grids etc. Such structures suffer from memory locality problem causing large number of cache misses and thereby proving detrimental to overall performance. As a result, improving memory locality by optimizing cache memory use of the algorithm and its data structures can yield impressive performance gains. A positive side-effect of improving locality of data layout/access is that TLB misses also decrease because most data will be accessed from locations in the same page. 

All optimizations which can be applied to keep data in on-chip level caches are classified as \emph{Cache Optimizations}. Many cache optimizations methods have been proposed in literature \cite{Srikant2002, Drepper2007, Garg.Sharapov2001} that try to address the memory latency issue. In essence, they basically try to do the following:

\begin{itemize}
\item Reduce cache misses(cold, capacity and conflict)
\item Reduce latency to serve a cache miss
\end{itemize}

Like all software optimizations, cache optimizations can be both aware or oblivious to underlying processor architecture. If cache optimizations require detailed knowledge of underlying cache hierarchy(levels, block or line sizes, alignment) then it is called \emph{Cache Aware}, otherwise \emph{Cache oblivious} \cite{Demaine2002, Kumar2003, Arge2005}. Most cache optimizations rely on code-restructuring, data-layout and data access optimizations. Table~\ref{tab:memoryoptimizations} shows the summary of these memory optimizations.

\begin{table}[t]
  \centering
  \caption{Memory hierarchy optimization summary}
  \begin{tabular}{ | p{6cm} | p{7cm} | }
  \hline
  \textbf{Optimization} & \textbf{Performance improvement}\\ \hline
  Cache blocking(linked lists) & Fewer cold misses\\ \hline
  Cache blocking(arrays) & Fewer capacity misses\\ \hline
  Sorting linked lists in sequential order of node addresses & Fewer cold misses, better prefetching, amortized average load latency\\ \hline
  Prefetching using auxiliary structures(linked lists) & Reduces cold misses, aids prefetching\\ \hline
  Reduce cache pollution and pack caches with useful data & Better cache utilization and reduced bandwidth wastage\\ \hline
  Cache aligned data structures \cite{Fry2008} & Lower cache footprint, halves cache misses due to misalignment\\ \hline
  Keep hot fields together and access in layout order & Higher cache utilization, reduced capacity misses due to lower cache footprint, more efficient prefetching\\ \hline
  Loop interchange & Fewer conflict misses ( cache thrashing )\\ \hline
  Loop fusion & Fewer conflict misses\\ \hline
  Loop fission & Fewer conflict and capacity misses\\ \hline
  Loop unrolling & Increases ILP, lowers loop overhead\\
  \hline
  \end{tabular}
  \label{tab:memoryoptimizations}   
\end{table}
We will now discuss in some detail the techniques in the table~\ref{tab:memoryoptimizations}.

\begin{enumerate}
\item \emph{Cache blocked linked lists}: Its basics have been discussed before. There is plenty of material\protect\footnote{\href{http://software.intel.com/en-us/articles/cache-block-size-on-processors-that-support-hyper-threading-technology}{http://software.intel.com/en-us/articles/cache-block-size-on-processors-that-support-hyper-threading-technology}} available on how to implement blocked linked lists and tune block sizes \cite{Frias2010}.

\item \emph{Use dynamic arrays for small lists}: Its basics have been discussed before. Dynamic arrays should be preferred over linked or unrolled lists when number of elements is small(exact size depends) and most insertions are at end or beginning rather in the middle.

\item \emph{Sort the linked list layout}( \cite{Madduri2009d} ): Nodes of a linked list are normally allocated in non-contiguous memory locations. In the worst case, a cache miss occurs on every load operation. By sorting the elements of the linked list in a sequential memory address order, cache misses can be reduced as consecutive nodes may occupy the same cache line. Further, this gives PE's in Multicore processor(with out-of order capability), opportunity to hide memory latency for successive node accesses by overlapping multiple load operations by issuing memory loads in advance.

\item \emph{Software prefetching}: Also called data level speculation, prefetching is a  technique to reduce memory latency by advancing load instructions in anticipation that results of such loads are very likely to be required soon. Prefetching in software requires ISA support and is usually done by compiler by inserting non-blocking load instructions. Programmer can also implement prefetching in software by using inline prefetch assembly instructions or API specific prefetching functions. Different studies \cite{KarlStens2000, ZhangTorrellas1995, AMD2008} have been done on data prefetching for irregular data structures. They basically either try to estimate prefetch distances or use an auxiliary data structure. Prefetching techniques can be hard to implement for irregular applications due to obvious reasons: difficulty in predicting traversal-orders and prefetch distances and many times not enough work to hide memory latency. Still, If feasible and done carefully, prefetching can yield good performance improvements.

\item \emph{Avoid wasting cache space}: Choose member data type according to range of actual data values so that more data can be read or written per cache line access.

\item \emph{Reduce cache pollution}\cite{Park.etal2004}: \emph{Cache pollution} is related to repetition of cold misses for same memory location. As long as relevant data is there in the cache, use it there and then only as much as possible before it gets replaced. This also reduces the bandwidth overhead due to reduced data transfer round trips between memory and processor

\item \emph{Cache aligned data structures}(\cite{Fry2008}):
All memory accesses which are not aligned to boundary of cache line size can be twice slower than aligned memory access because extra cache lines need to be fetched to load all data. As extra cache lines are fetched, conflict misses also increase. To avoid this, pack data structures with members such that they fit evenly on cache line boundaries. One can also force cache aligned memory allocation but forcing aligned allocation can create problem of memory fragmentation leading to higher memory consumption and possibly performance degradation.

\item \emph{Hot fields should be placed together and accessed in order of layout }\cite{Srikant2002, Fry2008}: Arrange layout of member items of a structure in order of frequency of access and also order in which they will be accessed. This will divide the memory layout of structure in groups of hot fields(higher access frequency) and cold fields(rarely accessed). As a result, group temporal/spatial re-usability increases as clubbed items are more likely to be accessed together. With this, the chances that clubbed items occupy the same cache line increases. Also by accessing the items within hot field in order of their layout, the prefetch mechanism can work more accurately.

\item \emph{Loop optimizations}\cite{Srikant2002}: Many loop specific optimizations can be used for structures other than linked lists. \emph{Loop-blocking or strip-mining }makes tiles or blocks of data size related to cache line size and reduces capacity misses. \emph{Loop interchange} improves spatial locality by interchanging nested loops to traverse data in continuous dimension(innermost array dimension in C). \emph{Loop fusion} amortizes cache misses by fusing two independent loops with compatible indexes. \emph{Loop fission} reduces splits a loop into many loops to reduce the memory footprint of loop iteration thereby reducing capacity misses. \emph{Loop unrolling} is a technique that reduces loop overheads, increases ILP, reduces control flow and improves cache utilization by data reuse. Unrolling is effective when loops enclose low to moderate amount of computation and contain very few function calls.
\end{enumerate}

\section{Multicore Platform}
\label{sec:multicore}
As explained in earlier sections, many architecture aware optimizations exists which can be effectively utilized by understanding details of the underlying processor architecture. With the aim of getting efficient parallel implementations of Triad census algorithm for Multicore processors, in this section we present details of architectural and micro-architectural features of the Intel Multiprocessors used for evaluating Triad Census algorithm. The reader is encouraged to read \textbf{Appendix~\ref{app:concpar} \nameref{app:concpar}} and \textbf{Appendix~\ref{app:procarchmicro} \nameref{app:procarchmicro}}, as they cover significant state of the art concepts and terminology relevant for this section and the following sections.
\\
\\
\subsection{Platform Overview}
\label{subsec:multicoreplatformoverview}
For implementing and testing single-threaded and multi-threaded implementation of triad census algorithm, we used Intel Core 2 Duo(64 bit) processor. For the main optimizations, experiments and evaluation of different triad census algorithm implementations, we used high performance Dell 11G PowerEdge R510 server containing two \textbf{Intel Xeon E5630 Quad core} processors on two different sockets(Figure~\ref{fig:westmere}). The \emph{Westmere-EP} (Intel Xeon 5600 series) or \emph{Gulftown} is basically a die-shrink (32 nm) of previous Intel micro-architecture named \emph{Nehalem}(45 nm) and implements the \emph{Intel64 ISA}. Westmere is the "\emph{tick}" and Nehalem is the "\emph{tock}" in the Intel's tick-tock processor series. Tock is an entirely new micro-architecture while Tick is a new silicon process technology based on last Tock. Table~\ref{tab:multicoremachines} shows specifications of these two Multicore systems.

\begin{table}[!ht]
  \centering
  \caption{Details of Multicore Systems}
  \begin{tabular}{ | p{3cm} | p{4cm} | p{6cm} | }
  \hline
  \textbf{Machine} & \textbf{Core 2 Duo(64 bit)} & \textbf{Xeon E5630}\\ \hline
  \textbf{Type} & Mobile & Server\\ \hline
  \textbf{Code name} & Penryn &Westmere EP\\ \hline
  \textbf{Base Micro-arch}? &Core &Nehalem\\ \hline
  \textbf{Tick(Die shrink) or Tock(Micro})? &Die shrink &Die shrink\\ \hline
  \textbf{Architecture style} &SMP &NUMA\\ \hline
  \textbf{Sockets /Core per socket /HW threads per core} & 1/2/1 &2/4/2\\ \hline
  \textbf{Core frequency} &2 GHz &2.53 GHz /2.8 GHz\\ \hline
  \textbf{Cache hierarchy} & L1 - 32 KB Data , 32 KB Inst per core L2 - 2048 KB shared Both 8-way set associative All caches write back, non-inclusive, 64-byte line size &L1 - 32 KB Data/core , 32 KB Inst/Core L2 - 256 KB per core L3 - 12 MB shared between cores / socket (Smart cache - fully inclusive) All caches write-back, 64-byte cache line size\\ \hline
  \textbf{DRAM} & 3 GB DDR2 RAM (2-ch) &8 GB DDR3@1066 MHz (3-ch) / socket\\ \hline
  \textbf{Bus Frequency /Bandwidth (bidirectional)} & FSB - 800 MHz &QPI with 2 links @2.93 GHz QPI - 5.86 GT/s or 23.4 GB/s Memory - 25.6 GB/s\\ \hline
  \textbf{Hyper-threading} ? & No & Yes\\ 
  \hline
  \end{tabular}
  \label{tab:multicoremachines}   
\end{table}


\subsection{Intel Nehalem(Westmere-EP) Micro-architecture}
\label{subsec:xeonmicro}
We will now explain in detail the organization and details of the processor-memory system of this dual socket server(Figure~\ref{fig:westmere})\cite{Westmere2010}. The server contains two sockets each holding one Intel Xeon Quad core processor chip. A Xeon Quad core chip is a CMP having mainly two parts on the chip: \textbf{core} and \textbf{uncore}. Intel says the core/uncore approach is mainly taken to make whole chip design modular and  depending upon the requirements of domain in which the processor will be used, this approach enables it to make reusable, flexible, power efficient and scalable multi-socket multicore platforms with different number of sockets, processor cores, cache sizes, memory controllers, QPI ports etc. 

\subsubsection{Core}
\label{subsubsec:core}
The \textbf{core} part of Xeon CMP has four processor cores(Figure~\ref{fig:westmere}) and can run at maximum frequency (turbo) of 2.8 GHz. Each processor core has following features:

\begin{itemize}
\item Implements Intel64 ISA and supports two hardware threads based on Intel's implementation of Simultaneous Multithreading called Hyper-threading (HT)

\item Has separate L1 Data and Instruction cache and a combined L2 Data/Instruction cache. L2 is an exclusive cache w.r.t L1. Both L1 and L2 have write-back write policy.

\item Consists of following units: in-order \emph{front-end}, \emph{out-of-order execution-engine} and \emph{in-order retirement unit}.

\begin{itemize}
\item \textbf{Front-end} can decode four macro-instructions per cycle and issues four micro-instructions per cycle to the out-of-order execution engine.  Instruction stream from two SMT threads are decoded in alternate clock cycles. There is a second level branch predictor which improves branch prediction compared to earlier architectures. In comparison, recovery from misprediction is also faster. Similar to Core Micro-architecture, Macro-fusion and micro-fusion is used to improve front-end throughput to the out-of-order engine. 

\item \textbf{Out-of order Execution Engine} can receive up to four micro-ops per cycle from Front-end. Micro-ops go through register renaming, resource allocation and reservation station for re-ordering before up to six micro-ops per cycle can be dispatched through six dispatch ports to various pipelined execution units. Appropriate entry is made in 128-size re-order buffer for in-order retirement. So upto six operations can be executed per cycle which includes 1 load, 1 store address, 1 store data and three Arithmetic-Logic operations. Also, window for out-of-order execution is increased by increasing capacity of various buffers like reservation station (36 entries instead of 32), load buffers (48 instead of 32 ) and store buffers (32 instead of 20)

\item \textbf{In-order Retirement Unit} does retirement and write-back. It can retire upto four micro-ops per cycle. A reorder buffer with 128 entries allows retirement of completed instructions.
\end{itemize}

\end{itemize}
Overall length of processor pipeline is two stages longer than Intel Core Micro-architecture which enables higher exploitation of ILP. In addition, instructions for synchronizing concurrent access like lock, xchg execute much faster than Intel Core micro-architecture.

\subsubsection{Uncore}
\label{subsubsec:uncore}
The \textbf{uncore}\cite{DerekBachandSelimBilginRobertGreinerPerHammarlundDavidLHillThomasHuff2010} part of the chip mainly combines \textbf{L3 cache}, \textbf{QPI interconnect} and a \textbf{Northbridge circuit} all integrated on same die as the processor cores. Following are important features of Uncore:

\begin{itemize}
\item 12 MB of \textbf{L3 cache} as LLC. L3 cache is fully shared between four processors cores which means any single core may fill or access the full L3 cache if other cores are not using it. Also it is an inclusive cache i.e. it contains data from all L1/L2 caches in all processor cores. One advantage of the inclusive L3 cache is that, if a core knows that a data is not present in L3, then this data will not be present on other cores of the same chip effectively reducing snooping traffic between cores of the same chip.

\item An \textbf{Integrated Memory Controller} (IMC) that supports three 8-byte wide  DDR3-1066 bidirectional memory channels. Each channel supports memory bandwidth of 8 GB/s. This is unlike older Intel processor implementations where memory controller was always outside the processor chip ( on Northbridge chip ) and frequently was a source of bottleneck in multicores based on such configuration. In effect, also separates the the cache coherence traffic and memory access traffic enabling higher memory bandwidth compared to previous architectures.

\item Two \textbf{QPI controllers} along with QPI( Intel Quick path interconnect) port interfaces

\item Auxiliary circuits for system management and performance monitoring

\item \textbf{Global Queue(GQ)}: This component is a crossbar that acts as the sole interface between LLC(Last level cache), QPI Controller, Integrated Memory Controller and cores inside the Core part. In addition , it provides a Power control unit to dynamically control voltage and frequency of processor cores according to workload demands thereby enabling Intel's Turbo boost technology mechanism. GQ component also implements other features such as LLC cache controller, Cache coherency mechanism in conjunction with QPI controller. All data handled by GQ is at granularity of cache line size. Cache line requests come to GQ due to on-chip L2 cache misses, from remote socket and from I/O Hub. GQ maintains three different queues with different capacities to handle these requests: Write queue for memory store from local cores, Load queue for memory load requests from local cores and QPI queue for off-chip requests coming from remote socket or I/O hub.

\item \textbf{Cache coherence protocol} implemented in QPI is called \textbf{MESIF} and has an additional "\emph{Forward}" cache line state in addition to usual four MESI states which allows it to handle socket-to-socket coherence.
\end{itemize}

\subsubsection{Memory and Socket Communication}
\label{subsubsec:memorysocket}
\begin{figure}[!ht]
	\centerline{\includegraphics[scale=0.5]{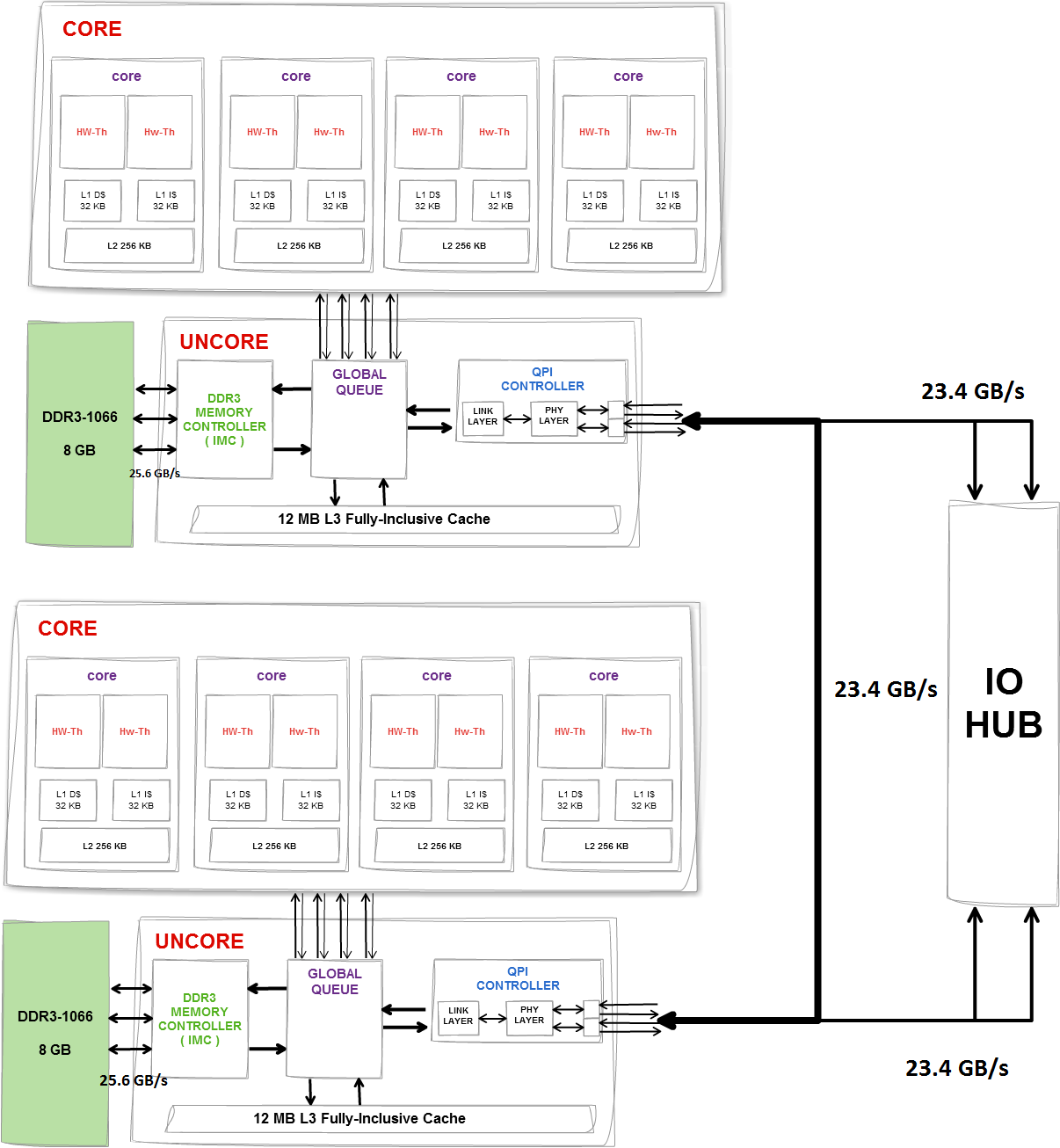}}
	\caption{Dual socket configuration of two Quad Core Intel Xeon E5630 processors with two hardware threads per core. Each Xeon chip can access its local DDR RAM directly through its IMC. This organisation makes the system a  2-way cc-NUMA with shared address space. Each Xeon processor chip is organised as Core and Uncore}\label{fig:westmere}
\end{figure}

In the server, each processor chip is connected to an 8GB DDR3 RAM module through its IMC interface. Maximum memory bandwidth per socket is 25.6 GB/s ( 3 ch x 8 GB/s per ch). As our Westmere-EP processor has two QPI ports,  first of the two QPI ports is used to connect the processor to its sibling Xeon Quad processor chips on the remote socket. The second of the two QPI ports is used to connect the processor socket to the I/O Hub. We will now discuss architectural features of Intel QPI
\\
\\
\textbf{Intel QPI Interconnect}\cite{Maddox2009} is a full-duplex, point-to-point and packetized high bandwidth low latency interconnect. At physical level, a QPI port is pair of unidirectional links - Tx and Rx link - that allows connecting a component to another component with a QPI port. Each Tx/Rx link is made up of 20 differential signal pairs plus a differential forwarded clock. Each differential signal pair carries a 1-bit of information, this implies 20-bits information can be carried by one link. Out of 20 bits, 16 bits are used to carry data payload per forwarded clock edge and remaining 4 bits are used for protocol/error information.
\\
\\
In Westmere-EP processors, each QPI link is operated at a clock frequency of 2.93 GHz. Also, the data links are operated at double-data rate, such that the data signals are asserted on both rising and falling edge of the clock. As a result, each QPI link can make 5.86 Giga Transfers per second ( 2 transfers/Hz x 2.93 GHz ). For a QPI link pair, this translated to aggregate 11.72 GT/s ( 2 links x 5.86 GT/s per link).  In terms of theoretical aggregate bandwidth, this comes out to be  23.44 GB/s (11.72 GT/s x 2 bytes per transfer ) for QPI link pair or unidirectional 11.72 GB/s per link. 
\\
\\
It can be seen in Figure~\ref{fig:westmere}, that when a chip accesses remote DRAM, this will incur higher latency and lesser bandwidth compared to accessing its local DRAM. This distributed shared memory makes this two CMP-DRAM arrangement a 2-socket cc-NUMA shared address space system. In the figure we can see the two NUMA nodes or memory domains.
\\
\\

\section{Tesla GPGPU Platform}
\label{sec:gpgpu}
GPGPU's(General Purpose GPGPU's) have come forward as the most promising widespread SPMD model based performance accelerators. Their ubiquitous presence in desktops, laptops and servers means availability of massive Manycore multithreaded multiprocessing that generally excel in highly data-parallel and compute intensive applications. The philosophy of GPGPU based computing is to use a heterogeneous CPU-GPGPU co-processing approach for accelerating performance for applications having a varying mixture of serial and parallel fraction. The approach is to accelerate highly data parallel part of the application using the fine-grained(hw threads) GPGPU processor and accelerate the sequential part of the application using the latency-optimized coarse-grained(hw threads) Multicore CPU. In other words, exploit fine-grained thread level parallelism(TLP) for highly data parallel codes on GPGPU and exploit Instruction level parallelism(ILP) plus coarse-grained TLP in inherently sequential(irregular) codes on Multicore CPU's. 
\\
\\
In this section, we first present overview of our NVIDIA Tesla C1060 GPGPU(General Purpose GPGPU) based hardware platform on which Triad Census is evaluated. Thereafter, we concentrate on understanding in some detail the important hardware features of the NVIDIA Tesla C1060 GPGPU. The software programming model for NVIDIA GPGPU's - \textit{called CUDA} - is discussed in Appendix~\ref{app:cudamodel}. In the last part of this section, we mainly discuss important differences between GPGPU's and Multicore. 

\subsection{Platform Overview}
\label{subsec:gpgpuplatform}
\begin{figure}[hb!]
	\centering
	\centerline{\includegraphics[scale=0.5]{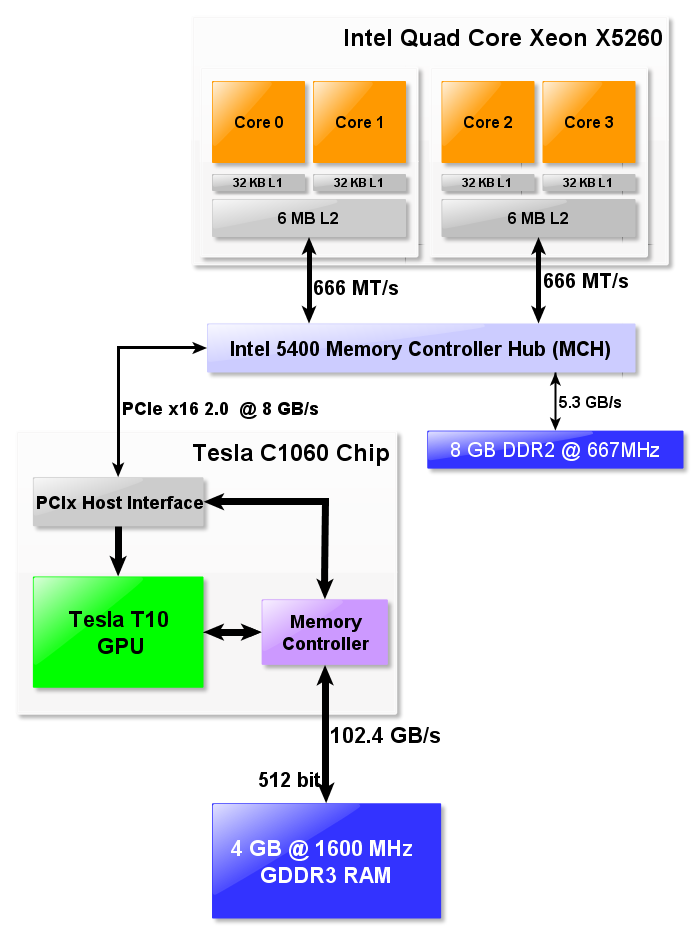}}
	\caption{Tesla C1060 GPGPU connected through PCIe 2.0 @ 8GB/s to Intel Quad Core(2core x 2socket) Xeon X5260 @3.33 GHz, per core L1-32 KB data cache, Per Socket 6MB L2 Unified Cache shared between two cores, 1333 MHZ FSB, and 8 GB DDR2@667 MHz RAM }\label{fig:gpgpucpusystem}
\end{figure}

For evaluating Triad Census on GPGPU, we used ICS workstation system which consists of a discrete NVIDIA Tesla C1060 GPGPU compute processor connected to Quad-Core Intel Xeon X5260 Multicore processor through a PCI-Express x16 2.0 interface. Intel Xeon X5260 Multicore processor is nicknamed Wolfdale-DP and is a die-shrink(tick) of Intel Core Micro architecture(Penryn). Figure~\ref{fig:gpgpucpusystem} shows the details of this platform.

\subsection{NVIDIA Tesla GPGPU Processor}
\label{subsec:teslagpgpu}
Tesla C1060 GPGPU belongs to 2nd generation Tesla architecture(T10/GT200b)\footnote{As of this writing, the state of the art GPGPU processor architectures are Fermi T20/GF100/GF104 \cite{Nickolls.Dally2010,Fermi2009} and Kepler GK104/GK11 \cite{Kepler2012} released in 2009 and 2012}. It is a 1.4 billion transistor chip made using 55nm technology node and consists of 240 CUDA processor cores(n) with processor core(\textit{shader}) frequency of 1300 MHz(f), graphics core(\textit{core engine}) frequency of 600 MHz, 1 TFLOPS (f x n x 3 flops/clock-cycle) of peak single-precision floating point rates, IEEE 754-2008 double precision 64-bit floating point arithmetic, 4GB of GDDR3 memory having clock rate of 1600 MHz with 512-bit (64-bit x 8) wide memory interface. It supports peak theoretical memory bandwidth of 102 GB/s (512/8 bytes x 1600 x $10^6$ cycles) between GDDR RAM and GPGPU device. CUDA compute capability of Tesla C1060 GPGPU is 1.3. \protect\footnote{CUDA compute capability describes CUDA features supported by the GPGPU. More advanced CUDA compute capabilities are 2.0 and 2.1 are available in Fermi GF100/104 GPGPU's}

\begin{figure}[hb!]
	\centering
	\centerline{\includegraphics[scale=0.65]{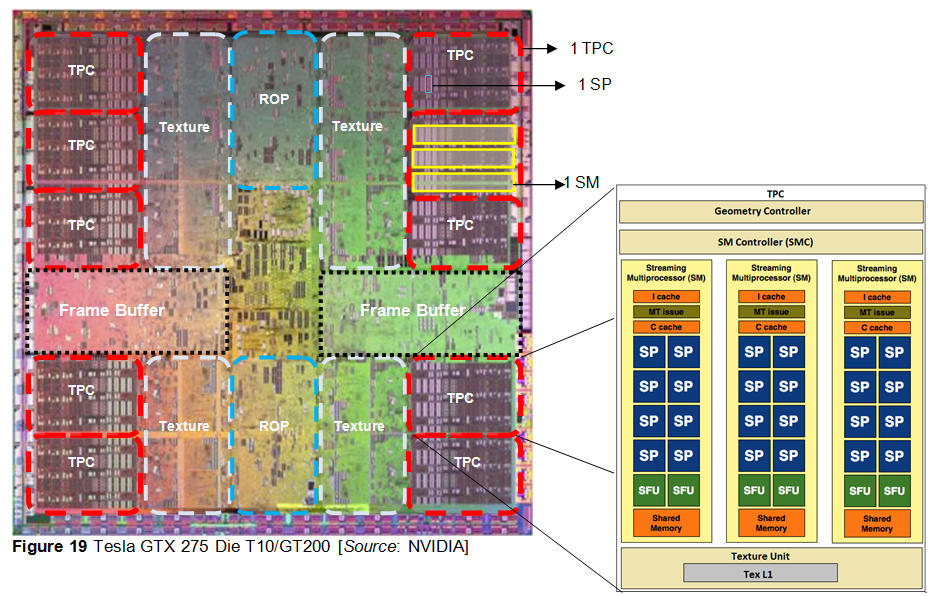}}
	\caption{Tesla C1060 processor die (T10/GT200 GPU Architecture). \textit{Source}: NVIDIA}\label{fig:gpucpudie}
\end{figure}

In terms of chip resource organization (see Figure~\ref{fig:gpucpudie}), Tesla C1060 GPGPU consists of a single \textbf{Streaming Processor Array }(SPA). The SPA consists of ten \textbf{Thread Processing Clusters }(TPC). Each TPC consists of three \textbf{Streaming Multiprocessors}(SM). Each SM consists of eight \textbf{Scalar Processors}(SP's or CUDA cores). Thereby each TPC consists of 3 SM x 8 SP/SM = 24 CUDA cores or SP's. As there are 10 TPC's, in total GT200 has 10 TPC x 24 SP/TPC = 240 CUDA cores. In total, 30720 hardware threads can be executed by the Streaming Processor Array(SPA), or in other words 3072 threads/TPC, 1024 threads/SM and 128 threads/SP.
\\
\\
\textbf{Each SM} has following additional features:
\begin{itemize}
\item 64 KB (32 bit x 16K) \textbf{Register file}
\item 16KB - \textbf{Shared Memory} which is connected to 8 SP's through a low-latency interconnect and is organised into 16 banks of 1KB each
\item \textbf{Instruction Cache} (I-Cache), 8 KB-Constant Cache (C-Cache)
\item Two \textbf{Special Function Units} (SFU's) each clocked at 1300 MHz used for transcendental functions, attribute interpolation and floating point MUL instructions. Each SFU contains 4 single precision floating point multipliers, so in total there are 8 single precision FP multiplier units per SM (1 FP Multiplier per SP).
\item One IEEE 754-2008 \textbf{Double-Precision floating point unit} multiplier supporting 1 Fused Multiply Add (FMA) per cc.
\item Multithreaded \textbf{Instruction Fetch and Issue Unit} (MT issue) supporting fine-grained interleaved multithreading (see Glossary). MT unit enables \textbf{SIMT} (Single Instruction Multiple Thread) by creating, managing, and scheduling hardware threads in group of 32 parallel threads called \textbf{Warp}
\item One \textbf{Warp Scheduler} (see later sections for more details)
\item Eight \textbf{Load/Store Units} (LD/ST) for executing Load/Store/Atomic Memory operations.
\end{itemize}

\textbf{Each SP} has following additional features:
\begin{itemize}
\item Fully pipelined in-order issue scalar processor with pipeline latency of 24 clock cycles.
\item One scalar Integer Unit (IU) supporting 32-bit precision multiply, 1 multiple-add (MAD) operation per cc and efficient 64-bit operations.
\end{itemize}

Total size of register file is 1920 KB (10 TPC x  3 SM/TPC x 64 KB/SM). Total size of shared memory is 480 KB (10 TPC x  3 SM/TPC x 16 KB/SM).

\subsection{Differences between Multicore and GPGPU processors}
\label{subsec:cpuvsgpgpu}
Before moving to the evaluation of Triad Census algorithm on our Tesla GPGPU, we point out some major differences between GPGPU's and Multicore CPU's
\begin{enumerate}

\item \textbf{Usage Context for Workload Processing}:  Primary usage of Multicore is as \textit{main processor} for running operating system and all software applications on top of it. Multicore processors are built on legacy single core ILP(Instruction Level Parallel) processors and can run both single threaded and multithreaded applications(exploiting coarse grained TLP). On the other hand, GPGPU's are used as \textit{co-processor} of the Multicore processor and are primarily meant for \textit{accelerating} highly multithreaded and compute intensive workloads. Although the GPGPU's focus on accelerating applications through Thread level parallelism(TLP), to an extent they can also extract ILP. 

\item \textbf{Hardware Threads}: GPGPU's provide massive hardware multithreading (fine-grained) in form of thousand of hardware threads(30720 Tesla T10/GT200). Multicores typically have few number (multiples of 2 say 2, 6, 8 etc.) of hardware threads (see SMT or CMT) per processor core. Hardware threads in GPGPU's are light-weight and require very few cycles to generate and schedule, whereas threads in Multicores are comparatively coarser and thus require more allocation/de-allocation cycles typically multiples of 10 cycles. 

\item \textbf{Chip resource utilization}: GPGPU's tend to use most part of chip area for in-order small processor cores, register file, thread scheduler and lesser area for on-chip memory hierarchy, control logic.  In Multicore processors, on-chip cache hierarchy and control logic like branch predictor, O-o-O scheduler occupies significant part of chip area whereas chip area dedicated for computation occupies relatively lesser area. ( Figure~\ref{fig:gpucpuarea} )

\begin{figure}
	\centering
	\includegraphics[scale=0.6]{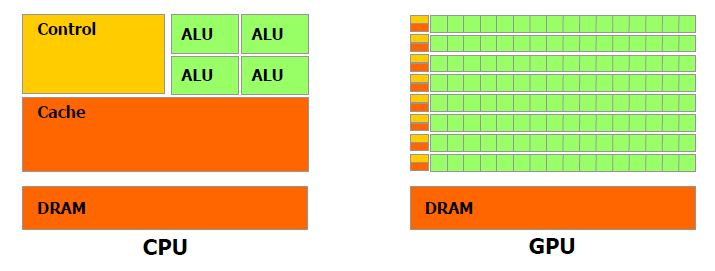}
	\caption{Single core CPU and GPGPU architecture comparison. \textit{Source}: NVIDIA CUDA C Programming guide}\label{fig:gpucpuarea}
\end{figure}

\item \textbf{SIMD vs. SIMT:} Multicores use SIMD (single instruction, multiple data) units for doing vector processing on vectorized registers by doing same operation on multiple data words. On the other hand GPGPU's employ SIMT (single instruction multiple thread) where bunch of threads(warps on NVIDIA GPU) execute same set of instructions in lock step, although it does permits arbitrary branching behaviour for individual threads.

\item \textbf{Memory Bandwidth}: for GPGPU's (>100 GB/S) memory bandwidth is many times (~10 or more) higher than memory bandwidth in Multicores (10-30 GB/s).
\item \textbf{Floating-point Throughput}: GPGPU's provide massive floating point operations throughput of the order of 1 TFLOP (Tesla T10) in single precision and more (Fermi T20) whereas Multicore processors have much lower maximum FLOPS. 

\item \textbf{Memory latency hiding}: GPGPU's use massive Hardware Multithreading along with ILP, combined with high memory bandwidth of GDDR memory to hide memory latency. Multicore processors employ sophisticated cache hierarchy, branch prediction, hardware prefetching, out of order processing and hardware multithreading(coarse grained) to hide memory latency.

\item \textbf{Programmability}: Multicores can be programmed much easily with little or almost no knowledge of the underlying processor hardware. On the other hand, programming GPGPU's requires learning a vendor specific programming framework (example CUDA) and requires programmer to have deeper knowledge of the  processor architecture.

\item \textbf{On-chip memory management}: In Multicores, on-chip cache hierarchy and register allocation is mostly handled by compiler and hardware with programmer having little to no control over on-chip memory allocation. In GPGPU's, allocation of on-chip memory such as registers can be controlled through compiler and usage of on-chip scratch pad memory(shared memory) can be explicitly managed by programmer.
\end{enumerate}

\chapter{Related Work}
\label{chap:relatedwork}
This chapter presents related work on sequential and parallel graph algorithm implementation on CMP's along with some discussion on various graph data structure representations as used in them. In particular, we focus on comparing our work with the only known previous work where Triad Census algorithm was accelerated on a high-end CMP(Cray XMT supercomputer).
\\
\\
There exists considerable amount of literature which describes work on algorithm, design, implementation and optimization of sequential/parallel irregular/graph applications for commodity Multicore and high-end CMP systems. \cite{Park.etal2004} studied cache-aware and cache-oblivious based data structure design techniques for various sequential graph algorithms. They demonstrate good execution speedups for Dijkstra's single-source shortest path algorithm and Prim's minimum spanning tree using cache-aware adjacency array representations. Further, they obtain significant speedups using cache-oblivious recursive implementation for Floyd-Warshall algorithm. They also combine adjacency array based approach and cache-oblivious approach and apply it to matching algorithm for bi-partite graphs.
\\
\\
Many authors have presented work on effective parallelization of irregular algorithms on commodity Multicore and Manycore processors. \cite{Runger2009} investigate number of parallelization strategies for mixed regular-irregular applications. The main focus of their work is to improve utilization of performance critical resources(cache memory, NUMA specific thread/data mapping) of the commodity Multicore processors(AMD Opteron, Intel Xeon). Authors in \cite{Xia2009} study effect of graph topology (random, sparse, high data locality) on performance of Parallel BFS performance on shared memory Multicores(Dual socket Quad core Intel Xeon and AMD Opteron). They propose an improved Parallel BFS algorithm which dynamically adapts itself(number of threads) to the variations in graph topology. Authors in \cite{Madduri2009c} evaluate number of basic dynamic parallel graph kernels(st-connectivity, BFS, betweenness centrality) on SUN Ultra Sparc T2 and IBM p5 570 SMP and obtain good speedups (22x,11x, 23x). As such algorithms involve run-time parallel insertion/deletions in large scale graphs of small-world networks, a number of novel graph representation techniques are proposed. To get high performance for runtime structural updates to graph data structure, they implement adjacency lists as resizeable adjacency arrays called \emph{Dynamic Arrays}. Another data structure that they experiment for adjacency representation is \emph{treap} which allows faster deletions than dynamic arrays. They use similar re-sizing strategy for treaps and dynamic arrays. To handle irregularity in graph topology, they make a hybrid adjacency representation which uses dynamic arrays for low degree vertices and treap for high degree vertices. 

A number of graph representations have been studied on GPGPU processor architectures. \cite{Harish2007} implemented and evaluated number of graph algorithms using \emph{Adjacency Array and Adjacency Matrix} based representations namely breadth first search, st-connectivity, single-source shortest path, all-pairs shortest path, minimum spanning tree, and maximum flow algorithms. Authors \cite{Pradesh2010} use \emph{Edge list} representations for implementing graph connectivity algorithms. Authors in \cite{Dehne2010} used optimized linked list representation for irregular graph algorithms like list ranking and connected components. They use arrays and sub-lists to improve performance bottlenecks associated with obtaining good performance on GPGPU's. Further they make some comparisons of speedups obtained on SMP's and GPGPU's. Some recent work\cite{HongKimOguntebiOlukotun2011, HongOlukotun2011} evaluates and compares performance of parallel graph algorithms(BFS) on different multicore processors such as Intel Xeon Multicores(Nehalem), SUN Ultra Sparc T2 and NVIDIA Tesla/Fermi GPGPU's 

There also exists number of studies\cite{Nieplocha2007, Madduri2009c, Chin2009, PingaliKulkarni2009, PingaliDonald2011} where irregular applications are evaluated on high-end CMP's such as SUN Niagara, SUN Ultra Sparc T2, Cray MTA-2 and Cray XMT. These studies demonstrate capabilities of such highly latency tolerant multithreaded processor architectures to effectively handle applications with high and varying degree of irregularities. Authors in \cite{Bader2006} demonstrate effectiveness of shared memory CMP called Cray MTA-2(40 core, no cache hierarchy but with 128 hw threads) to yield highly scalable performance for irregular graph algorithms(BFS and st-connectivity) even when processing graph networks with highly irregular topologies(random, scale-free, sparse). Similarly, authors in \cite{Madduri2009a} present an improved parallel algorithm of betweenness centrality graph kernel and obtain good parallel speedups(10x) on Cray XMT(16 PE). In the next section, we will present details of the only known previous work where Triad Census algorithm was accelerated.

\section{Acceleration of Triad Census on Cray XMT}
\label{relatedeval}
In their work\cite{Chin2009}, authors evaluated Multithreaded implementation of the original Subquadratic Triad census algorithm \cite{Batagelj2001} on Cray XMT supercomputer. Table~\ref{tab:thesissummary} summarizes the details of their work and also highlights(see Comments column) comparison of our work to it along with our contributions.

Authors evaluated four versions of Subquadratic triad census algorithm on Cray XMT. First, they implemented the original Subquadratic triad census algorithm with minor variations. Second, they implement a distributed task queue based implementation of their modified subquadratic triad census algorithm. In this implementation, each task was a canonical dyad and task queues were created sequentially. For synchronizing concurrent updates to census array, the implementation used fast atomics. For storing graph data, both implementations used compact data structures(no details). For parallelization, they relied on compiler to automatically parallelize loops(implicit parallelism). Third and fourth implementations basically augment loops in their first and second implementations with \textit{loop future} compiler pragma to designate each iteration of the loop to be executed by a thread. For experimentation and performance evaluation, they used one real world graph data set and two random graph data sets(generated using a tool). By iterating through phases of simulations--optimization (compiler directives, stream management, load balancing by interleaved loop iteration scheduling), they are able to obtain nearly linear speed-ups for most of their implementations on almost all data sets. They conclude that loop future versions scale better with high CPU utilization (load balancing)

\begin{center}
\begin{longtable}[c]{| p{2cm} | p{3.5cm} | p{4cm} | p{4.5cm} |}
\caption[Comparison of our work to previous work on accleration of Triad Cenus on Cray XMT]{Comparison of our work to previous work on accleration of Triad Cenus on Cray XMT} \label{tab:thesissummary} \\

\hline & \textbf{Previous work} & \textbf{Our work} & \textbf{Comments} \\ \hline 
\endfirsthead

\hline & \textbf{Previous work} & \textbf{Our work} & \textbf{Comments} \\ \hline 
\endhead

\hline \multicolumn{4}{|r|}{{Continued on next page}} \\ \hline
\endfoot

\hline \hline
\endlastfoot

\textbf{Hardware} & \textbf{Cray XMT} with 4 threadstorm multithreaded VLIW processors each supporting 128 hardware threads (streams) and 1 TB shared memory. Includes high-speed interconnect and network 3D-Torus topology. Fine-grained multithreading - thread management and synchronization. & \textbf{AlaRI server as Multicore} 
Dual socket Intel Quad Core Xeon E5630 with 2 hardware threads per core and 16 GB shared memory in cc-NUMA (8GB/socket) configuration. Socket-to-socket and socket-memory communication through Intel QPI interconnect.
\newline
\newline
\textbf{ICS Workstation with NVIDIA Tesla GPGPU}\newline ICS Workstation has NVIDIA Tesla C1060 GPGPU processor with 4GB of GDDR3 memory. The chip has 240-scalar processor cores arranged as 30 Streaming Multiprocessors(SM) and supports maximum 1024 hardware threads per SM. This GPGPU is connected to Intel Quad core Xeon X5260 processor with 8 GB memory through a PCIe 2.0 x16 interface. & Cray XMT belongs to class of Supercomputers. They do not rely on exploiting reference locality and instead leverage performance gains from latency hiding (highly latency tolerant) through fine-grained multithreading. Out of many Multithreaded processor based systems in the market, Cray XMT's have been found to fit well the requirements of latency sensitive irregular algorithms, particularly irregular graph algorithms. In the past, many data intensive irregular algorithms \protect\footnote{\href{http://cass-mt.pnnl.gov/docs/XMTOverviewSC08BOF.pdf}{Source: PNNL, US}} \protect\footnote{\href{http://user.cscs.ch/fileadmin/user_upload/customers/CSCS_Application_Data/Files/Presentations/Courses_Ws_2011/CRAY_XMT_Programming_Jun11/XMT_Tutorial_CSCS_1.pdf}{Source: CSCS, CH}} have performed well on this machine architecture.
\newline
\newline
Xeon Multicore is a server class machine. Multicores use cache hierarchy, branch predictor, hardware prefetching, out-of-order processing and coarse grained thread level parallelism for hiding memory latency. They are designed for general purpose applications.
\newline
\newline
Tesla GPGPU is a graphics hardware designed for use as co-processor. GPGPU's use massive fine-grained multithreading and high GDDR memory bandwidth for hiding and reducing memory latency. Designed for data parallel applications.\\ \hline

\textbf{Programming environment} & Distributed shared memory model(shared address space). Parallelizing C/C++ Cray compiler automatically parallelizes loops(also nested). Compiler generates special load store instructions and VLIW instructions(upto three operations - control, arithmetic, memory). Pragmas for controlling multi-threading and Canal profiling tool.
& \textbf{Multicore} \newline
- Implemented using C, Pthread and gcc compiler \newline
- For application profiling, Intel Vtune Amplifier was used
\newline
\newline
\textbf{Tesla C1060}\newline
- Implemented using C++/STL and C/CUDA\newline
- Application profiling was done using CUDA computeprof command line profiler & No comments \\ \hline

\textbf{Data sets} & Only one real world graph data set used - patents. All other graph sets were synthetic graphs generated using some tool. & Data processing on CMP's, Multicores and GPGPU's takes place on real world data and hence we choose real world graph data sets like Amazon, Google Webgraph, Slashdot, Patents, Actors(IMDB). All data sets obtained from from Stanford Large Network Dataset Collection and pajek collection	& Performance evaluation based on real world graph networks increases reliability of our speedup results \\ \hline

\textbf{Branch Optimizations} & No specific details in the work & We identified and optimized some branches (two if statement and two function calls) which were unnecessary and used precomputed values to improve performance (Version v0.4) & Improved speedup by upto 1.23x on Xeon Multicore (on GPGPU's we have not measured improvement as we already included this optimization in initial implementation) \\ \hline

\textbf{Graph data structures} & Linked list and compact data structures(exact details not specified) & \textbf{Multicores}\newline
- Linked list and cache blocked linked list(block size optimized after experimentation)\newline
- Implemented a novel Sorted insert algorithm for cache blocked linked lists\newline
\newline
\textbf{Tesla GPGPU}\newline
- All graph data was first constructed on CPU using dynamic arrays. This graph data is then transformed to CRS format for processing by GPGPU & \textbf{Multicores}\newline
With cache blocked linked list our sequential implementation ran 1.03x to 6.04x faster, with 3.1x average speedup\\ \hline

\textbf{Memory challenges} & \textbf{Cray XMT}\newline
1 TB memory with Cray specific memory allocator	& \textbf{AlaRI Xeon Multicore}\newline
- 16 GB (2-way NUMA) with standard non- NUMA-aware memory allocator\newline
\newline
\textbf{ICS Tesla GPGPU}\newline
- 4 GB GDDR3 with no support for dynamic memory allocation & In our GPGPU C++/STL/C/CUDA Triad Census code, we implemented technique (with almost zero overhead) to run triad census algorithm without needing any dynamic allocation. This enabled us to accommodate big graph data sets within GPGPU memory limits. Our changes can be ported to any class of GPGPU accelerators can give good performance in comparison to scenarios in which frequent dynamic memory allocation/de-allocation occurs at runtime. \\ \hline

\textbf{Synch. of concurrent updates to triad census algorithm} & \textbf{Cray XMT}\newline
Fast atomics used for synchronizing concurrent updates to shared triad census array	& \textbf{Intel Xeon Multicore}\newline
Experimented with two approaches:\newline
- shared triad census array updated through atomics\newline
- local triad census array per thread (completely decoupled threads)\newline
\newline
\textbf{Tesla GPGPU}\newline
We experimented with two approaches:\newline
- shared triad census array stored in global memory and protected with fast atomics\newline
- per thread block shared triad census array protected through atomics (completely decoupled thread blocks)	& \textbf{Multicores}\newline
We found that second approach i.e. Local census array per thread is generally faster\newline
\newline
\textbf{Tesla GPGPU}\newline
We found that the per thread block triad census array approach gave us on an average 2x speedup \\ \hline

\textbf{Load Balancing} & \textbf{Cray XMT}\newline
Distributed task queue approach	& \textbf{Multicores}\newline
Distributed task queues. We identified different load balancing strategies for triad census algorithm. Refined authors load balancing strategy and brought the implemented load balancing closer to ideal load balancing without any overhead\newline
\newline
\textbf{Tesla GPGPU}\newline
Centralized task queues in CRS format & Our refined load balancing strategy improves balancing of workload. \\ \hline

\textbf{Speedups} & \textbf{Cray XMT}\newline
Good linear or nearly linear speedups & \textbf{Multicore and GPGPU}\newline
Linear and Super linear speedups obtained due to sequential optimizations. Some of the speedups were sub-linear for highly sparse graphs where Multicore and GPGPU's memory couldn't beat Cray XMT's ability to hide memory latency	& \textbf{Multicore}\newline
Super-linear speedups included \textbf{56x(google) , Amazon(31x) and Slashdot(25x)}. Sub-linear speedups included \textbf{Patents(8.4x) and Actors(11x)}\newline
\newline
\textbf{Tesla GPGPU}\newline
Superlinear speedups of \textbf{58.4x(google), 38.5x(Amazon) and Slashdot(22x)}. Sublinear speedups of \textbf{4.2x(patents) and 8.5x(Actors)} \\ 
\end{longtable}
\end{center}

\chapter{Evaluation of Triad Census on Intel Multicores}
\label{chap:multicoreval}
In this chapter we discuss implementation and evaluation of the Subquadratic Triad Census algorithm on Intel Multicores. As said before, for implementation and testing we used Intel Core 2 Duo processor based system. Later for optimizations, experiments and evaluation we used a system containing two Intel Xeon processors.

\section{Methodology}
\label{sec:multicoremethod}
We briefly explain the overall methodology we adopted for accelerating triad census algorithm starting:
\begin{itemize}
\item Start with a sequential triad census implementation
\item Obtain baseline results on Intel core 2 duo processor and Intel Xeon processor based system
\item Profile the sequential implementation to find function hotsposts where the algorithm is spending most of the time. Analyse the profiling results and identify prospective sequential optimization's
\item Evaluate Sequential optimization's to accelerate sequential performance
\item Create a Multithreaded implementation of optimized sequential implementation
\item Profile Multithreaded implementation and identify potential sources of performance bottlenecks. Analyse the source of bottlenecks in multithreaded implementation and identify corresponding optimizations to alleviate them
\item Implement and evaluate the optimizations to further accelerate performance of the Multithreaded triad census algorithm
\end{itemize}

Table~\ref{tab:graphdatasets} shows the graph data sets that are used for evaluation of Triad Census algorithm on our Multicore and GPGPU based machines.

\begin{table}
  \centering
  \caption{Graph data sets used for evaluation}
  \begin{tabular}{ l | c | c | c | p{5cm} }
  \hline
\textbf{Data Set}& \textbf{Vertices}  & \textbf{Edges} & \textbf{Directed ?} & \textbf{Description} \\ \hline \hline
Actors\cite{NDactorwww2004} & 520223 & 2940808 & No & Actor network data based on imdb.com.\\ \hline
Patents \cite{Patent2001} & 3774768 & 16518948 & Yes & The NBER U.S. Patent Citations Data File, version 2001.\\ \hline
Amazon\cite{Amazon2003} & 403394 & 3387388 & Yes & Amazon product co-purchasing network from June 01 2003\\ \hline
Slashdot \cite{SlashdotGoogle2008} & 82144 & 549202 & Yes & Slashdot Zoo signed social network from February 21 2009\\ \hline
Google \cite{SlashdotGoogle2008} & 916428  & 5105039 & Yes & Webgraph from the Google programming contest, 2002\\ \hline
eatSR.net \cite{EatSR2003} & 23219 & 325589 & Yes & Pajek Edinburgh Associative Thesaurus, 2003\\ \hline
NDwww.net \cite{NDactorwww2004} & 325729  & 1497135 & Yes & Notre Dam web-page network\\ \hline
  \end{tabular}  
  \label{tab:graphdatasets}
\end{table}


\section{Sequential Triad Census}
\label{sec:triadseq}
We initially obtained a sequential implementation of Triad Census algorithm(see figure~\ref{fig:triadcensus} and figure~\ref{fig:TriadCode} in chapter~\ref{chap:background}) from authors \cite{Chin2009}. This implementation is based upon original Subquadratic Triad Census algorithm \cite{Batagelj2001}. The implementation is written in C programming language and uses linked lists based adjacency list data structure for graph representation.

\subsection{Implementation details}
\label{subsec:triadseqimpl}
The graph is read from an input graph file which contains list of vertices and edges(or arcs) of the undirected(directed) graph. The graph file normally lists \textbf{Order} of graph data set explicitly and optionally lists \textbf{Size} of the graph. Each vertex and edge(arc) has an optional text label. Graph data sets differentiate between directed and undirected graphs through different keywords: \textit{arcs} for directed  graphs and \textit{edge} for undirected graphs.

The read graph is stored as an \textbf{array of Vertex objects} (AOS or Array of structures). Each \textbf{Vertex object} has a text label and a stores pointer to linked list of \textbf{Edge objects} incident with it in the outward direction i.e. \textbf{Vertex object} is the \emph{tail} of all the arcs represented by its list of Edge objects. An \textbf{Edge object} represents a directed edge. It has a text label and stores index of destination vertex to which it is connected i.e. destination vertex is the \emph{head} of the arc represented by the \textbf{Edge object}. To access the directed edge list of a vertex, we simply index the array of \textbf{Vertex} objects by the vertex id.

\begin{figure}[ht!]
	\centerline{\includegraphics[scale=0.5]{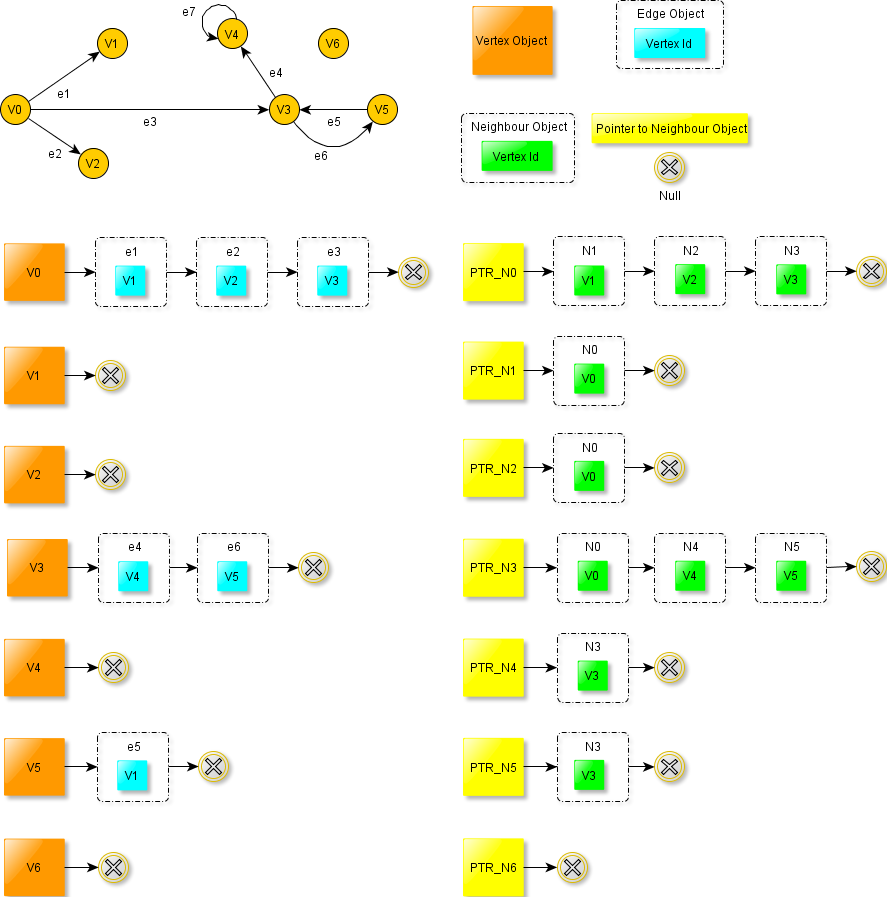}}
	\caption{Top Left: shows a digraph of order=7 and size=7.  Top Right: shows key to bottom two diagrams. Bottom Right: shows array of neighbour list pointers indexed by vertex id.  Bottom left: shows array of Vertex objects indexed by vertex id}\label{fig:seqtriadds}
\end{figure}

The Triad Census algorithm also requires usage of open neighbourhood set for each vertex of the graph. For this  an \textbf{array of neighbour lists} (array of list pointers) structure is used in which each element represents a pointer to open neighbourhood list of a given vertex. Each neighbour list is a \textbf{linked list of Neighbour objects}. Note that each \textbf{Neighbour object} still represents a vertex. To access the open neighbourhood list of a vertex, we can simply index the array of neighbour list pointers by the vertex id. Figure~\ref{fig:seqtriadds} shows the graph data structure.

\subsection{Baseline results}
\label{subsec:triadseqbase}
Sequential Triad Census code was instrumented to get wall elapsed time with millisecond resolution. The code was then compiled with settings shown in table~\ref{tab:compilesettings}.

\begin{table}[h!]
  \centering
  \caption{Compilation settings for Sequential Triad Census implementation}
  \begin{tabular}{ l | c | c | p{3cm} }
    \hline
    Machine & Compiler & Compilation flags\\ \hline \hline
    Intel Core 2 Duo(x86-64 ISA) & gcc 4.4.1  & -c -m64 -pipe -g Wall -O3 \\ \hline
    Dual-socket Intel Xeon E5630 & gcc 4.1.2  & -c -m64 -pipe -g Wall -O3\\
    \hline
  \end{tabular}  
  \label{tab:compilesettings}
\end{table}

The sequential implementation was then executed with many different graph data sets on both Intel Core 2 duo and Intel Xeon server to obtain baseline results(v0.1). Table~\ref{tab:seqoptimresspeedup}(\textbf{row with Version v0.1}) shows the initially obtained baseline results for different graph data sets on Intel Xeon E5630 Westmere-EP platform.

\subsection{Performance profiling and analysis}
\label{subsec:triadseqprofile}
To get insights into performance bottlenecks in sequential Triad Census implementation, we performed its code profiling on the Intel Core 2 duo machine. The profiling was done for many data sets. Due to space constraints, here we will present profiling data only for three data sets namely eatSR.net, NDwww.net and Actors. Although the profiling experiments presented here are specific to respective data sets, but the derived analysis results are general and are thus applicable to the other data sets.
\\
\\
Initially, we did profiling using \textbf{gprof} tool to identify hotspots at application level(function). Here, the Sequential Triad Census implementation was compiled with highest(-O3) optimization level. Below is a gprof execution time trace for data sets \emph{eatSR.net} and \emph{NDwww.net}.

\begin{verbatim}
eatSR.net
  %   cumulative  self               self    total
time  seconds     seconds  calls     s/call  s/call  name
52.45 40.77       40.77    189408983 0.00    0.00    is_edge
33.17 66.56       25.79    34415992  0.00    0.00    add_neighbor
9.08  73.62       7.06     20783877  0.00    0.00    is_neighbor

NDwww.net
  %   cumulative  self               self    total
time  seconds     seconds  calls     s/call  s/call  name
76.09 243.29      243.29   57670691  0.00    0.00    is_neighbor
21.00 310.44      67.15    7187328   0.00    0.00    add_neighbor
1.89  316.49      6.05     348535010 0.00    0.00    is_edge
\end{verbatim}

As can be observed above, in one of the traces, the function \emph{is\_edge} is the hotspot while in other trace the function \emph{is\_neighbor} is the hotspot. Also, \emph{add\_neighbour} function occupies the same relative position in terms of ranking of hotspot in both traces. These three hotspot functions basically traverse linked list either searching for an element(\textit{is\_edge} and \textit{is\_neighbor} functions have same code but traverse on different lists) or finding a position where element is to be inserted (\textit{add\_neighbor}). 
\\
\\
To have a deeper insight in the actual source of bottleneck, we profiled using another hotspot profiling tool named \textbf{Zoom} \cite{RotaterightLLC} which allows static and dynamic profiling of an application. Zoom allows collection of statistics at different levels (process, module, function, code line level, assembly instruction level). Static profiling statistics are processor architecture specific and include statistics like distribution of assembly instructions, nominal instruction timing(latency:timing) and likely resulting pipeline stalls during execution. Dynamic statistics are collected during runtime and include execution time(self, total) at multiple levels(module, function, code line level, assembly instruction). User needs to choose or create a profile configuration in order to get the dynamic statistics. Profile configuration provides what details on how the tool should sample(time based or event based). Event based profile configurations include cache miss and retired instruction profile
\\
\\
We profiled with \emph{unoptimized} (w/o -O3 compilation) Sequential Triad Census implementation on \emph{eatSR.net} data set using Zoom's event-based profile configuration called \textit{cache miss configuration}: A sample is taken every 1,000 occurrences of event "Last Level Cache Misses". The last level cache here represents the level L2 of cache in Intel Core 2 Duo processor. Below is a partial trace of statistics collected from Zoom.

\begin{center}
\begin{verbatim}
=====================================================
Hotspot - Process SubQuadTriads [20424]
=====================================================
Self         Total      Symbol          Module
2.5min       2.5min     is_edge         SubQuadTriads
13.3sec      14.3sec    inc_adj_union   SubQuadTriads
=====================================================
\end{verbatim}
\end{center}

Out of the total 2.9 min, function \emph{is\_edge} alone takes 2.5 min. Next, \emph{inc\_adj\_union} function takes 13.3 sec. Below tables show elapsed execution time at source and assembly code levels for the \emph{is\_edge} and \emph{inc\_adj\_union} functions.

\begin{verbatim}
is_edge code trace for eatSR.net data set
------------------------------------------------------
Source Line     Time     Source Code
------------------------------------------------------
line 0          2.4min   if(v2 < iptr->dest_node){
line 1          5.2sec      iptr = iptr->next ;
line 2                   }
------------------------------------------------------

is_edge assembly instruction trace for eatSR.net data set
-------------------------------------------------------------------------------
Time       Address    Stalls   Instruction                     Source Line
-------------------------------------------------------------------------------
            + 0x41              mov rax,[rbp-0x8]               line 0
            + 0x45              mov rax, [rax+0xc8]             line 0
2.4min      + 0x4c    2         cmp rax, [rbp-0x28]             line 0
80.8ms      + 0x50              jbe 0x401960 <is_edge + 0x63>   line 0
99.3ms      + 0x52              mov rax, [rbp-0x8]              line 1
22.9ms      + 0x56    2         mov rax, [rax+0xd0]             line 1
3.8sec      + 0x5d    2         mov [rbp-0x8], rax              line 1
1.3sec      + 0x61              jmp 0x40197f <is_edge + 0x82>   line 1
-------------------------------------------------------------------------------

inc_adj_union code trace for eatSR.net data set
-------------------------------------------------------------------
Source Line     Time       Source Code
-------------------------------------------------------------------
line 0          1.2msec    while (iptr) {
line 1          13.3sec    if (iptr->dest_node != exc_neighbor1) {
-------------------------------------------------------------------

inc_adj_union assembly instruction trace for eatSR.net data set
--------------------------------------------------------------------------------
Time       Address    Stalls    Instruction                         Source Line
--------------------------------------------------------------------------------
566.4us    + 0x9e               mov rax, [rbp-0x20]                 line 1
4.4ms      + 0xa2     2         mov rax, [rax]                      line 1
13.3sec    + 0xa5     2         cmp rax, [rbp-0x28]                 line 1
99.8us     + 0xa9               jz 0x40205d <inc_adj_union + 0xdc>  line 1
--------------------------------------------------------------------------------
\end{verbatim}

In the above \emph{code traces}, one can see that we are basically traversing linked list(loop) and making a decision(if statement) based on data value read from current node of the linked list. For both functions, the first four lines of \emph{instruction traces} are similar, as they contain two move - \textit{mov} instructions followed by a \textit{cmp} instruction and then a conditional jump - \textit{jbe/jz} instruction. 
\\
\\
Consider for the time being the source code \textit{line 0} of function \textit{is\_edge} and four lines of assembly instruction trace corresponding to this source code line. Among the four assembly instructions, the first is a move instruction which makes a memory read request(\textit{iptr}) and writes it to \textit{rax} register. Based on the value in rax register, the second move instruction  makes a memory read request(value \textit{iptr->dest\_node}) and writes it to rax register. Clearly, there is a RAW dependency and also WAW dependency between these two move instructions. A WAW dependency also exists between the second move instruction and following compare instruction leading to pipeline stalls. Further, due to poor locality of linked lists, it is very likely that cache misses occur when each of these three instructions(two move and one compare) make a memory read request (\textit{iptr}, value \textit{iptr->dest\_node}, value \textit{v2}). There is also a RAW dependency between compare \textit{cmp} and the jump instruction \textit{jbe}, which likely causes jump instruction to stall as it cannot resolve its target branch until the compare instruction writes its result to the appropriate flag registers(ZF or CF). The same behaviour can be observed for \textit{inc\_adj\_union} function.
\\
\\
Later, we ran hotspot analysis on Actor graph data set using \textbf{Intel Vtune Amplifier} tool \cite{IntelVtune}. This data set takes a total 104.6 sec CPU execution time. From the trace, the main bottleneck function is \textit{triad\_code} which consumes nearly 71 sec. 

\begin{verbatim}
Thread / Function / Call Stack            CPU Time  Function (Full)
main                                      104.570s
  triad_code                               70.193s  triad_code
  make_triad_census                        15.457s  make_triad_census
  inc_adj_union                             7.692s  inc_adj_union
  add_neighbor                              6.950s  add_neighbor

triad_code code trace for Actors data set
Line    Source                           CPU Time
773     if (is_edge (verts, u, w)) {      9.121s
774         edges[2] = 1;                 0.015s
775     }
776
777     if (is_edge (verts, w, u)) {     42.568s
778         edges[3] = 1;                 0.062s
779     }

triad_code instruction trace for Actors data set
Address   Line  Assembly                       CPU Time
0x401df1   773  cmp edi, dword ptr [eax+0xcc]    0.651s
0x401df7   773  jnbe 0x401e0f <Block 30>         5.091s
0x401df9   773  jb 0x401e03 <Block 28>           1.868s
0x401e38   777  cmp ebx, dword ptr [eax+0xcc]    0.093s
0x401e3e   777  jnbe 0x401e56 <Block 40>        24.164s
		Block 36:
0x401e40   777  jb 0x401e4a <Block 38>          15.955s
\end{verbatim}

From the above traces, we think there are mainly two factors that contribute to the bottleneck in function \textit{triad\_code}. First is the irregularity in control flow which arises due to one function call(\textit{is\_edge}) and enclosing data dependent branch instruction(conditional jump based on result of is\_edge function). Second factor contributing to poor performance is the performance of \textit{is\_edge} function itself which we already explained earlier. 
\\
\\
The profiling experiments were also done on other data sets and similar results were observed for most of these functions. As a result, we think the following three factors either individually or in combination, leads to poor performance of sequential Triad Census implementation:
\begin{itemize}
\item Occurrence of cache miss due to irregular memory access during linked list traversal
\item Irregular(data dependent) control flow
\item RAW/WAW dependencies leading to possible pipeline stalls during execution
\end{itemize}

All these sources of bottlenecks, limit the ILP of the Sequential Triad Census algorithm implementation. In summary, we note down following important conclusions from code profiling exercise:
\begin{itemize}
\item We obtained base line results of Sequential Triad Census algorithm on Xeon Multicore processor;
\item Most execution time is spent during linked list traversal for searching or manipulation which causes irregular memory access(cache misses), irregular(data-dependent) control-flow and pipeline stalls due to RAW/WAW dependencies;
\item We have identified the hotspot functions, code and instructions which are candidates for further optimizations. The following are the hotspot functions: \textit{is\_edge, add\_neighbor, is\_neighbor, inc\_adj\_union, triad\_code, make\_triad\_census}.
\end{itemize}

\subsection{Optimizing Sequential Triad Census}
\label{subsec:triadseqoptim}
To address the source of bottlenecks in on sequential implementation of Triad Census algorithm, we tried different optimizations to improve its execution time. Some of the optimizations resulted in substantial improvements while other optimizations produced results contrary to expectations and of course were not used. Although the experimental results presented here are for Actors data set, but we selected only those optimizations which also worked on other data sets. We did incremental optimizations which means that any presented optimization includes all previously applied optimizations. These incremental optimizations are marked as versions v0.2, v0.3, v0.4, v0.5. We refer to baseline results as v0.1. After evaluating these optimizations on Intel Core 2 Duo for Actors Data, we evaluated these different versions for different graph data sets on dual-socket Intel Xeon Multicore. Finally we choose only those optimizations which worked for most of the data sets. If any optimization that worked on Intel Core 2 duo but lead to bad results on Intel Xeon Multicore for some or many graph data sets, we decided not to use it in the final combinations of optimizations.
\\
\\
Throughout this optimization phase, we compiled our triad census application with Level 3 optimization level of GCC compiler. At this level, gcc automatically performs most of the optimizations. The baseline(v0.1) execution time for Actors graph data set on Intel core 2 duo was 103 sec. We will now  explain the optimizations we experimented with on Intel Core 2 Duo.
\\
\\
\textbf{Cache fitting the data structures (v0.2)}\\
Clearly, the adjacency based structures implemented using linked-lists make the algorithm sensitive to memory latency. In order to reduce the cache misses incurred while traversing the adjacency linked lists, we tried an optimization which aligns the structure to cache line boundary. In this cache alignment optimization, we basically try to fit size of data structure to a nearest multiple of cache line size. This is not the same as aligning dynamic allocation of memory block to cache line boundary. We first check the size and layout of each adjacency structure(\emph{Edge, Node, Neighbour}) using utility called \textbf{pahole} \protect\footnote{\href{http://www.ohloh.net/p/pahole}{http://www.ohloh.net/p/pahole}}and also used gcc compiler optimization "-Wpadded" to check which of the structures were padded by compiler to force alignment to structure boundaries. We rearranged the structure elements according to member size and access frequency(\emph{hot fields}) to reduce padding and any memory holes. The Edge and Node structures were then fitted to cache line multiples of 3 or 4 by decreasing/increasing the length of label strings in edge and node structures and accordingly. As we see in table~\ref{tab:optimcacheper}, in both cases Fitting (inc) and Fitting (dec) performance improvement is \textbf{1.14x} speedup w.r.t baseline, but we decided to choose Fitting (dec) optimization as it lowers the memory footprint.
\\
\\
\begin{table}[ht!]
  \centering
  \caption{Baseline structure characteristics}
  \begin{tabularx}{\textwidth}{c|c|X|c}
    \hline
    Structure & Size(bytes) & Cache lines(64 byte cache line) & Wasted(bytes)\\ 
    \hline
    Edge      & 208         & 4                               & 16 \\ \hline
    Neighbour & 216         & 4                               & 24 \\
    \hline
  \end{tabularx}  
  \label{tab:basecachefit}
\end{table}

\begin{table}[ht!]
  \centering
  \caption{After structure fitting (decreasing struct size, increasing struct size)}
  \begin{tabularx}{\textwidth}{c|c|X|c}
    \hline
    Structure & Size(bytes) & Cache lines(64 byte cache line) & Wasted(bytes)\\ \hline \hline
    Edge (dec, inc)      & 192, 256         & 3, 4                 & 0, 0 \\ \hline
    Neighbour (dec, inc) & 192, 256         & 3, 4                 & 0, 0 \\
    \hline
  \end{tabularx}  
  \label{tab:optimcachefit}
\end{table}

\begin{table}[ht!]
  \centering
  \caption{Performance improvement after applying optimization(Actors)}
  \begin{tabular}{ l | c | c }
    \hline
                  & Time(sec) & Memory(MB) \\ \hline \hline
    Baseline(v0.1)& 103 & 829\\ \hline
    Fitting (dec) &	90 & 776 \\
    Fitting (inc) &	90 & 989 \\
    \hline
  \end{tabular}  
  \label{tab:optimcacheper}
\end{table}


\textbf{Function inlining (v0.3)}\\
It's generally a good idea to inline the functions that are frequently called and have small code size. This results in overall reduction of function call-return overhead and regularizes the control-flow thereby reducing instruction cache misses. In addition, this gives compiler a bigger windows of instructions to extract ILP. At gcc optimization level 2 and 3, this optimization is automatically applied. Nevertheless, we still inlined the possible functions by giving compiler hints and compiled with "-Winline" compiler option to see which of the functions could not be inlined. Resulting performance improvement is \textbf{1.02x} speedup w.r.t previous optimization or \textbf{1.17x} speedup w.r.t baseline. Execution time now reduced to 88 sec.
\\
\\
\textbf{Eliminating branches by using Pre-computed Triad Types (v0.4)}\\
The Triad Census function Figure~\ref{fig:triadcensus} ) uses TriadCode function ( Figure~\ref{fig:TriadCode} ) for calculating all triads except null triads (type-1 triads calculated in line 29) and dyadic triads (type-2 and type-3 triads counted in lines 9-14). Also Lines 9-13 of the Triad Census algorithm already pre-compute the part of Line 1-2 calculations of TriadCode algorithm i.e. IsEdge(u,v)+2*(IsEdge(v,u). Instead of recalculating this sum again in TriadCode function, we can simply pass this pre-computed value to it(See Figure~\ref{fig:TriadCodeMod} ). This reduces the number of \emph{IsEdge} function calls in the TriadCode function from 6 to 4 i.e 33\% lesser function calls. This results in performance improvement of approximately \textbf{1.18x} speedup w.r.t previous optimization or \textbf{1.36x} speedup w.r.t baseline. Execution time now reduced to 76 sec.
\\
\\
> \emph{Changes in Subquadratic Triad Census Algorithm}
Between lines 8-9 we insert three new lines with the following code:
\begin{verbatim}
    IsEdge[0] = IsEdge(u, v)
    IsEdge[1] = IsEdge(v, u)
    precomputed_triad_type = IsEdge[0] + 2 * IsEdge[1];    
\end{verbatim}
Line 9 is modified to following:
\begin{verbatim}
    if IsEdge[0] && IsEdge[1] then
\end{verbatim}
Line 17 is replaced with following:
\begin{verbatim}
    TriadType = TriadCode(u, v, w, precomputed_triad_type)
\end{verbatim}
\begin{figure}
\fbox{\begin{minipage}{13 cm}
\begin{flushleft}
\textbf{Function}: TriadCode\\
\textbf{Input}: vertices u, v, w and pre-computed triad type value\\
\textbf{Output}: isomorphic or non-isomorphic triad code
\end{flushleft}
\begin{algorithmic}[1]
\State $TriadCodeVal = \textbf{precomputed\_triad\_type}$
\State $TriadCodeVal += 4 * IsEdge(u,w)$
\State $TriadCodeVal += 8 * IsEdge(w,u)$
\State $TriadCodeVal += 16 * IsEdge(v,w)$
\State $TriadCodeVal += 32 * IsEdge(w,v)$
\State $return ( IsIsomorphic ? TriadTable[TriadCodeVal]:TriadCodeVal ) + 1$
\end{algorithmic}
	\caption{Modified TriadCode Algorithm\label{fig:TriadCodeMod}}
\end{minipage}}
\end{figure}

\textbf{Cache blocking, faster searching and AOS-to-SOA (v0.5)}
As already discussed in sub-section~\ref{subsec:effadjlst} of chapter~\ref{chap:background}, unrolled lists or cache blocked linked lists is a good alternative to linked lists when it is critical to obtain good cache hit ratio. To further improve performance of Triad Census implementation, we replaced linked lists structures with cache blocked linked lists. The replaced list structures includes edge and neighbour lists. One notable algorithm named \emph{\textbf{Sorted Insert}} required significant implementation and optimization effort based on heuristics. This algorithm was required because of the requirement of merging(\emph{set union and set difference algorithm required while calculating set S}) two neighbour lists which mandated that both the neighbour lists participating in the merging(union) must be sorted. We tested cache blocked implementation with various block sizes for edge and neighbour lists. The optimal blocking value found was found to be 8 (edge blocking) and 16 (neighbour blocking).The resulting performance improved from 76 sec to 60 sec.
\\
\\
To take full advantage of the Streaming SIMD Extensions (SSE) based capabilities in Intel processors, the Intel optimization manual recommends to convert or \emph{swizzle} data layout from \textbf{AOS}(Array of structures) format to \textbf{SOA} (Structure of arrays) format \cite{IntelOptimManual}. Our cache blocked implementation initially used SOA approach for blocking edge lists:

\begin{lstlisting}[label=aosedge,caption=Original cache blocked edge structure in AOS format]
#define BLOCK_EDGES 6
typedef struct edge
{
   char edge_name[MAXLABEL];
   unsigned long long int dest_node;
}EDGE, *EDGE_PTR;

typedef struct edge_entry
{
   EDGE array_edges[BLOCK_EDGES];
   struct edge_entry *next;
   int last_index;
} EDGE_SET, *EDGE_SET_PTR;
\end{lstlisting}
We replaced AOS with SOA based implementation
\begin{lstlisting}[label=soaedge,caption=Modified cache blocked edge structure in SOA format]
#define BLOCK_EDGES 6
typedef struct edge_entry
{
   char arr_edge_names[BLOCK_EDGES][MAXLABEL];
   unsigned long long int arr_dest_nodes[BLOCK_EDGES];
   int last_index;
   struct edge_entry *next;
} EDGE_SET, *EDGE_SET_PTR;
\end{lstlisting}
As destination nodes are accessed more often than the edge names (example IsEdge(v1,v2) ), SoA format saves memory and prevents unnecessary fetching of edge names, thus resulting is more efficient use of the the caches and avoiding bandwidth wastage. 
\\
\\
Also, in some of the functions, searching was required. This includes following functions: \textit{add\_neighbor, inc\_copy\_neighbor\_set, is\_edge, is\_neighbor}. The conventional wisdom \protect\footnote{\href{http://en.wikipedia.org/wiki/Binary\_search\_algorithm}{http://en.wikipedia.org/wiki/Binary\_search\_algorithm}} says that for a pre-sorted array of length smaller than 100 elements, linear search would beat binary search. However in our cache blocked implementation, we that binary search resulted in slightly better performance. The execution time improved further from 60 sec to 50 sec. Overall, the three optimizations in v0.5 resulted in performance speedup of \textbf{1.52x} w.r.t previous optimization or \textbf{2.06x} speedup w.r.t baseline.
\\
\\
Besides, the above optimizations, we had tried few more optimizations which either proved detrimental to performance or didn't improve any performance. This includes

\begin{itemize}
\item Cache alignment optimization that was tried would allocate memory blocks on cache line boundaries by replacing function \textit{malloc} with \textit{posix\_memalign}. 
\item Software prefetching to reduce memory latency due to pointer chasing in linked list, implemented through assembly instructions(\textit{PREFETCHNTA}) and compiler based directives (\textit{\_\_mm\_prefetch})
\end{itemize} 

\subsection{Summary of sequential optimizations}
\label{subsec:seqsummary}
\begin{figure}[ht!]
    \centerline{\includegraphics[scale=0.8]{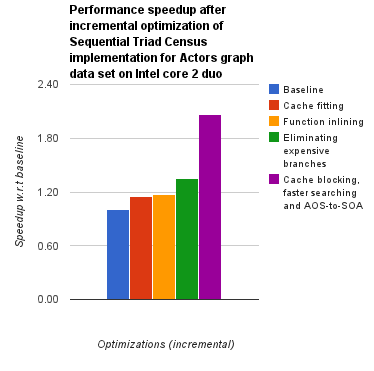}}
	\caption{Performance speedup of optimized sequential triad census for actors graph network on Intel Core 2 Duo with baseline execution time of 103 sec.}\label{fig:seqtriadoptimcore2}
\end{figure}

Figure~\ref{fig:seqtriadoptimcore2} shows the resulting performance improvement(speedup) from the above optimizations(incremental) for the Actors graph data set on Intel Core 2 duo machine. 

We then ran this optimized code on our Intel Xeon Multicore for different graph data sets. Table~\ref{tab:seqoptimresspeedup} shows the resulting speedups and Table~\ref{tab:seqoptimresper} shows the execution time as observed on the Intel Xeon Multicore. What we observe is that Actors and Patents data sets benefit most from optimizations v0.2, v0.3 and v0.4 while Amazon, Slashdot and Google data sets benefit most from v0.5 optimization. Overall the above sequential optimizations resulted in \textbf{average 3.3x speedup, maximum 6x speedup and minimum 1.3x speedup}.

\begin{table}[htbp]
  \centering
  \caption{Sequential Triad Census performance speedup and memory overhead results on Intel Xeon Multicore}
  \begin{tabularx}{\textwidth}{|X|c|c|c|c|c|c|}
  \hline
  \multicolumn{7}{|c|}{\textbf{Overall performance speedup w.r.t baseline(v0.1)}}\\
  \hline
  \textbf{Optimization} & \textbf{Version} & \textbf{Actors} & \textbf{Patents} & \textbf{Amazon} & \textbf{Slashdot} & \textbf{Google}\\ 
  \hline 
  Baseline & v0.1 & 1.00 & 1.00 & 1.00 & 1.00 & 1.00\\
  \hline
  Cache fitting & v0.2 & 1.13 & 1.16 & 0.99 & 1.01 & 1.00\\ 
  \hline
  Function inlining & v0.3 & 1.13 & 1.18 & 1.00 & 1.01 & 1.00\\
  \hline
  Eliminating expensive branches & v0.4 & 1.40 & 1.32 & 1.00 & 1.05 & 1.00\\
  \hline
  Cache blocking, faster searching, ASO-to-SOA & v0.5 & 2.00 & 1.37 & 4.58 & 2.91 & 6.06\\
  \hline
  \multicolumn{7}{|c|}{\textbf{Overall memory overhead w.r.t baseline(v0.1)}}\\
  \hline
  Baseline & v0.1 & 1.00 & 1.00 & 1.00 & 1.00 & 1.00\\ 
  \hline
  Cache fitting & v0.2 & 0.94 & 0.94 & 0.94 & 0.94 & 0.94\\ 
  \hline
  Function inlining & v0.3 & 0.94 & 0.94 & 0.94 & 0.94 & 0.94\\ 
  \hline
  Eliminating expensive branches & v0.4 & 0.94 & 0.94 & 0.94 & 0.94 & 0.94\\ 
  \hline
  Cache blocking, faster searching, ASO-to-SOA & v0.5 & 1.40 & 1.08 & 1.05 & 1.06 & 1.17\\
  \hline 
  \end{tabularx}
  \label{tab:seqoptimresspeedup}
\end{table}

\begin{table}[htbp]
  \centering
  \caption{Sequential Triad Census execution time(sec) and memory consumption(MB) results on Intel Xeon Multicore}
  \begin{tabularx}{\textwidth}{|X|c|c|c|c|c|c|}
  \hline
  \multicolumn{7}{|c|}{\textbf{Overall Execution Time(seconds)}}\\
  \hline
  \textbf{Optimization} & \textbf{Version} & \textbf{Actors} & \textbf{Patents} & \textbf{Amazon} & \textbf{Slashdot} & \textbf{Google}\\ 
  \hline
  Baseline & v0.1 & 60 & 324 & 394 & 393 & 12203 \\ 
  \hline
  Cache fitting & v0.2 & 53 & 280 & 396 & 389 & 12183 \\ 
  \hline
  Function inlining & v0.3 & 53 & 274 & 394 & 390 & 12180 \\ 
  \hline
  Eliminating expensive branches & v0.4 & 43 & 245 & 394 & 373 & 12160 \\ 
  \hline
  Cache blocking, faster searching and AOS-to-SOA & v0.5 & 30 & 237 & 86 & 135 & 2013 \\
  \hline
  \multicolumn{7}{|c|}{\textbf{Overall memory consumption(MBytes)}}\\
  \hline
  Baseline & v0.1 & 830 & 5320 & 960 & 170 & 1550\\ \hline
  Cache fitting & v0.2 & 778 & 5010 & 903 & 160 & 1457\\ \hline
  Function inlining & v0.3 & 778 & 5010 & 903 & 160 & 1457\\ \hline
  Eliminating expensive branches & v0.4 & 778 & 5010 & 903 & 160 & 1457\\ \hline
  Cache blocking, faster searching and AOS-to-SOA & v0.5 & 1163 & 5762 & 1007 & 181 & 1813\\
  \hline 
  \end{tabularx}
  \label{tab:seqoptimresper}
\end{table}

\section{Multithreaded Triad Census}
\label{sec:triadmulti}
In order to harness hardware multi-threading capabilities of the dual Intel Quad Core Xeon processors(2 sockets x 4 core/processor x 2 HT/core = 16 Hardware threads), we created multithreaded implementation of the optimized sequential triad census algorithm using \textbf{PThread} multi-threading framework. \textbf{PThread} is a POSIX standard for explicit threads implemented in form of a C based API on many platforms(Unix, Linux, Mac, sun and windows). It provides data structures, algorithms and constructs for explicit thread management and synchronization. Besides PThread, there are number of libraries and frameworks available for implementation of parallel algorithms on shared memory CMP's. For implementations written in C/C++, other popular libraries besides PThread are OpenMP, Boost threads, Intel TBB and Cilk. These libraries have different features and advantages/disadvantages. We choose PThread as it enables programmer to exercise comparatively finer control over many different aspects of multithreading.

\subsection{Graph partitioning and Load Balancing strategy}
\label{subsec:multitriadloadbalance}
Parallelization of many graph algorithm problems broadly requires three steps:
\begin{itemize}
\item \textbf{Graph partitioning}: This step involves finding ways to effectively decompose the graph structure into substructures that can be processed in parallel. A good graph partitioning strategy would decompose it in such a way so as to \emph{decouple} graph partitions from each other and minimize communication \& synchronization required between concurrent computations. At the same time, the partitioning procedure should be efficient in terms of performance and memory overheads.

\item \textbf{Balancing workload}: The amount of computation required to process an input data set for producing the desired output is defined as the \textbf{workload}. When workload is shared equally between processors or hardware threads, then each one of them will spend roughly equal amount of time executing the assigned workload.  In Multiprocessor systems, good load balancing is essential for obtaining good scalability with increasing number of processors and increasing size of data set. Many a times graph partitioning step is not able to effectively address the load balancing problem. This is possible because the amount of work involved in processing a graph partition may depend upon properties of the parts ( vertices, edges, neighbourhoods, other active regions etc. ) of the graph partition such as in-degree, out-degree, vertex/edge weights, neighbourhood size, cycles, distance, eccentricity, radius, diameter, center, circumference etc. So there could be two same sized graph partitions which may or may not have same workload present in them. To load balance, the partitions of the graph should be sized to balance the workload.

\item \textbf{Mapping load balanced partitions to threads}: Once the graph is partitioned according to the load balancing strategy, different graph partitions can be allocated to different threads for processing. If the partitions are completely decoupled i.e. independent, threads may independently process their respective graph data and write the results to a common sink(concurrent write to common destination is protected by a lock or atomic). In case the partitions are not completely decoupled, the threads need to use some thread communication and synchronization mechanism to produce the final results
\end{itemize}

In the case of Triad Census algorithm, the ideal partitioning and load balancing strategy is not immediately clear. To understand this better, we will first analyse the distribution of workload in the algorithm. For any further discussions, we will refer to the Triad Census Algorithm (See Figure~\ref{fig:triadcensus} ). We will first define some units of workloads. A unit of $W_{2}$ work is amount of computation involved to increment $T_{2}$ census count i.e. all computation required until Line 14 is executed once. A unit of $W_{3}$ work is amount of computation involved to increment $T_{3}$ census count  i.e. all computation required until Line 18 is executed once. Note that, there is dependency between $T_{2}$ and $T_{3}$ ( $T_{2}$ requires $|S|$ while $T_{3}$ requires set $S$ and pre-computed triad type calculated under $T_{2}$)
\\
\\
Referring to the Triad Census algorithm, it is clear that $|N[u]|$ is the measure of $W_{2}$ workload w.r.t vertex $u$. Note that $W_{2}$ workload doesn't depend upon $|N[v]|$. A higher or lower value of $|N[v]|$ will not impact number of times computations in line 9-14 are executed ( only value of  $|N[u]|$ will ). By aggregating $|N[u]|$ $\forall u \in V$, we can find total $W_{2}$ workload for the graph. If there are \textbf{T} threads, good load balancing of $W_{2}$ workload will assign $(\sum_{u \in V} |N[u]|)/T$ work to each thread. Similarly, measure of $W_{3}$ workload w.r.t canonical dyad $<u,v>$ is $|S|$. Lower or higher value of $|S|$ will increase or decrease the number of times computation in lines 16-19 are executed. By aggregating $|S|$ $\forall <u,v>$, we can find total $W_{3}$ workload for the graph. If there are $T$ threads, ideal load balancing of $W_{3}$ workload will assign $(\sum_{u \in V}\sum_{v \in N[u], u<v} |N(u) \cup N(v) \setminus \{u,v\}|)/T$ work to each thread.
\\
\\
In order to partition the graph, we must first define as to what constitutes an independent piece of work - or a \textbf{task abstraction}.  Table~\ref{tab:loadbalancingstrat} lists possible task abstraction and the corresponding graph partitioning, load balancing approaches for Triad Census algorithm. Ideally, one would like to load balance both $W_{2}$ and $W_{3}$ workload precisely. As we can see from table~\ref{tab:loadbalancingstrat}, a perfect load balancing doesn't come for free as overheads increase considerably. In other words, there is a trade-off between load balancing and the associated overhead. From scalability point of view, keeping performance overheads low is important and so \textbf{Canonical Dyad(uniform distr.)} and \textbf{Canonical Dyad(non-uniform distr.)} approaches should be good candidates and we decided to evaluate these two strategies. We implemented these two chosen graph partitioning strategies by using the task queue based approach as used by authors \cite{Chin2009}. Figure~\ref{fig:multitriadtaskqueuedyad3} and Figure~\ref{fig:multitriadtaskqueuedyad2} shows the respective task queue generation algorithms based on Canonical Dyad(non-uniform distr.) and Canonical Dyad(uniform distr.) strategy. Figure~\ref{fig:multitriadcensus} shows the corresponding modified Sequential Triad Census algorithm that uses output(array of task queues) of these task queue generation algorithms.

\begin{table}[htbp]
  \centering
  \caption{Task abstractions, graph partitioning, load balancing strategies. Assuming $T$ threads. $n$ vertices, $N(u)$ neighbours of vertex u \label{tab:loadbalancingstrat}}
  \begin{tabularx}{1.1\textwidth}{|X|X|X|X|}
  \hline
  Task abstraction & Partitioning strategy & Load balancing strategy & Comments \\ \hline \hline
Vertex & Distribute vertices among threads & n/T vertices per thread & Simple to implement, No overheads, No or poor load balancing \\
  \hline
  Dyad(uniform distribution) & Distribute dyads uniformly among threads &  $(\sum_{u \in V} |N[u]|)/T$ dyads allocated per thread & Good load balancing of $W_{2}$ workload but not $W_{3}$ workload. Non-canonical dyads will unbalance the load distribution. Some performance and memory overheads required for assigning dyads to each thread\\
   \hline
   Canonical Dyad(uniform distribution) & Distribute canonical dyads uniformly among threads &  $(\sum_{u \in V}\sum_{v \in N[u], u<v} 1)/T$ canonical dyads allocated per thread & Fully balanced $W_{2}$ workload. $W_{3}$ workload is subsumed under canonical dyads allocated to each thread and is thus partially balanced. Overhead increases as non-canonical dyads need to be eliminated\\
   \hline
   Canonical Dyad(non-uniform distribution) & Distribute canonical dyads among threads keeping $W_{3}$ load same for each thread & Number of dyads are allocated to each thread so that $W_{3}$ load across all threads is equal i.e.$(\sum_{u \in V}\sum_{v \in N[u], u<v} |N(u) \cup N(v) \setminus \{u,v\}|)/T$ workload per thread & Good load balancing of $W_{3}$ workload and only partial load balancing of $W_{2}$. Non-canonical triads will unbalance the $W_{3}$ load distribution\\ 
   \hline
   Canonical Dyads and Triads(uniform distribution) & Distribute canonical dyads and triads among threads uniformly & $(\sum_{u \in V}\sum_{v \in N[u], u<v} 1)/T$ canonical dyads allocated per thread and $(\sum_{u \in V}\sum_{v \in N[u], u<v} |N(u) \cup N(v) \setminus \{u,v\}|)/T$ canonical triads allocated per thread & Load balances $W_{3}$ and $W_{2}$. Calculating $T_{2}$ census requires that dyads and triads be stored separately and possibly processed separately. Non-canonical triads will unbalance the $W_{3}$ load distribution. Overheads increase further as now triads need to be stored along with dyads. Distribution of triads increases performance overheads\\
  \hline 
  \end{tabularx}
\end{table}

\begin{figure}[t!]
\fbox{\begin{minipage}{13 cm}
\begin{flushleft}
\textbf{Function}: Distributed Task Queue generation algorithm that generates an array of task queues based on Canonical Dyad(non-uniform distr.) load balancing strategy\\
\textbf{Input}: Graph \(G = [V, E]\) where V is set of vertices, E is array of edge lists and N is array of neighbour lists, MaxNsetSize is maximum size of neighbour set of a bunch of dyads\\
\textbf{Output}: Array of task queues - TQ[] \\
\end{flushleft}
    \begin{algorithmic}[1]
    \State $thid \gets 1$
    \State $NsetSize \gets 0$
	\\
	\For{ $u \in V$} 
		\For{ $v \in N[u]$}
			\If {$u$ \textless $v$}
			    \State $TQ[thid] \gets TQ[thid] \cup <u,v>$
			    \State $S \gets N[u] \cup N[v] \setminus \{u,v\}$
			    \State $NsetSize \gets NsetSize + |S|$
				\If { $NsetSize > MaxNsetSize$ }
				    \State $thid \gets thid + 1$
				    \State $NsetSize \gets 0$
				\EndIf
			\EndIf	
		\EndFor
	\EndFor
	\end{algorithmic}
	\caption{Distributed Task Queue generation algorithm based on Canonical Dyad(non-uniform distr.) load balancing}
	\label{fig:multitriadtaskqueuedyad3}
\end{minipage}}
\end{figure}

\begin{figure}[t!]
\fbox{\begin{minipage}{13 cm}
\begin{flushleft}
\textbf{Function}: Distributed Task Queue generation algorithm that generates an array of task queues based on Canonical Dyad(uniform distr.) load balancing strategy\\
\textbf{Input}: Graph \(G = [V, E]\) where V is set of vertices, E is array of edge lists and N is array of neighbour lists, MaxNsetSize is maximum size of neighbour set of a bunch of dyads\\
\textbf{Output}: Array of task queues - TQ[] \\
\end{flushleft}
    \begin{algorithmic}[1]
    \State $thid \gets 1$
    \State $NsetSize \gets 0$
	\\
	\For{ $u \in V$} 
		\For{ $v \in N[u]$}
			\If {$u\textless v$}
			    \State $TQ[thid] \gets TQ[thid] \cup <u,v>$
			    \State $NsetSize \gets NsetSize + |N[u]| + |N[v]| - 2$
				\If { $NsetSize > MaxNsetSize$ }
				    \State $thid \gets thid + 1$
				    \State $NsetSize \gets 0$
				\EndIf
			\EndIf	
		\EndFor
	\EndFor
	\end{algorithmic}
	\caption{Distributed Task Queue generation algorithm based on Canonical Dyad(uniform distr.) load balancing}
	\label{fig:multitriadtaskqueuedyad2}
\end{minipage}}
\end{figure}

\begin{figure}[t!]
\fbox{
\begin{minipage}{13 cm}
\begin{flushleft}
\textbf{Function}: Sequential Subquadratic Triad Census Algorithm based on distributed task queues\\
\textbf{Input}: Graph \(G = [V, E]\) where V is set of vertices, E is array of edge lists, N is array of neighbour lists, TQ is array of task queues\\
\textbf{Output}: Census array with frequencies of isomorphic triadic types\\
\end{flushleft}
    \begin{algorithmic}[1]
	\For{$i = 1 \to 16$} 
	    \State Census[$i] \gets 0$
	\EndFor
	\\
	\For{ $taskqueue \in TQ[]$} 
		\For{ $task \in taskqueue$}
	        \State $u \gets task.u$
   		        \State $v \gets task.v$
		    \State $S \gets N(u) \cup N(v)  \setminus \{u,v\}$
   		    \State $IsEdge[0] \gets IsEdge(u,v)$
   		    \State $IsEdge[1] \gets IsEdge(v,u)$
            \State $precomputed\_triad\_type  \gets IsEdge[0] + 2 * IsEdge[1]$
			\If { $IsEdge[0] \wedge IsEdge[1]$ }
			    \State $TriadType\gets 3$
			\Else
                \State $TriadType\gets 2$
			\EndIf
			\State $Census[TriadType] \gets Census[TriadType] + n - |S| - 2$
			\For{ $w \in S$ }
				\If{$v$ \textless $w \vee ( w$ \textless $v \wedge u$ \textless $w \wedge \neg IsNeighbour(u,w) )$}
	                \State $TriadType \gets TriadCode(u, v, w, precomputed\_triad\_type)$
	                \State $Census[TriadType] \gets Census[TriadType] + 1$
	            \EndIf
			\EndFor
		\EndFor
	\EndFor
	\\
	\State $sum \gets 0$
	\For{$i = 2 \to 16$} 
	    \State $sum \gets sum + Census[i]$
	\EndFor
	\State $Census[1] \gets n*(n-1)*(n-2)/6 - sum$
	\end{algorithmic}
	\caption{Subquadratic Triad Census Algorithm based on distributed task queues}
	\label{fig:multitriadcensus}
\end{minipage}}
\end{figure}

\begin{figure}[ht!]
	\centerline{\includegraphics[scale=0.5]{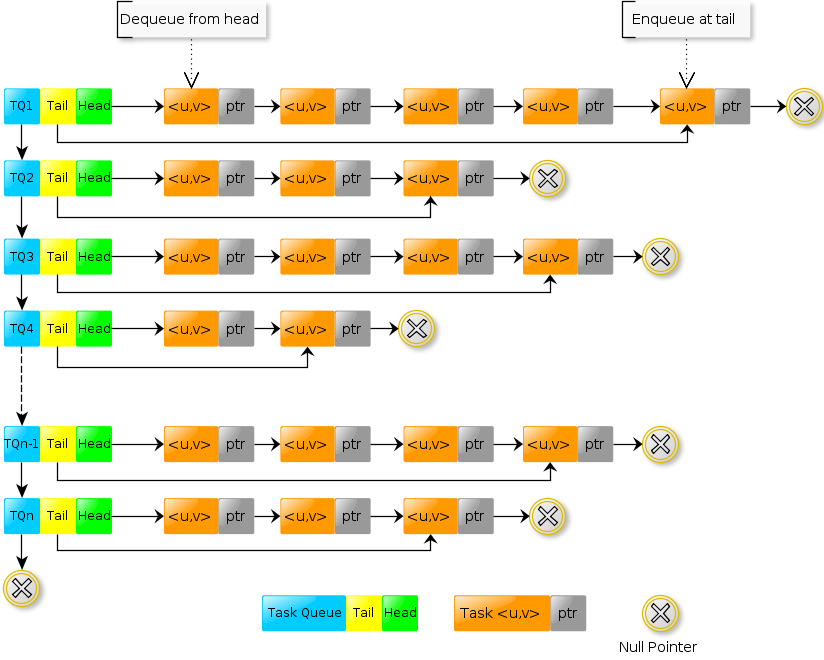}}
	\caption{Distributed task queue data structure}\label{fig:distrtaskqueue}
\end{figure}

Both of these task queue generation algorithms are based on \emph{\textbf{distributed task-queue}} data structure \cite{TaskQueuePattern}. A \textbf{distributed task-queue} data structure consists of an array of \textbf{task queues} where each \textbf{task queue} is a queue of \textbf{tasks}. Both algorithms use the same task abstraction(\textbf{canonical dyad} $<u,v>$ where $u<v$) and create task queues in the same way(see Figure~\ref{fig:distrtaskqueue}). They differ only in the manner in which workload of a task is calculated (which depends upon the respective load balancing strategy). The algorithms create task queues iteratively by appending a task to the existing task queue.  After appending a task to current task queue, the workload of current task is calculated according to the load balancing strategy and then this value is accumulated in the workload of current task queue (\textit{NsetSize}). The unit of workload for a task(canonical dyad) is size of its neighbour set. When workload of the current task queue reaches a pre-defined maximum workload(\textit{MaxNsetSize}), a new task queue is created and the above procedure repeats. 
\\
\\
The value of maximum workload per task queue - \textit{MaxNsetSize} - can be varied to change the number of created task queues. More workload per task queue means overall lesser number of task queues and vice versa. Depending upon required number of task queues, we calculated value of this maximum workload for each graph data set separately. In our Multithreaded implementations, we decided to schedule one task-queue per thread i.e. required number of task queues equals the required number of software threads. In this case, maximum workload per task queue(or thread) equals the corresponding value defined in load balancing strategy column in Table~\ref{tab:loadbalancingstrat}. Note that alternate \emph{task-queue to thread mapping} strategies (for example scheduling multiple task queues per thread) are possible. Also, Canonical Dyad(uniform distr.) strategy implemented in Figure~\ref{fig:multitriadtaskqueuedyad2} is nearly the same approach as taken by authors \cite{Chin2009}, only difference being that we modified the expression which calculates workload of current task queue(\textit{NsetSize}). The reason for this slight modification will be explained in next section.

\subsection{Using Canonical Dyad(non-uniform distr.) load balancing(v0.6)}
\label{sebsec:multitriadcensusatomics}
Based on \textbf{Canonical Dyad(non-uniform distr.)} load balancing approach, we created our initial Multithreaded implementation using PThread multithreading framework. Each created thread is assigned one task queue from the array of task queues(see Figure~\ref{fig:multitriadcensus} and Figure~\ref{fig:distrtaskqueue}). All threads execute the same triad census algorithm but each thread processes tasks in its own task queue. To build final triad census array, we mainly experimented with following two approaches and found that second approach is slightly better as it decouples the threads completely.

\begin{enumerate}
\item \textbf{Single census array shared between threads}: In this approach, a single census array is shared among all the threads. In order to synchronize concurrent updates to census array from multiple threads, we used fast atomic primitive \textit{\_\_sync\_fetch\_and\_add} provided by gcc. 

\item \textbf{Local census array per thread}: In this approach, we assigned each thread its own census array. Now there is no writeable data shared between the threads. All threads execute asynchronously and each thread produces its own census array corresponding to the task queue allocated to it. When a thread has finished processing tasks in its task queue, it waits at a barrier for other threads to finish. When all threads are finished processing their respective task queues, local arrays of all threads are accumulated in a global census array.
\end{enumerate}

If we compare the above two approaches, the first approach has advantage only in terms of reduced memory as there is just one census array and not per thread copies of census array. In second approach we waste some memory but we improve performance by avoiding threads waiting for exclusive access to shared census array. Also, in case there are too many threads, the second approach should scale better as all decoupled threads can proceed independently. Experimentally we found that second approach fairs better in terms of performance and so decided to use it. We ran our implementation with 16 threads(each software thread per hardware thread) Table~\ref{tab:multitriadcensusspeedup} shows the speedup obtained from this implementation w.r.t baseline sequential triad census(v0.1). Table~\ref{tab:multitriadcensusper} shows the absolute values of time in seconds. From the results we see that the an average speedup (w.r.t baseline) of \textbf{10x} is obtained for Amazon, Slashdot and Google data sets. Also, Actors(5x) and patents(3.2x) graph data sets fair real badly in terms of speedup. 
\\
\\
To understand why we couldn't get good scaling, we analysed the overall breakup of execution time( see Table~\ref{tab:multitriadcensusbreakup} ). In terms of percentage execution time and absolute number, we can see that the most time is spent creating task queues during pre-processing. As a result, sequential pre-processing dominates the overall execution time and is thus the source of bottleneck leading in poor parallel scaling. Consequently, we conclude that although Canonical Dyad(non-uniform distr.) approach is precise, but its has very high overheads causing serious performance bottlenecks and should this be replaced by a simpler and maybe less precise load balancing strategy like Canonical Dyad(uniform distr.)

\begin{table}[hb!]
  \centering
  \caption{Execution Time(seconds) breakup of Multithreaded Triad Census algorithm(v0.6) on Intel Xeon Multicore \label{tab:multitriadcensusbreakup}}
  \begin{tabular}{ p{5cm} | c | c | c | c | c } 
  \hline
  \textbf{Version v0.6} & \textbf{Actors} & \textbf{Patents} & \textbf{Amazon} & \textbf{Slashdot} & \textbf{Google} \\ 
  \hline
  \textbf{\%Time - Read Graph}  & 11.67 & 5.49 & 6.29 & 0.85 & 0.24\\
  \textbf{\%Time - Create Neighbour sets} & 5.50 & 6.18 & 2.00 & 0.55 & 0.13\\
  \textbf{\%Time - Create Task queues} & 55.00 & 63.40 & 70.86 & 70.75 & 83.86\\ 
  \hline
  \textbf{\%Time - Total Pre-processing} & 72.17 & 75.07 & 79.14 & 72.15 & 84.23\\
  \textbf{\%Time - Triad Census execution}  & 27.83 & 24.93 & 20.86 & 27.85 & 15.77 \\
  \hline
  \textbf{Total Pre-processing time (sec)} & 8.66 & 76.5 & 27.69 & 28.86 & 1069.72\\
  \textbf{Triad Census exec time (sec)} & 3.33 & 24.5 & 7.3 & 11.14 & 200.28\\ 
  \hline
  \textbf{Total Time (seconds)} & 12.00 & 102.00 & 35.00 & 40.00 & 1270.00\\  
  \hline
  \end{tabular}
\end{table}

\begin{table}[htbp]
  \centering
  \caption{Multithreaded Triad Census speedup(v0.6) w.r.t baseline(v0.1) results on Intel Xeon Multicore}
  \begin{tabularx}{\textwidth}{|X|c|c|c|c|c|c|}
  \hline
  \textbf{Optimization} & \textbf{Version} & \textbf{Actors} & \textbf{Patents} & \textbf{Amazon} & \textbf{Slashdot} & \textbf{Google}\\ 
  \hline
  Baseline & v0.1    & 1.00   & 1.00    & 1.00   & 1.00     & 1.00 \\ 
  \hline
  Multithreaded with per thread local array and Canonical Dyad(non-uniform distr.) load balancing & v0.6 & 5.00 & 3.18 & 11.26 & 9.83 & 9.61 \\
  \hline 
  \end{tabularx}
  \label{tab:multitriadcensusspeedup}
\end{table}

\begin{table}[htbp]
  \centering
  \caption{Multithreaded Triad Census Execution Timing(seconds) results(v0.6) on Intel Xeon Multicore}
  \begin{tabularx}{\textwidth}{|X|c|c|c|c|c|c|}
  \hline
  \textbf{Optimization} & \textbf{Version} & \textbf{Actors} & \textbf{Patents} & \textbf{Amazon} & \textbf{Slashdot} & \textbf{Google}\\ 
  \hline
  Baseline     & v0.1 & 60 & 324 & 394 & 393 & 12203\\ 
  \hline               
  Multithreaded with per thread local array and Canonical Dyad(non-uniform distr.) load balancing & v0.6 & 12 & 102 & 35 & 40 & 1270\\ 
  \hline 
  \end{tabularx}
  \label{tab:multitriadcensusper}
\end{table}

\subsection{Using Canonical Dyad(uniform distr.) load balancing(v0.7)}
\label{sebsec:multitriadcensusfa}
The Multithreaded Triad Census implementation involves number of pre-processing tasks before Parallel Triad Census algorithm can be executed. This involves reading graph, making edge/neighbour sets and creating task-queues. As most of this pre-processing is done sequentially, according to Amdahl's law, a sequential bottleneck will limit the \emph{strong scalability} possible for our Multithreaded implementation. Removing sequential bottlenecks thus remained our top priority. 
\\
\\
In order to improve \emph{strong scaling} of Multithreaded Triad Census algorithm, we decided to replace \textbf{Canonical Dyad(non-uniform distr.)} load balancing with \textbf{Canonical Dyad(uniform distr.)} load balancing approach (see Figure~\ref{fig:multitriadtaskqueuedyad2}). This load balancing approach is nearly the same as authors \cite{Chin2009} approach except that while calculating \emph{NsetSize}(line 8), we subtract two from overall expression. The logic behind this subtraction is as follows: As $<u,v$> is a dyad, the set $N[u]$ will always contain $v$ and set $N[v]$ will always contain $u$. By subtracting two from the expression $|N[u]| + |N[v]|$, we are bringing the value of this expression more closer to ideal value $|S| = N[u] \cup N[v] \setminus \{u,v\}$. 
\\
\\
Although a factor of two may seem trivial, but the Table~\ref{tab:multitriadloadbalanceimproved} shows the difference in aggregate neighbour set size(all dyads) when we use Canonical Dyad(uniform distr.) approach. Experimentally we found that our approach significantly improves the load balancing as changing the \emph{MaxNsetSize} from $M1$ to $M2$ leads to nearly proportional change($~M2/M1$) in number of threads being created, suggesting the fact that load being proportionally divided between threads. When we used authors \cite{Chin2009} approach, this was not the case and we had to use trial and error to find MaxNsetSize values for getting good load balancing.

\begin{table}
  \centering
  \caption{Comparison of load balance strategy based on aggregate NsetSize(all dyads)\label{tab:multitriadloadbalanceimproved}}
  \begin{tabular}{ l | c | c }
  \hline
  \textbf{Data Sets} & \textbf{Author's \cite{Chin2009} Approach} & \textbf{Our Canonical Dyad(uniform distr.) approach} \\ \hline
  Actors    & 81908982 & 78968174 \\ \hline
  Patents	& 704600440 & 671562549\\ \hline
  Amazon    & 149306190 & 144419374\\ \hline
  Slashdot  & 148238520 & 147237558\\ \hline
  Google    & 1463477820 & 1454834448\\
  \hline
  \end{tabular}
\end{table}

Finally, we ran simulations with our Canonical Dyad(uniform distr.) load balancing strategy and found that we were able to substantially lower the task queue creation part of pre-processing time and overall pre-processing time. Table~\ref{tab:multitriadloadbalanceimprovedres16} shows the results with 16 software threads (1 software thread per hardware thread) and also the breakup of execution time. Table~\ref{tab:multitriadloadbalanceimprovedresmax} shows best results obtained through simulations which corresponds to substantially higher number of software threads per hardware thread. Figure~\ref{fig:datasets-scaling-graphs} shows the graphs for speedup and time scaling w.r.t number of threads.
\\
\\
With 16 software threads, we get good speedups for \textbf{Amazon(15x), Slashdot(19x), Google(35x)} data sets(Table~\ref{tab:multitriadloadbalanceimprovedres16}). Speedups for \textbf{Patents(7.3x) and Actors(9.4x)} has also improved but still not linear. Thus with 16 threads, the \textbf{average speedup comes out to be 17x} w.r.t baseline sequential(v0.1). 
\\
\\
Also, as we increase the number of number of software threads w.r.t hardware threads (Table~\ref{tab:multitriadloadbalanceimprovedresmax}), the speedup continues to improve. Somewhere between 600-700 software threads, we get maximum speedups for \textbf{Amazon(31x), Slashdot(25x), Google(56x), Patents(8.4x), Actors(11x)}. Above this thread count threshold, the speedup improvement saturates. This improvement in speedup with higher number of threads is but natural with Simultaneous Multithreaded Xeon processors because even when 16 threads stall due to a memory load operation for example, the processor still has enough threads to keep its pipeline filled with useful instructions(reduced vertical waste).
\\
\\
The speedup for networks Patents and Actors networks has also improved a little bit in comparison to speedups with 16 software threads but the scaling is still not as effective as for Amazon, Slashdot and Google networks. At this point, we analyse as to why there is discrepancy between scaling of Amazon, Slashdot Google networks and Patents/Actors network. If one compares Triad census arrays generated for Amazon, Slashdot and Google networks with Triad census arrays generated for Actors/Patents, we see that in Actors and Patents network, most of the census count belongs to triad type 1,2, and 3 and all other census counts are mostly zeros. This means we are hardly executing the innermost for loop( line 19-24 in Figure~\ref{fig:multitriadcensus} ). In case of Amazon, Slashdot and Google networks, the distribution of census count is more uniform and we get some census count for most of the triad types if not all. This suggests that in these networks we are running the innermost for loop lot more number of times than in patents and actors network.
\\
\\
From these observations, firstly we can see that in Amazon, Slashdot and Google networks there are more number of instructions that are executed. With higher number of executed instructions per thread, processor has more opportunities to extract ILP per thread. Effectively this means that processor pipeline utilization is higher(lower vertical waste) for Amazon, Slashdot and Google networks. Also as we increase the number of threads, the overall vertical waste(pipeline stalls due to hazards) reduces further as now we have still have enough useful instructions from waiting threads to keep the processor pipeline busy. Secondly, in Triad census algorithm, there are very few number of ALU operations in comparison to Memory load/store operations. This ratio of useful ALU operations to memory load/store operations i.e. Arithmetic intensity is also higher for Amazon, Slashdot and Google networks in comparison to Actors and Patent networks.  With higher arithmetic intensity, the processor does more useful computational work(reduced pipeline horizontal waste) per load/store operation. Thus with higher number of executable instructions in Amazon, Slashdot and Google networks, utilization of ALU's also increases. This also gives processor more opportunities to hide memory latency as now it has more useful work to do while another thread is stalled on a expensive memory operation. As we increase the number of threads, the overall horizontal waste reduces further as now we have enough useful ALU instructions from threads waiting to be executed to keep the ALU resources busy.

\begin{sidewaystable}
  \centering
  \caption{Multithreaded Triad Census with improved load balancing [Canonical Dyad(uniform distr.)] (v0.7) - performance speedup w.r.t baseline(v0.1) and Execution Timing(seconds) breakup with 16 threads on SMT Intel Xeon Multicore\label{tab:multitriadloadbalanceimprovedres16}}
  \begin{tabularx}{\textwidth}{X||X|X|X||X|X|X||X}
  \hline 
  \hline
  \textbf{Data Sets} & \textbf{Total Execution Time(sec)} & \textbf{Parallel Triad Census Time(sec)} & \textbf{Pre-processing Time(sec)} & \textbf{Graph Read Time (sec)} & \textbf{Time to make Neighbour Sets(sec)} & \textbf{Time to create TaskQueues and Load balancing(sec)} & \textbf{Speedup w.r.t baseline}\\
  \hline
  \hline
  Actors & 6.36 & 3.99 & 2.13 & 1.35 & 0.68 & 0.1 & \textbf{9.42} \\ 
  \hline
  Patents & 44.13 & 27.23 & 15.17 & 5.43 & 7.81 & 1.93 & \textbf{7.34}\\ 
  \hline
  Amazon  & 25.69 & 22.53 & 2.91 & 2.01 & 0.73 & 0.17 & \textbf{15.34}\\ 
  \hline
  Slashdot & 20.38 & 19.88 & 0.43 & 0.27 & 0.14 & 0.02 & \textbf{19.28}\\ 
  \hline
  Google & 339.82 & 333.96 & 5.25 & 3.12 & 1.74 & 0.39 & \textbf{35.91} \\
  \hline 
  \hline
  \end{tabularx}
  \caption{Multithreaded Triad Census with improved load balancing [Canonical Dyad(uniform distr.)] (v0.7) - best speedup w.r.t baseline(v0.1) and Execution Timing(seconds) breakup with maximum threads(found experimantally) on SMT Intel Xeon Multicore\label{tab:multitriadloadbalanceimprovedresmax}}
  \begin{tabularx}{\textwidth}{X||X||X|X|X||X|X|X||X}
  \hline 
  \hline
  \textbf{Data Sets} & \textbf{Threads} & \textbf{Total Execution Time(sec)} & \textbf{Parallel Triad Census Time(sec)} & \textbf{Pre-processing Time(sec)} & \textbf{Graph Read Time (sec)} & \textbf{Time to make Neighbour Sets(sec)} & \textbf{Time to create TaskQueues and Load balancing(sec)} & \textbf{Speedup w.r.t baseline}\\
  \hline 
  \hline
  Actors & 672 & 5.5 & 3.21 & 2.09 & 1.31 & 0.68 & 0.1 & \textbf{10.9}\\ 
  \hline
  Patents & 688 & 38.76 & 21.9 & 14.94 & 5.19 & 7.81 & 1.94 & \textbf{8.36}\\ 
  \hline
  Amazon & 624 & 12.49 & 9.52 & 2.73 & 1.82 & 0.74 & 0.17 & \textbf{31.52}\\ 
  \hline
  Slashdot & 631 & 15.85 & 15.19 & 0.6 & 0.33 & 0.21 & 0.06 & \textbf{24.78}\\ 
  \hline
  Google & 688 & 217.87 & 212.07 & 5.22 & 3.1 & 1.73 & 0.39 & \textbf{56}\\ 
  \hline 
  \hline
  \end{tabularx}
\end{sidewaystable}

\begin{figure}
	\subfloat[Actors social network speedup][]
	{
	    \label{fig:actors-speedup}
	    \centerline{\includegraphics[scale=0.7]{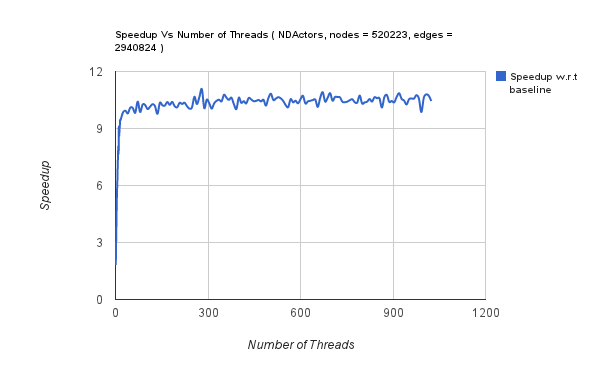}}
	}
	\\
	\subfloat[Actors social network time scaling][]
	{
	    \centerline{\includegraphics[scale=0.7]{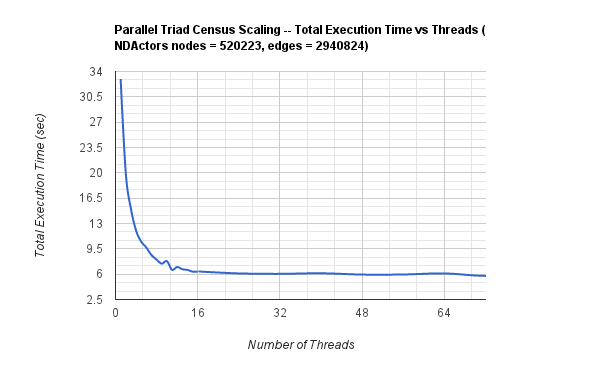}}
	}	
	\caption[]{Performance Speedup and Execution time scaling for Actors network}
	\label{fig:datasets-scaling-graphs}
\end{figure}

\begin{figure}
    \ContinuedFloat
	\subfloat[Amazon network speedup][]
	{
	    \label{fig:amazon-speedup}
	    \centerline{\includegraphics[scale=0.7]{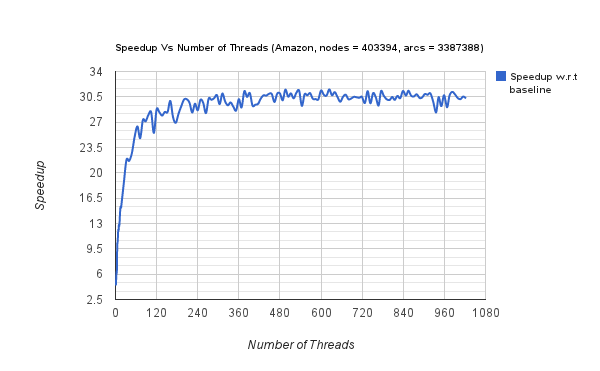}}
    }
    \\
	\subfloat[Amazon network time scaling][]
	{
	    \centerline{\includegraphics[scale=0.7]{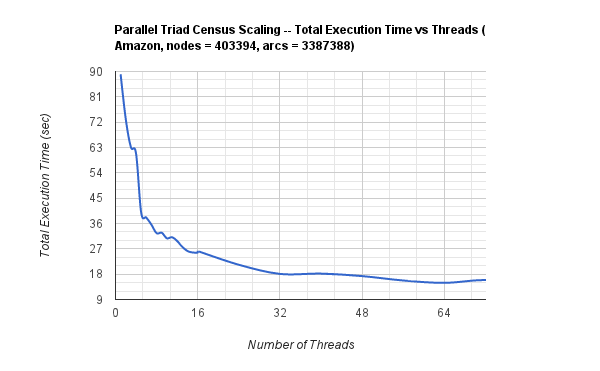}}
    }     
	\caption[]{Performance Speedup and Execution time scaling for Amazon network}
	\label{fig:datasets-scaling-graphs}
\end{figure}

\begin{figure}
    \ContinuedFloat
	\subfloat[Google web graph speedup][]
	{
	    \label{fig:google-speedup}
	    \centerline{\includegraphics[scale=0.7]{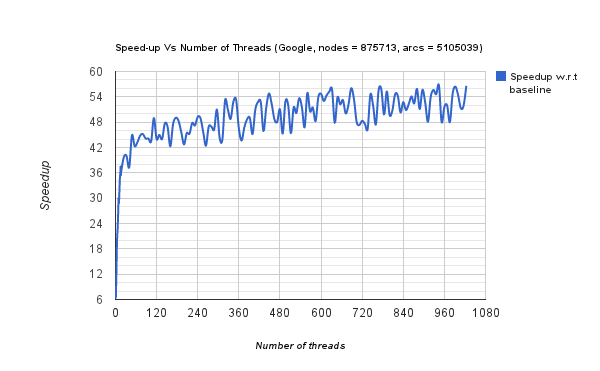}}
	}
	\\
    \subfloat[Google web graph time scaling][]
	{
	    \centerline{\includegraphics[scale=0.7]{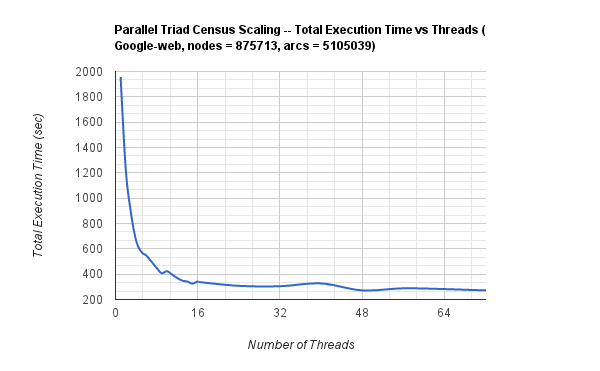}}
	}   
	\caption[]{Performance Speedup and Execution time scaling for Google web network}
	\label{fig:datasets-scaling-graphs}
\end{figure}

\begin{figure}
    \ContinuedFloat
	\subfloat[Patents network speedup][]
	{
	    \label{fig:patents-speedup}
	    \centerline{\includegraphics[scale=0.7]{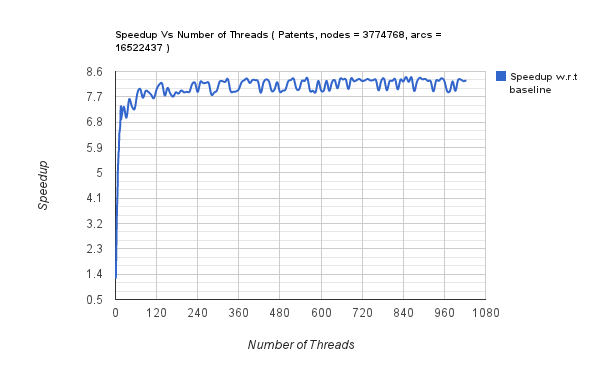}}
    }
    \\
	\subfloat[Patents network time scaling][]
	{
	    \centerline{\includegraphics[scale=0.7]{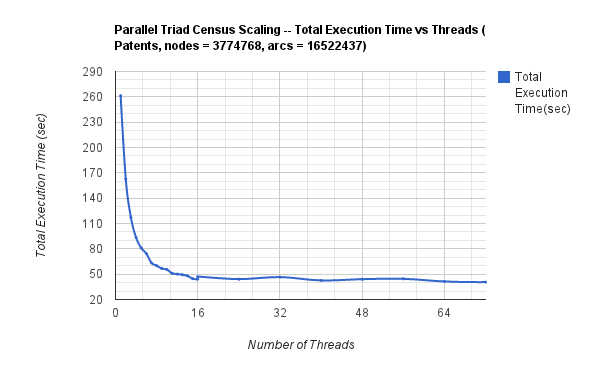}}
    }     
    \caption[]{Performance Speedup and Execution time scaling for Patents network}
	\label{fig:datasets-scaling-graphs}
\end{figure}

\begin{figure}
    \ContinuedFloat
	\subfloat[Slashdot network speedup][]
	{
	    \label{fig:slashdot-speedup}
	    \centerline{\includegraphics[scale=0.7]{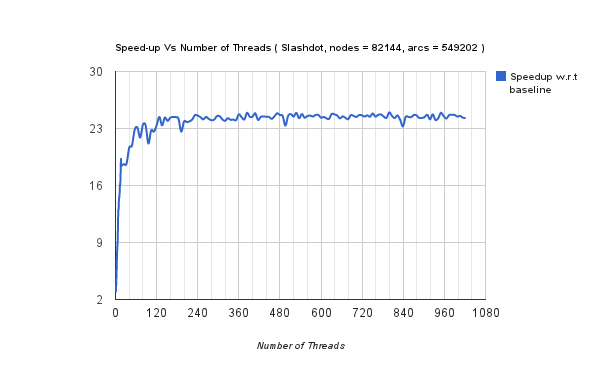}}
    }
    \\
	\subfloat[Slashdot network time scaling][]
	{
	    \centerline{\includegraphics[scale=0.7]{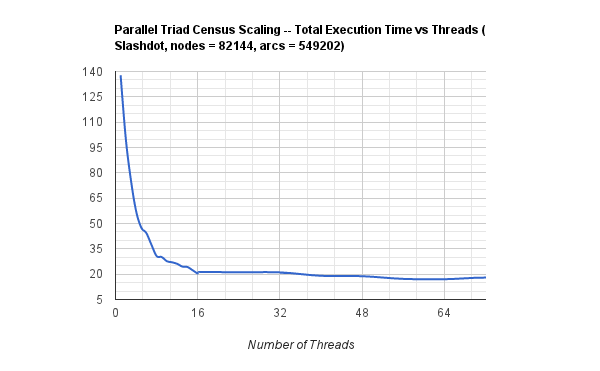}}
    }     
    \caption[]{Performance Speedup and Execution time scaling for Slashdot network}
	\label{fig:datasets-scaling-graphs}
\end{figure}

\chapter{Evaluation of Triad Census on Tesla GPGPU}
\label{chap:gpgpueval}
In the last chapter, results from Multithreaded implementation of Triad Census algorithm gave us good insights into complexities involved in efficient parallelization of Triad Census algorithm on Multicore processors. We now follow this up with efficient parallel implementation of Triad Census algorithm for NVIDIA Tesla GPGPU. In this chapter, we discuss the approach, implementation and evaluation of Multithreaded Triad census algorithm on \textbf{NVIDIA Tesla C1060 GPGPU}.

\section{Design and Implementation}
\label{sec:gpgpuappimpl}
Storing and processing data based on pointer chasing data structures like linked list is not advisable for GPGPU's. Instead, linear contiguous arrays which can be read and written in parallel result in best performance. Below we give a detailed description of the multithreaded triad census implementation written using \textbf{C++/STL and C/CUDA for Tesla C1060 GPGPU}. For implementation and testing purposes we used \textbf{CUDA nvcc 3.0 compiler driver }in \textbf{device emulation mode} and at times also used \textbf{ocelot emulator}\protect\footnote{\href{http://code.google.com/p/gpuocelot/}{http://code.google.com/p/gpuocelot/}}. For experiments and simulations we used \textbf{Tesla C1060 GPGPU and CUDA nvcc 4.0 compiler driver}. 


\subsection{Graph Data Pre-processing on host CPU}
\label{subsec:gpgpuimplpreprocess}
We started with C++/STL implementation of graph data pre-processing on host CPU. The pre-processing stage basically constructs all the graph data structures from graph data files. Once this data is constructed, it is transferred to the Tesla GPGPU for processing by Triad Census kernel implementation. Both undirected or directed graph data sets can be read. In addition, we extended support for processing zero-index or one-index based graphs i.e. graphs where vertex id can start with 0 or 1.
\\
\\
\textbf{Edge and Neighbour Data Structures}\\
To store the Edge and Neighbour data information, we used \textbf{Compressed Row Storage (CRS)} format as outlined in section~\ref{subsec:effadjlst} on page~\pageref{fig:adjarray}. This format usually uses three arrays for storing a sparse graph or matrix: \textbf{value, column indexes and row indexes}. For our purposes, value array was not needed and only column indexes and row indexes were used for storing adjacency information of Edges and Neighbours. Column array stores the adjacency(neighbour) information for each and every vertex while the row array (indexed on vertex id) stores the indexes of column array from where adjacency(neighbour) information of a vertex starts.
\\
As not all graph information( number of edges etc.) was available beforehand, we had to construct the graph dynamically from graph data sets. For this we used \textbf{Standard Template Library(STL)} implementation of dynamic arrays popularly known as \textbf{Vector container }. This container is guaranteed  by C++ standard to be contiguous \protect\footnote{\href{http://herbsutter.com/2008/04/07/cringe-not-vectors-are-guaranteed-to-be-contiguous/}{http://herbsutter.com/2008/04/07/cringe-not-vectors-are-guaranteed-to-be-contiguous/}}.  Using vectors containers and associated STL algorithms/iterators, we constructed adjacency array information for all vertices by reading and inserting one edge(neighbour) at a time. Whenever it was possible, we would reserve some space in the vector to make sure that subsequent push backs to the vector internal array do not have to do frequent reallocation-copy-insert operations.  Following are the data structures constructed for reading the graph:

\begin{itemize}
\item Edge vectors of all vertices
\item Neighbour vectors of all vertices
\end{itemize}

All Edge(Neighbour) vectors were then concatenated into a single big edge(neighbour) array that represents the column array of CRS format. While concatenating the edge(neighbour) vectors, we also constructed its corresponding index array(row indexes) to locate adjacency(neighbour) array of each vertex.
\\
\\
\textbf{Graph Partitioning and Load Balancing}\\
For graph partitioning, we used a different approach then the one used for Multicore. Instead of using a distributed task queue, we created a \textbf{centralized task queue} based on CRS format where column array stores a single big task queue and row array to store indexes. Similar to Triad Census implementation on Multicores, we used \textbf{dyad $<u,v>$} as the \emph{task abstraction}. The load balancing strategy remained the same as used in our earlier Multicore implementation. While constructing this centralized task queue, the corresponding row index array(indexed on thread id)  is created to store the index information of each sub-queue that can be assigned to a GPGPU thread.
\\
\\\
\textbf{Managing graph data under Tesla Memory Constraints}\\
In the Triad census algorithm, the set \textbf{$S = N[u] \cup N[v] \setminus \{u, v\}$} needs to be dynamically created at run-time for each and every dyad task $<u,v>$ begin processed. As NVIDIA Tesla C1060 GPGPU is based on 1.3 compute capability, it doesn't support dynamic memory allocation or de-allocation inside kernel or the device code. Even if device with 2.x capability is used, frequent heap memory allocations/de-allocations on device can substantially deteriorate performance.  We thought about mainly two alternatives to solve this problem which basically explored ways to allocate required memory buffers on CPU and then copying it to GPGPU for creating S sets at run-time:

\begin{itemize}
\item First approach is to create all the possible $S$ sets for each and every possible  canonical dyad on CPU and then transform all of the $S$ sets into CRS format suitable for GPGPU processing. This are two main issues with this approach. Firstly, sequentially creating set $S$ on CPU will make pre-processing stage very slow(this was the exact pre-processing sequential bottleneck in Canonical Dyad(non-uniform distr.) load balancing approach) creating the sequential bottleneck. Secondly, our Tesla GPGPU has a limited memory capacity of 4 GB. Throughout our Multicore and CUDA implementation we used 64-bit  unsigned integers for storing graph data. We calculated that for big graphs like Wikitalk we needed at least 192720 MB of memory (see Table~\ref{tab:gpgpumemreq}) if this approach is used. One way to deal with memory exhaustion problem was to instead use 32-bit unsigned integers for storing graph data  and do up-casting to 64 bit unsigned integer only while calculating out of range triad census sums. This pretty much halves the memory requirement but still we found that some graph sets couldn't fit in GPGPU GDDR memory.

\begin{table}
  \centering
  \caption{Initial GPGPU memory requirements for different graph data sets for 64 bit unsigned integer type data i.e. range of 0 to 18,446,744,073,709,551,615}
  \begin{tabular}{ l | p{3cm} | p{3cm} | p{3cm} }
  \hline
  \textbf{Data Set} & \textbf{$\sum |S|$ Memory requirement (MB)} & \textbf{Other Memory requirement (MB)} & \textbf{Total Memory requirement (MB)}\\  \hline \hline
  Actors & 630 & 64 & 704 \\ \hline
  Patents & 5372 & 688 & 6060 \\ \hline
  Amazon & 1155 & 110 & 1265 \\ \hline
  Slashdot & 1200 & 20 & 1220 \\ \hline
  Google & 11100 & 185 & 11285 \\ \hline
  Wikitalk & 192500 & 220 & 192720 \\ 
  \hline \hline
  \end{tabular}  
  \label{tab:gpgpumemreq}
\end{table}

\item Second approach - \textit{the one we decided to use} - is based on observation that each GPGPU thread will process only a subset of dyad tasks of the centralized task queue and these tasks will be processed iteratively. Also, different dyad tasks allocated to a thread have different memory requirements to build the neighbour set |S|. However if we can assign to each CUDA thread a pre-allocated buffer(in global memory) whose size equals the maxim neighbour set size among all tasks of its task queue, then the thread can reuse this buffer to process different tasks in its task queue iteratively.
\\
\\		
For Example: Thread-1 is responsible for dyadic tasks: tk1, tk2 and tk3. Memory requirements to create sets S1, S2 and S3 for each of these tasks is 24, 40 and 8 bytes. If a memory buffer of maximum(24,40,8) = 40 bytes is provided to
Thread-1, then it can iteratively reuse it to create set S1 in first iteration, set S2 in second iteration and set S3 in third iteration.
\end{itemize}

We discarded the first approach due to mentioned issues and decided to use the second approach. In terms of implementation, we had to make modifications in our load balancing implementation to store maximum neighbour set size requirement of each and every task queue (or thread). Based on these maximum values, we created an empty CRS column array whose size equals sum of maximum neighbour set sizes of all threads. A corresponding CRS row array is also created to store the index of starting location of neighbour buffer of each thread. From the resulting implementation, we found that all graph data sets were able to fit well within 4 GB memory limit of GPGPU (Table~\ref{tab:gpgpumemreqref}). Note that these memory requirement depends upon the number of threads or task queues created. When we increase number of task queue or threads, the memory requirement increases. As of now these requirement were calculated for 30720 thread limit of Tesla C1060 (except for Wikitalk data set for which we reached the 4 GB limit with 10000 threads).

\begin{table}
  \centering
  \caption{Final GPGPU memory requirements for different graph data sets for 64 bit unsigned integer type data}
  \begin{tabular}{ l | p{4cm} }
  \hline
  \textbf{Data Set} & \textbf{Overall Memory requirement (MB)} \\  \hline \hline
  Actors & 100\\ \hline
  Patents & 750\\ \hline
  Amazon & 350\\ \hline
  Slashdot & 230\\ \hline
  Google & 1230\\ \hline
  Wikitalk & 3910\\ 
  \hline \hline
  \end{tabular}  
  \label{tab:gpgpumemreqref}
\end{table}
The \textbf{single-threaded pre-processing stage was optimized} to make it as fast as possible so as to keep its overall time consumption much lower w.r.t triad census computation time. Figure~\ref{fig:adjarray} in section~\ref{subsec:effadjlst} shows an example graph and its corresponding CRS arrays. 

\subsection{CUDA Kernel implementation of Triad Census}
\label{subsec:gpgpuimplkernel}
All data structures constructed in pre-processing stage were wrapped in device pointers and corresponding device memory was allocated. Host data was then copied to device global(GDDR) memory using appropriate CUDA C language primitives. We utilized constant memory for storing the mapping table that translates non-isomorphic triads(64 types) to corresponding isomorphic triads(16 types). An empty Triad census array on host and device were allocated. The device Triad census array was subsequently transferred to the global memory for storing Triad Census results of execution of Triad Census kernel.
\\
\\
We implemented the Triad census kernel and required device function in form of C++ Function templates. Following functions were implemented: \emph{IsEdge} (if u-v is an edge), \emph{IsNeighbour} ( if u and v are neighbours ), \emph{TriadCode}, \emph{Set Union} and \emph{Set Difference}( for calculating S sets ). All the code was written using CUDA C/C++ and compiled for Tesla C1060 using nvcc compiler driver version 4.0.
\\
\\
\textbf{Thread hierarchy configuration parameters}\\
For the thread hierarchy configuration parameters ( \emph{kernel<<<thread\_hierarchy\_config>>>} ) of the triad census kernel, we made the number of threads per thread block ( \emph{threadblock\_config }) configurable at compile time. The number of thread blocks per grid( \emph{grid\_config} ) is then accordingly calculated at run-time. We kept the the block and grid dimensions to 1-D as we are dealing with 1-D arrays.
\\
\\
\textbf{Shared Census Array}\\
The Triad census array stored in the device global memory is shared by all threads running the same kernel on their respective task queues. To synchronize concurrent updates to the census array stored in global memory by different CUDA threads, we used CUDA atomic primitive function named \textbf{atomicAdd}. 
\\
\\
\textbf{Timing Kernel Execution}\\
For timing the total triad census kernel execution correctly, before measuring the elapsed kernel execution time, we used \emph{cudaThreadSynchronize()} after kernel function call to make sure that host CPU waits for the previous asynchronous kernel execution to finish.
\\
\\
\section{Initial experiments and Speedup Results}
\label{sec:gpgpuinitresults}
While implementing the device code, we incorporated most of the Multicore optimizations which we thought also made sense on GPGPU ( e.g.: pre-computing triad type). We compiled the code with following nvcc compiler driver options:
\\
\\
-arch=sm\_13

To compile the device code for NVIDIA Tesla C1060 GPGPU device of compute capability 1.3
\\
\\
-O3

To be passed to the host g++ compiler for host code optimization
\\
\\
--ptxas-options=-v

Specify -v option to the ptx optimizing assembler to get the details of  register, local memory, shared memory, and constant memory usage of the kernel
\\
\\
Initial compilation revealed the following per thread memory usage:
\\
\\
\textbf{Number of 32-bit registers per thread}: 31\\
\textbf{Shared memory Per thread block }( allocated by nvcc compiler ): 128 bytes\\
\textbf{Constant memory}: 536 bytes\\
\textbf{Local memory per thread}: 0 bytes\\
\\
\\
We used this memory utilization data with the spreadsheet based CUDA occupancy calculator to explore possible thread hierarchy parameters. CUDA occupancy calculator allows to calculate the SM occupancy. The SM occupancy gives a measure of Thread level parallelism in terms of ratio of active warps to the maximum number of warps supported on the SM of the GPGPU. 
\\
\\
NVIDIA recommends higher occupancy for latency bound applications so that if few of the warps stall due to slow loads, the SM scheduler still has enough warps that are waiting for execution. However this approach might not always give best results as demonstrated by authors \cite{Volkov2010}. There is a trade-off between granularity of thread block and the resource usage (registers, shared memory ) per thread. Higher occupancy translates into fine grained TLP resulting in lower number of resources per thread but requires large number of threads to lower hide memory or ALU latency. Alternatively, lower occupancy translates into coarse grained TLP resulting in more resources per thread. Lower occupancy is effective only when number of threads are lower and each thread does more work ( ALU and memory instructions ) in order to effectively use the extra resources.
\\
\\
During this initial phase of experimentation, we kept our options open for exploring granularity of Thread blocks. Tesla C1060 supports maximum 1024 threads or 32 warps or 8 thread blocks(with 4 warps per thread block) per SM. 
The thread occupancy \% in Tesla GPGPU is limited mainly by per thread block usage of shared memory and registers. In our case, the occupancy calculator calculates that with 31 registers per thread, our implementation can achieve maximum occupancy of 50\% per SM i.e. 16 active wraps per SM. Table~\ref{tab:gpgpuoccup} shows all possible thread hierarchy configurations to get this maximum achievable occupancy of 50\%. All other threads per block configuration lowers the achievable occupancy.

\begin{table}[!h]
  \centering
  \caption{Thread hierarchy configuration suggested by Occupancy calculator for 50\% occupancy(maximum with 31 registers)}
  \begin{tabular}{ p{4cm} | p{4cm} | p{4cm} }
  \hline
  \textbf{Threads per block(we configure this at compile time)} & \textbf{Thread Blocks per SM(chosen by compiler based on threads per block)} & \textbf{Occupancy (\%)} \\  \hline \hline
  48 or 64 & 8 & 50\\ \hline
  112 or 128 & 4 & 50\\ \hline
  240 or 256 & 2 & 50\\ \hline
  496 or 512 & 1 & 50\\
  \hline \hline
  \end{tabular}  
  \label{tab:gpgpuoccup}
\end{table}

\begin{figure}[htbp]
	\subfloat[][]
	{
	    \label{fig:gpu-base-speedup}
	    \centerline{\includegraphics[scale=0.7]{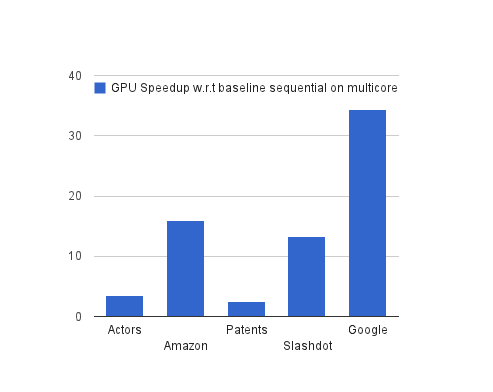}}
    }
    \\
   	\subfloat[][]
	{
	    \label{fig:gpu-base-multicore}
	    \centerline{\includegraphics[scale=0.7]{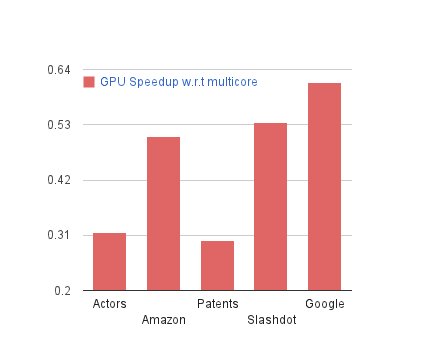}}
    }
	\caption[]
	{ Initial speedup results obtained for CUDA implementation of triad census algorithm
	    \subref{fig:gpu-base-speedup} Speedup w.r.t baseline sequential
	    \subref{fig:gpu-base-multicore} Speedup w.r.t fastest multicore
	    \label{fig:gpu-base-results}
	}
\end{figure}

Initial trial runs revealed that varying threads per block had little effect on performance when number of threads is kept at near maximum of 30720 threads. Performance deteriorates when we launch with lesser number of threads. Table!\ref{tab:gpgpuinitres} describes initial results with 256 threads/block and 120 thread blocks thread hierarchy configuration. As we can see from Figure~\ref{fig:gpu-base-results}, we get good speedup w.r.t baseline sequential for \textbf{Amazon(16x), Slashdot(13x) and Google(34x) }data sets. However for patents(2.5x) and Actors(3.45x) networks the speedup results is relatively poor. We suspect that the reason for discrepancy in speedups has similarities to the reasons we discussed during acceleration on Multicores. Only this time we see the discrepancy on the Tesla GPGPU. If we compare the GPGPU results with fastest Multicore results, we see that on an average GPGPU results are 2.4x slower.

\begin{table}[ht!]
  \centering
  \caption{Initial performance results of Triad Census on Tesla GPGPU with 256 threads/block and 120 thread blocks}
  \begin{tabularx}{\textwidth}{c|c|X|X|X|X}
  \hline
  \textbf{Data sets} & \textbf{Total time(sec)} & \textbf{Algorithm exec time(sec)} & \textbf{Pre-process time(sec)} & \textbf{GPGPU Speedup w.r.t multicore} & \textbf{GPGPU Speedup w.r.t sequential}\\ \hline \hline
  Actors & 17.38 & 16.17 & 1.21 & 0.32 & 3.45\\ \hline
  Amazon & 24.7	 & 21.25 & 3.45	& 0.51 & 15.95\\ \hline
  Patents & 129.73 & 115.06	& 14.67	& 0.3 & 2.5\\ \hline
  Slashdot & 29.69 & 29.22 & 0.47 & 0.53 & 13.24\\ \hline
  Google & 354.71 & 348.64 & 6.07 & 0.61 & 34.4\\  
  \hline \hline
  \end{tabularx}  
  \label{tab:gpgpuinitres}
\end{table}

\section{Profiling Triad Census CUDA kernel}
\label{sec:gpgpuprofiling}
To get better insight into execution of Triad census kernel on Tesla GPGPU in terms of performance, memory and branching, we profiled its using command-line based profiler tool called \textbf{computeprof}. The \textbf{computeprof} tool enables user to gather different runtime information statistics(\textit{profile counters}) for a CUDA kernel. Depending upon the profile settings, the profiler collects various profile counters like executed instructions, occupancy, warp\_serialize, global coherent load/stores, branches, divergent\_branches etc. Table~\ref{tab:computeprofcounters} in Appendix~\ref{app:comprof} shows description of basic profile counters and also some of the derived counters that we created for profiling Triad Census kernel. For more detailed description of various CUDA compute profile counters see \protect\footnote{\href{http://developer.download.nvidia.com/compute/DevZone/docs/html/C/doc/Compute\_Command\_Line\_Profiler\_User\_Guide.pdf}{computeprof user-guide}} 
\\
\\
We first configured the profiler with profile counters in a profile configuration file and then enabled it. After this, we ran our Triad Census CUDA application. The compute profiler collects the statistics while the application runs. When the CUDA application has finished its execution, the profiler dumps the collected profile counter values in configured output file (CSV format). In order to target potential bottlenecks in Triad Census Kernel, we created different types of profiles. Using basic profile counters, we derived some new profile counters like instruction throughput, global memory read/write/overall throughput, percentage divergent branches.
\\
\\
We will now describe the analysis of the performance and memory statistics obtained by profiling our initial CUDA implementation of Triad Census algorithm:

\begin{enumerate}
\item Firstly we see in Table~\ref{tab:gpgpucoherstats} that all the global load and store transactions are coherent or coalesced which is ideally what one needs for good global memory performance. Also, we can see in Table~\ref{tab:gpgpuloadstore} that the number of 32 byte global memory load and store transactions are much higher compared to number of 64 bytes and 128 byte global memory load/store transactions. Increasing the count of 128 byte global memory transactions and reducing 32 byte transactions should improve results.
\\
Another way could be to distribute the single centralized task queue into two centralized task queues. First task queue would contain $u$ values of task and the second task queue would contain $v$ values of task. This optimization would be basically synonymous with AOS (Array of structures ) to SOA (Structure of Arrays) optimization.

\begin{table}
  \centering
  \caption{Memory Coherency statistics}
  \begin{tabularx}{\textwidth}{c|c|c|c|c}
  \hline \hline
  \textbf{Data set} & \textbf{gld\_incoherent} & \textbf{gld\_coherent} & \textbf{gst\_incoherent} & \textbf{gst\_coherent}\\ \hline \hline
  Actors & 0 & 337167983 & 0 & 16894132\\ \hline
  Amazon & 0 & 355220513 & 0 & 26933520\\ \hline
  Patents & 0 & 2894521697 & 0 & 130256154\\ \hline
  Slashdot & 0 & 1111485764 & 0 & 29259288\\
  \hline \hline
  \end{tabularx}  
  \label{tab:gpgpucoherstats}
\end{table}

\begin{table}
  \centering
  \caption{Global Memory Load/Store Transactions}
  \begin{tabularx}{\textwidth}{c|c|c|c|c|c|c}
  \hline 
  \hline
  \textbf{data set} & \textbf{gld\_128b} & \textbf{gst\_128b} & \textbf{gld\_64b} & \textbf{gst\_64b} & \textbf{gld\_32b} & \textbf{gst\_32b}\\ 
  \hline 
  \hline
  Actors & 1155633 & 108 & 464835 & 14 & 335547515 & 8446606\\ 
  \hline
  Amazon & 2600465 & 3 & 1150847 & 0 & 377896253 & 13192445\\ 
  \hline
  Patents & 693274 & 49 & 221158 & 3 & 2944903315 & 65148117\\ 
  \hline
  Slashdot & 3407550 & 0 & 1555610 & 0 & 1198046922 & 14502431\\
  \hline 
  \hline
  \end{tabularx}  
  \label{tab:gpgpuloadstore}
\end{table}

\begin{table}
  \centering
  \caption{Global Memory Throughput}
  \begin{tabularx}{\textwidth}{c|X|X|X}
  \hline 
  \hline
  \textbf{data set} & \textbf{glob mem read throughput}(GB/s) & \textbf{glob mem write throughput}(GB/s) & \textbf{Total global memory throughput}(GB/s)\\ 
  \hline 
  \hline
  Actors & 8.59 & 0.21 & 8.8\\ 
  \hline
  Amazon & 5.87 & 0.2 & 6.07\\ 
  \hline
  Patents & 8.19 & 0.18 & 8.37\\ 
  \hline
  Slashdot & 13.27 & 0.16 & 13.43\\
  \hline 
  \hline
  \end{tabularx}
  \label{tab:gpgputhrough}
\end{table}

\item The instruction throughput(Table~\ref{tab:gpgpuinstrstats}) and overall global memory throughput(See Table~\ref{tab:gpgputhrough}) is very low compared to the peak hardware values(102.4 GB/s theoretical GDDR memory bandwidth for Tesla C1060) , again reiterating the fact the the triad census kernel is not compute bound but rather memory latency bound. There are number of ways to hide memory latency on CUDA devices. Increasing warp occupancy is one factor that can reduce memory latency and this requires reducing register pressure by using shared memory instead. Alternatively, we could increase per thread register usage with lower occupancy by either storing some per thread global memory data in registers or by increasing ILP somehow \cite{Volkov2010}. Other approaches to hide memory latency: Use shared memory for doing computations per thread block, data prefetching and reducing register dependent RAW stalls. 

\begin{table}
  \centering
  \caption{Instruction profile}
  \begin{tabular}{ c | c | c }
  \hline \hline
  \textbf{Data sets}	& \textbf{instructions} & \textbf{Instruction throughput}\\ \hline \hline
  Actors	&650649633	&0.03\\ \hline
  Amazon	&516564153	&0.02\\ \hline
  Patents	&4294967295	&0.03\\ \hline
  Slashdot	&2168289924	&0.06\\ 
  \hline
  \hline
  \end{tabular}  
  \label{tab:gpgpuinstrstats}
\end{table}  

\item \textbf{Divergent branches}: The average percentage of divergent branches in our triad census kernel is around 3\% (See Table~\ref{tab:gpgpubranchstats}). In addition to triad census kernel, there are number of device functions which carry branches :  \emph{IsEdge}, \emph{IsNeighbour} and \emph{set\_adj\_union}. In comparison to 1\% average divergence in many CUDA benchmark kernels \protect\footnote{\href{http://divmap.wordpress.com/home/divergence-data/}{http://divmap.wordpress.com/home/divergence-data/}}, our percentage divergence seems little high indicating that serialization of threads within a warp taking different paths. In worst case warp serialization can slow down by 32 times if all threads in a warp take 32 separate paths. In our case, as most branches have two paths, so a factor of 2x slowdown is possible.
\\
\\
There are different approaches to reduce divergence. We have already discussed some approaches(see section~\ref{subsubsec:cntrloptim}) for regularizing control flow. Other suggested approaches \cite{CudaPractices} to reduce branch divergence include using signed integers as loop counters, creating smaller branch-controlled compound statements so that compiler optimizes using predication, grouping threads according to branch path they tend to take and lowering total number of CUDA threads. The nvcc compiler driver can replace the instructions in different branch paths with predicated instructions, provided number of such instructions is lower than a threshold value. 

\begin{table}
  \centering
  \caption{Branch profile}
  \begin{tabularx}{\textwidth}{X|c|X|X|c|X}
  \hline \hline
  \textbf{Graph Network} & \textbf{Branches} & \textbf{Divergent Branches} & \textbf{\% Divergent Branches} & \textbf{instructions} & \textbf{\% Branches}\\ \hline \hline
  Actors	&105456730	&3368243	&3.10	&650649633	&16.73\\ \hline
  Amazon	&86825816	&3714570	&4.10	&516564153	&17.53\\ \hline
  Patents	&1163523028	&29532915	&2.48	&4294967295	&27.78\\ \hline
  Slashdot	&341669397	&7379273	&2.11	&2168289924	&16.10\\ 
  \hline
  \hline
  \end{tabularx}  
  \label{tab:gpgpubranchstats}
\end{table}

\item Looking at the memory load store (Table~\ref{tab:gpgpuloadstorestats}), we see that there are considerable number of warps whose execution serialized on address conflicts to either shared or constant memory. 

\begin{table}
  \centering
  \caption{Memory load store profile}
  \begin{tabular}{ l | c | c | c }
  \hline \hline
  \textbf{Data set} & \textbf{gld\_request} & \textbf{gst\_request} & \textbf{warp\_serialize}\\ \hline \hline
  Actors	&49017268	&419723	&0\\ \hline
  Amazon	&52952146	&705510	&2098473\\ \hline
  Patents	&549967365	&5281342	&3546663\\ \hline
  Slashdot	&163110847	&673969	&2285670\\
  \hline
  \hline
  \end{tabular}  
  \label{tab:gpgpuloadstorestats}
\end{table}

So far in our implementation we have used constant memory for storing mapping table with 64 entries of size 8-byte each to map non-isomorphic triads to isomorphic triads. Below is the relevant code fragment in which the different locations in  mapping table are accessed depending upon the value returned by \emph{TriadCode} function and also \emph{preTriadType} value

\begin{verbatim}
if( v < w || 
   ( u < w && w < v && !IsNeighbor( Neighbors, NeighborIndexes, u, w) ) )
{
 triadType = dIsoTriadMappingTable[preTriadType - 1 + 
                TriadCode(Edges, EdgeIndexes, u, v, w)];
 atomicAdd( &TriadCensus[triadType-1], 1 );
}
\end{verbatim}

Looking at the above code fragment, we can see that the mapping table in constant memory is being read by all the CUDA threads. CUDA C best practices guide\cite{CudaPractices} suggests the following approach to obtain best constant memory performance:
\\
\\
\emph{For all threads of a half warp, reading from the constant cache is as fast as reading from a register as long as all threads read the same address. Accesses to different addresses by threads within a half warp are serialized, so cost scales linearly with the number of different addresses read by all threads within a half warp.}
\\
\\
As different threads within a warp or half wrap will be possibly accessing different locations of the mapping table, all such reads to constant memory will be serialized. If all threads of a warp were to access the same location in mapping table, this implies we need to find a mechanism to group threads according to the location they are going to access. As there are 64 such locations, we need a mechanism to categorize all threads in 64 groups. But how do we know which thread is going to access which location ? This basically depends upon the value of following expression

\begin{verbatim}
TriadCode(IN Edges,IN EdgeIndexes,IN u,IN v,IN w) - 1 + preTriadType
\end{verbatim}

and which in turn depends upon whether \textit{u-v, v-u, u-w, w-u, v-w} and \textit{w-u} pairs are edges or not. Due to the very random nature of graph data, it is very difficult to predict the answers to these questions beforehand. Pre-computing these value may seriously  deteriorate overall performance. Therefore we rule out a mechanism to group these threads according to the locations they are going to access.
\\
\\
One way to avoid this performance bottleneck is to simply calculate non-isomorphic triads. By doing that, we don't need to do any mapping using the mapping table. 

Second approach is to replicate the mapping table for each thread block in shared memory. Shared memory is much suitable for concurrent access to different locations as it is physically organized in separate memory banks which can be accessed in parallel. Effectiveness of shared memory can reduce in case of bank conflicts though. Bank conflicts in shared memory can occur when simultaneous memory accesses from threads to different memory locations map to same memory bank. Bank conflicts lead to serialization of accesses to it, thereby reducing obtained performance. Shared memory conflicts \cite{CudaToolkit} can be prevented in two ways:

\begin{itemize}
\item Simultaneous memory access from threads map to different memory banks(16 banks in our GPGPU device)
\item Simultaneous memory load requests made by threads of a half-wrap map to a shared memory bank but all such memory loads are made for the \textit{same memory address location}. This leads to broadcast of loaded value to half-warp threads. More precisely, even if one of the memory loads is for a different memory address(mapped to same memory bank), this will lead to a bank conflict.
\end{itemize}

\end{enumerate}

\section{Optimizing Triad Census CUDA implementation}
\label{sec:gpgpuoptimiz} 
Based on our analysis in last section, we tried many optimizations. Some optimizations worked and some didn't. Below we give description of all the tried optimizations.

\subsection{Using shared memory to calculate Triad Census per Thread block}
\label{subsec:gpgpusharedmemoptim}
In this optimization we try to hide memory latency by using low latency and high throughput shared memory. Instead of all threads of all thread blocks updating the global triad census array through atomic instructions, we provide each thread block an empty Triad census array in shared memory where it can calculate its share of triad census array. Now only the threads local to a thread block can update their shared private triad census array in shared memory. When all the thread blocks are done creating their own Triad census, we merge the individual triad census results of different thread blocks to the global census array.
\\
\\
One issue we faced here was that for 1.3 compute capability devices, the \textbf{atomicAdd} function can work only on 64-bit integer values in global memory but not on 64-bit integer values in shared memory(works only on 32-bit integers). As some of the Triad census values for big graphs could easily overflow beyond the range provided by 32-bit unsigned integers, this could lead to incorrect results. Statistically speaking, triad types 1,2 and 3 tend to have very high census values in comparison to census values of remaining 13 (4 to 16) triad types. To resolve this issue, we calculated Triad census for triad types 2 and 3 using 64-bit global census array and not shared memory. In terms of code modifications it means that only the innermost atomicAdd function operated on shared memory (Line 22 in Figure~\ref{fig:multitriadcensus}) while the outermost atomicAdd function continued to operate on global triad census array (Line 18 Figure~\ref{fig:multitriadcensus}). 
\\
\\
After running experiments on this optimization, we saw that there was almost negligible performance impact  (< 1\%) of calculating triad types 2 and 3 in global census array instead of calculating it using shared triad census array. Consequently the shared memory consumption per thread block reported by nvcc compiler increased from 128 bytes to 192 bytes. 
\begin{verbatim}
ptxas info: Used 31 registers, 176+16 bytes smem, 512 bytes cmem[0], 
            24 bytes cmem[1]
\end{verbatim}
With this optimization, we obtained a maximum, average and minimum speedups of \textbf{2.45x, 2.05x, 1.64x w.r.t our baseline GPGPU implementation and 1.22x, 0.84x, 0.5x speedups w.r.t fastest Multicore implementation}. 
\\
\\
\textbf{With respect to sequential baseline results, the resulting maximum, average and minimum speedups were 58x(Google), 32x, 4.2x(patents)}. Table~\ref{tab:gpgpuoptimizedresults} shows speedups and absolute time elapsed values for different graph data sets. 
\\
\\
Comparing the speedup results of Triad Census CUDA implementation and Fastest Multicore Triad Census implementation(see column \emph{GPGPU Speedup w.r.t multicore} in Table~\ref{tab:gpgpuoptimizedresults}), we see that Tesla GPGPU on an average was 1.1x times faster. Only for patents graph data set we see that GPGPU speedup ran half the speed of Multicore. However, for most graph data sets, our CUDA implementation was equally fast as fastest Multicore results and for some graph data sets(Amazon, Google) it bettered the Multicore results.

\begin{table}
  \centering
  \caption{Speedups obtained by using shared memory to calculate Triad Census per Thread block}
  \begin{tabularx}{\textwidth}{c||X|X|X||X||X||X}
  \hline 
  \hline
  \textbf{Data sets} & \textbf{Total time(sec)} & \textbf{Algorithm exec time(sec)} & \textbf{Pre-process time(sec)}	 & \textbf{Speedup w.r.t GPGPU baseline} & \textbf{GPGPU Speedup w.r.t multicore} & \textbf{GPGPU Speedup w.r.t sequential baseline}\\ 
  \hline 
  \hline
  Actors	&7.08	&5.86	&1.22	&2.45	&0.78	&8.47\\ 
  \hline
  Amazon	&10.24	&6.76	&3.48	&2.41	&1.22	&38.48\\ 
  \hline
  Patents	&77.33	&62.63	&14.7	&1.68	&0.50	&4.19\\ 
  \hline
  Slashdot	&18.09	&17.62	&0.47	&1.64	&0.88	&21.72\\ 
  \hline
  Google	&209.03	&203	&6.03	&1.70	&1.04	&58.38\\ 
  \hline
            &	    &	    &\textbf{Average} & 2.05 & 1.10	& 32.81\\
  \hline 
  \hline
  \end{tabularx}  
  \label{tab:gpgpuoptimizedresults}
\end{table}

\subsection{Discussion on other optimizations}
\label{subsec:gpgpuotheroptimiz}
CUDA occupancy calculator shows that 50\% warp occupancy can be maintained with upto 8192 bytes of shared memory per thread block. So far we had used just 192 bytes per thread block, we can still utilize 8000 remaining bytes. With 8 byte integers, we can load maximum 1000 integer elements more in shared memory of each thread block. Total amount of shared memory in Tesla GPGPU is 480 KB with each SM having 16KB of shared memory. So maximum, we can store 60 K values in shared memory of type 8 byte integer.
\\
\\
We analysed different data structures placed in global memory that could possible be accommodated in shared memory. Data structures like \emph{Edges}, \emph{EdgeIndexes}, \emph{Neighbour}, \emph{NeighbourIndexes}, Task Queue were not feasible to be accommodated in available shared memory. We tried many combinations, data sharing and data reuse approaches to try and fit these data structures but all of these data structures either seemed too big to fit or they would fit by seriously lowering the total thread count or thread block size causing serious under-utilization of GPGPU hardware resources.
\\
\\
We found that we could only place two data structures completely in shared memory without causing any serious performance deteriorating trade-off's: 
\begin{itemize}
\item CRS row array storing indexes of centralized task queue
\item CRS row array storing indexes of maximum S set sizes vector
\end{itemize}
This optimization, however resulted in no performance improvements.
\\
\\
\textbf{Experimental optimizations that either reduced the performance or yielded insignificant performance improvement}

\begin{itemize}
\item \textbf{Reducing per thread register pressure}: We first tried to increase occupancy by forcing the compiler to use lesser number of registers per thread using the nvcc compiler flag --maxrregcount. In such cases, the compiler starts storing the thread local data in local memory. As we reduced the number of registers per thread from 31 to 16, 8 and 4, the amount of per thread  local memory increased correspondingly. Using occupancy calculator we tried different thread blocks corresponding to 100 \% occupancy. What we found was that even with 100\% occupancy, we couldn't hide memory latency and performance actually reduced as we reduced the number of registers per thread. 

\item \textbf{Using shared memory for mapping table}: We already saw that irregular access to mapping table stored in constant memory lead to warp serialization(caused due to address conflicts). In order to reduce warp serialization, we instead tried using share memory for storing mapping table for each thread-block. The resulting performance improvement was almost negligible , suggesting that warp serialization reduced just slightly. This also suggests that the irregular access of indexes now started causing some bank conflicts in shared memory.

\item \textbf{AOS to SOA conversion}: We tried replacing centralized task queue with two task queue by splitting the task <u, v> into two components. Now one queue just stores $u$'s in CRS column array while the other task queue stores corresponding $v$'s in another CRS column array. Also, as row index array remained the same, we can easily extract a task pair (u, v) from these two individual queues. When we tested this optimization, there were no improvements in performance. 

\item \textbf{Loop unrolling}: Referring to the work by \cite{Volkov2010}, unrolling deep nested loops could improve ILP per thread though this would mean more per thread register usage and decreased warp occupancy. If the benefits(memory latency hiding) of improved ILP and fast register memory are more than cost paid for reduction in warp occupancy(TLP), performance can be improved. In our Triad Census kernel, the innermost for-loop(Line 19--24 in Figure~\ref{fig:multitriadcensus}) couldn't be unrolled through CUDA unrolling pragma or hand-coded unrolling as the loop bounds are known only at run-time. Only loop that we could manually unroll was the outermost loop (line 6--26). By unrolling this outer for-loop X times, we started processing X tasks per loop iteration instead of one task per loop iteration with no unrolling. We tried loop unrolling factors of 2 and 3. With these unroll factors, the register per thread increased from 31(no unroll) to 42(2x unroll) and 51(3x unroll). However, we found that we couldn't gain any improvements in performance and at 3x unroll factor, performance actually reduced. However in our final implementation, we still decided to keep a 2x unroll factor as it had no negative impact on performance.

\item \textbf{Reducing Branch divergence}: In our initial GPGPU implementation, we already optimized branches in different device functions. This included branch elimination optimization that we had also successfully employed in our Multicore sequential triad census. Although Loop unrolling was also employed as explained previously, but we don't think it yielded any significant improvements to branch divergence. In addition, we tried hand-coded predication in triad census kernel and other device functions and implemented only the ones which gave us some benefits.
\end{itemize}

After unsuccessfully trying many optimizations, we carefully re-looked if any other optimizations might still improve performance but found that we had tried most of the optimization ideas that could possibly improve performance.  At this point, we are convinced that we have all the reasons to conclude that we have achieved the maximum obtainable performance of Triad Census algorithm on our Tesla GPGPU.

\begin{figure}[hb!]
    \centerline{\includegraphics[scale=0.6]{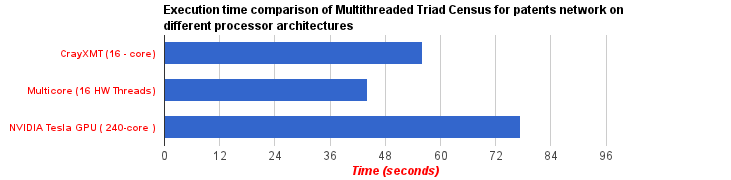}}
	\caption{Comparison of our results on Intel Xeon Multicore and Tesla GPGPU with previous work on Cray XMT}\label{fig:execcomp}
\end{figure}

\section{Summary}
\label{sec:gpgpusummary}
\begin{figure}[ht!]
    \caption{Comparison between GPGPU and Multicore w.r.t baseline sequential\label{fig:gpu-final-speedup}}
    \centerline{\includegraphics[scale=0.55]{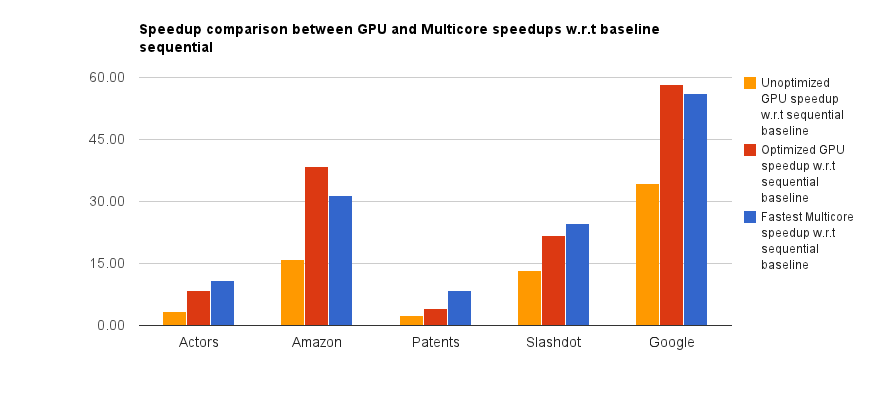}}
\end{figure}

To summarize the final results, figure~\ref{fig:gpu-final-speedup} shows the overall speedup results obtained which shows comparison between GPGPU and Multicore. Further, in previous work \cite{Chin2009} on accelerating Triad Census, authors used the same patents data set and accelerated it for 16 PE's of Cray XMT(128 PE's support). From previous work, we extracted the execution time for processing of patent data set. Figure~\ref{fig:execcomp} shows the comparison of our execution time results on Intel Xeon Multicore(with 16 threads) and GPGPU (30K threads) with previous results on Cray XMT(16 PE's). As we see, our results for patents data set on Intel Xeon Multicore are slightly better (1.27x times faster) than Cray XMT results while performance on GPGPU is comparatively slower(0.72x)

\chapter{Conclusion and future work}
\label{chap:conclusion}

The main contribution of this work is to accelerate an Irregular Graph Algorithm for Social Network Analysis called \emph{Triad Census algorithm} on Intel Xeon Multicore and NVIDIA Tesla GPGPU processor architectures. The main challenges in this thesis were firstly to address the processor - memory speed gap for algorithms which are bound by memory latency and secondly to evaluate and accelerate such an irregular algorithm on two orthogonal commodity processor architectures - Multicore and GPGPU. 
\\
\\
We started the work by focusing on the state of the art in graph theory, social network analysis and parallel graph algorithms. Then we concentrated on understanding the Triad Census algorithm and previous results on it. Thereafter we built good foundations on state of the art solutions in areas such as efficient graph data structures, processor architecture aware optimizations and reducing control-flow irregularities. After this, we spent effort on understanding the internals of Dual socket Quad core Intel Xeon based system(used for running experiments) and Advanced Processor architecture concepts. We started with a Sequential Triad Census algorithm and through profiling at application and instruction level, we found the targets for sequential optimization. We successfully applied many different optimization techniques such as branch elimination, cache fitting, Cache blocking etc. and obtained sequential speedups of maximum 6x, average 3.3x and minimum 1.3x for different real world social network graphs.
\\
\\
With good sequential speedup results, we started analysis for implementation of Multithreaded Triad Census algorithm for our Intel Xeon processor based Multicore system. We carefully analysed different ways of partitioning the graph data and balancing the workload across the threads. Based on this understanding, we implemented Multithreaded Triad census using distributed task queue approach. Through experiments and observations we deduced ways to decouple threads by reducing the amount of shared data between them, and also improved load balancing through experimentation and analysis. Overall, on Intel Xeon Multicores we were able to achieve \textbf{56x maximum, 33x average and 8.3x minimum speedups} for real world social network graphs.
\\
\\
By keeping speedup results on Multicores as a reference, we changed our focus to NVIDIA Tesla C1060 GPGPU. We spent significant effort on understanding many aspects of GPGPU's processor hardware and programming - differences in Multicore and GPGPU architectures, architecture details of Tesla GPGPU Architecture, CUDA Programming and memory model, CUDA thread scheduling, Thread Synchronization, state of the art techniques for optimizing CUDA applications. Based on our experiences on Multicore and good knowledge base developed on GPGPU's, we carefully crafted a CUDA implementation of Triad census algorithm. Through profiling and shared memory optimizations, initial results demonstrated speedups as competitive as Multicore results. We continued experimenting with all possible optimizations such as Loop unrolling, reducing branch divergence etc, exploring trade-off between register pressure and war occupancy etc. However after a lot of experiments and analysis we found that we could no longer extract any more performance out of Tesla GPGPU. Still, overall on the NVIDIA Tesla C1060 GPGPU, we were almost able to match the Xeon Multicore results - \textbf{58.4x maximum, 32.8x average and 4.2x minimum speedups}. 
\\
\\
In terms of raw performance, we found that for the graph data set called Patents network, our results on Intel Xeon Multicore(16 hw threads) were slightly better (1.27x times faster) than previous results on Cray XMT(16 hw threads) while resulting performance achieved on GPGPU was comparatively slower(0.72x). To the best of our knowledge, this algorithm has only been accelerated on a supercomputer class machine named Cray XMT and no work exists that demonstrates performance evaluation and comparison of this algorithm on relatively lower-cost Multicore and GPGPU based platforms.

\section{Future work}
Having said all this, there is always room to try to explore new paradigms and platforms in the Chip Multiprocessor landscape. We now briefly give an outline of our perspective on prospective future work

\begin{itemize}
\item \textbf{On Multicores}: As of now, the state of the art Multicore processor architecture from Intel is Ivy Bridge which is an entirely new micro-architecture. Also, all future architectures tend to be cc-NUMA style. In our current work we didn't explore NUMA specific optimization's but we think that such optimizations will become necessary as number of cores scale on the CMP's. On the algorithmic side, analysing the effect of graph topology on performance scalability of the application and doing cache oblivious analysis of the algorithm should open more avenues for improving application performance. Other algorithmic/implementation areas which can be explored on Multicores are: Exploring work or task stealing approach to improve load balancing and evaluating performance with threading libraries/frameworks such as OpenMP, Cilk and Intel TBB.
\item \textbf{On GPGPU's}: Since the release of NVIDIA Tesla GT200 GPU processor architecture, many more advanced GPGPU devices have been released by NVIDIA, AMD(Radeon) and more recently by Intel(Xeon Phi). From NVIDIA Fermi to NVIDIA Kepler, the GPGPU processor architectures have mainly seen a shift towards making them more memory-latency and programming friendly. For example dynamic nested parallelism is now possible on the latest NVIDIA GPGPU's and also cache hierarchy for on-chip shared memory is introduced to target applications sensitive to memory-latency. The number of threads, size and speed of on-chip and off-chip memory as well as SIMD resources have been improved, so as the speed of context switching and atomic instructions. It will be a good exercise to evaluate Triad Census or a similar irregular algorithm on latest GPGPU processor and a Multi-GPU system.
\end{itemize}

\appendix

\chapter{Concurrency and Parallelism}
\label{app:concpar}

\textbf{Amdahl's Law} \cite{SutterLarus2005, Grama.etal2003, Hennessy.Patterson2006}\\
It is a simple observation which quantifies the overall possible speedup in execution that can be gained when a given part of the execution is accelerated. Normally this speedup is provided by doing a part of the total work in parallel. 

\begin{verbatim}
        Speedup = ( S + P ) / ( S + P/N )
        
where,        
    T = S + P = Total execution time units
    S = Execution time units of serial part of T
    P = Execution time units of parallelizable part of T
    N = Factor by which P can be accelerated
\end{verbatim}
    
Amdahl's law assumes that total workload is fixed and thus focuses on strong scaling. Also see Gustafson's Law. 
\\
\\
\textbf{Asynchronous Execution} \cite{Herlihy.Shavit2008,Breshears2009}\\
Tasks are said to be executing asynchronously if they may execute at different speeds and each of
it can halt any time for an unpredictable amount of time due to unpredictable events (cache miss,
pre-emption, page-fault etc.). Different tasks thus may make progress independently of the each other.
\\
\\
\textbf{Atomicity}\\
A computation is said to be atomic, if once it starts executing, it is allowed to execute to completion
without any interruption. Atomicity comes in flavours of atomic variables, atomic statements, functions etc. An atomic computation cannot be subdivided further into smaller computations and for this granularity of computation even the operating systems has to wait for atomic computation to complete its execution
before it can interrupt it.
\\
\\
\textbf{Barrier Synchronization} \cite{Chapman2007, Herlihy.Shavit2008, Rauber.Ruenger2010}\\
A synchronization mechanism in which all threads or tasks must meet at a point called Barrier before any of them is allowed to make any further progress. Each thread waits for all other threads by blocking itself. A blocked thread waiting for other threads will unblock only if it is sure that all other threads have reached the barrier. Each thread thus must execute all operations or code before the barrier. Some barriers
are implicit while others can be specified explicitly in implementation. A barrier is thus a mechanism
to enforce synchronization of asynchronously executing independent threads.
\\
\\
\textbf{Blocking} \cite{Li.Yao2003,Buttazzo2004,Ben2006}\\
We say a task is blocked or enters a blocked state when it is waiting for a resource and cannot
continue execution until it gets access to that resource. The resource could be semaphore, response from an I/O device or memory, an interrupt (say from timer) or event or message etc. As soon as the resource becomes
available, the task is unblocked. An unblocked task can become a ready task by moving to a priority queue of ready tasks or can directly move to running state(by pre-empting current running task) if it is the task with highest priority compared to all ready tasks and the currently running task.
\\
\\
\textbf{Concurrent Program and Computation} \cite{Ben2006}\\
A concurrent program consists of a finite set of (sequential) processes. Each process is written using a finite set of atomic statements. The execution of a concurrent program proceeds by executing a sequence of the atomic statements obtained by arbitrarily interleaving the atomic statements from the processes. A \emph{computation} is an execution sequence that can occur as a result of the arbitrary interleaving of atomic statements of the processes of the program. For a set of N-concurrent tasks $T_{1}$, $T_{2}$,..,$T_{N}$ executing on a single PE, each having $S_{1}$, $S_{2}$, ..., $S_{N}$ number of atomic computations, the total number of possible interleaving are:
\\
\\
\textit{Number of Possible Interleaving} = $( S_{1} + S_{2} + ... + S_{N} )! / ( S_{1} * S_{2} * ... * S_{N} )$
\\
\\
\textbf{Concurrency} \cite{SutterLarus2005, Breshears2009, Magazine2009,Sutter2010,Hughes.Hughes2003}\\
As set of tasks or threads are said to be concurrent iff
\begin{itemize}
\item They execute and make progress at the same time or within the same time interval. By saying same time or same time interval we mean that they make progress together but they may or may not be executing at the same time instant.
\item They can be executed sequentially through interleaved execution (non-deterministic ordering) on a single
threaded uniprocessor but they can also be executed in parallel if multiple PE's are available as in Multicore, Manycore or Multithreaded processor.
\end{itemize}

Concurrency thus can imply either interleaved (pseudo-parallel) or parallel execution and is more of property of the algorithm. Concurrency as a concept is more general than parallelism.
\\
\\
\textbf{Data Level Parallelism} \cite{Reinders2007}\\
In this type of parallelism, same task or operation is executed in parallel on different data streams. In essence, they are similar to SIMD or SPMD as they perform same operations on different data sets. Vector processing is yet another form of data level parallelism. 
\\
\\
Decomposition of problem into tasks is usually based on data partitioning followed my static or semi-static mapping of tasks to threads or processes. Also the task interactions are generally insignificant in data parallel applications making them more amenable to coarse grained parallelism.
\\
\\
\textbf{Data reuse}\\
Data re-usability is central to obtaining high performance on high performance processor architectures. It cane be categorized in four categorize which is Cartesian product between two sets: \textbf{Locality set} - \emph{\{spatial, temporal\}} and \textbf{Data usage mode set} - \emph{\{self, group\}}:

\begin{itemize}
\item \emph{Self temporal use}: Accessing same memory location in consecutive iterations of the loop. Example: accessing a variable, array element, structure member in consecutive iterations.
\item \emph{Self spatial reuse}: Accessing consecutive memory locations in the same cache line in the same loop iteration. Example: access of consecutive array elements in same loop iteration
\item \emph{Group temporal reuse}: Accessing group of memory locations in consecutive iterations of a loop. Example: accessing same set of array elements, structure members in consecutive iterations.
\item \emph{Group spatial reuse}: Accessing a group of memory locations that belong to the same cache line in the same iteration.
\end{itemize}~\\
\\
\\
\textbf{Distributed Memory}\cite{Grama.etal2003}\\
Some parallel architecture's have memory which is \emph{physically distributed} among PE's such that each PE has its own cache-hierarchy and main memory. Depending upon whether the address space is local or globally shared, inter-processor communication occurs through explicit message passing or through shared address space.
\\
\\
\textbf{Granularity} \cite{Grama.etal2003, Breshears2009}\\
It is defined as the amount of concurrent computation is done before synchronization or communication is needed between concurrently executing computations. The longer the time between synchronizations or computation, the coarser is the granularity and vice versa. In simple terms it is just ratio of computation to communication. The number and size of tasks into which a problem is decomposed determines the granularity of the decomposition. Decomposition into a large number of small tasks is called \textbf{fine-grained} and decomposition into a small number of large tasks is called \textbf{coarse-grained}.
\\
\\
\textbf{Gustafson's Law} \cite{Reinders2007}\\
This was an observation made by John Gustafson from Sandia National Labs. It basically complements the Amdahl's law. Gustafson observed that speedup scales with number of cores only if the parallel workload increases at a faster rate relative to serial workload. It this focuses on \emph{weak scaling}.
\\
\\
The difference between Amdahl's and Gustafson's law lies in whether one wants to make an existing program run faster with the same workload(see strong scaling) or whether one envisions working on a larger workload. The scalability of an application comes down to increasing the work done in parallel and minimizing the work done serially.
\\
\\
Amdahl motivates us to reduce the serial portion, whereas Gustafson tells us to consider larger problems, where the parallel work is likely to increase relative to the serial work. Making today's application run faster by switching to a parallel algorithm without expanding the problem size is harder than making it run faster on a larger problem.
\\
\\
\textbf{Latency}\\
It is another term used for delay. It is the time elapsed between the instant an item of an input data set enters a processing system to the time instant when this item comes out after processing.
\\
\\
\textbf{Multitasking}\\
Multitasking allows more than one task to execute at the same time on a single processor by interleaving their execution. Each task executes for a designated time interval. When the time interval has expired or some event occurs, the task is removed from the processor and another task is assigned to it. Multitasking can be implemented through multithreading
\\
\\
\textbf{Multithreading (Software)}\\
Operating systems allow user to achieve multi-tasking through Multithreading by providing API or programming language constructs. Essentially Multithreading involves concurrent execution of threads belonging to a multithreaded process. For communication and synchronization, specific constructs are provided.
\\
\\
\textbf{Mutex}\\
A Mutex is a programming abstraction for implementing the concept of mutual exclusion.
\\
\\
\textbf{Mutual Exclusion} \cite{Ben2006, Hughes.Hughes2003}\\\
It is the process of preventing concurrently executing threads or tasks to simultaneous access a shared resource or critical section. In other words, atomic statements from the critical sections of two or more processes must not be interleaved.
\\
\\
\textbf{Non-determinism} \cite{Chapman2007, Herlihy.Shavit2008, Breshears2009, Reinders2007}\\
Same program or a system for same input set gives different output at different times of execution. Non-deterministic behaviour of multi-threaded programs occurs due to races.
\\
\\
\textbf{Parallelism} \cite{SutterLarus2005, Breshears2009, Magazine2009, Sutter2010, Hughes.Hughes2003}\\
Two or more tasks or threads are said to be executing in \emph{parallel} if they execute \emph{simultaneously}. Simultaneous execution means making progress together at every instant of time. Parallelism can be achieved by running each thread of the process on a separate processor core or hardware thread.
\\
\\
Parallelism thus implies \emph{simultaneous or overlapped execution} and requires separate processor core or hardware thread for each software thread. Also, it is a subset of Concurrency.
\\
\\
\textbf{Pipelining} \cite{Reinders2007, Kaeslin2008}\\
It is a fundamental technique to improve throughput of a processing system where a stream of data is processed in stages such that output of each stage is fed as input to the next stage. Every pipeline stage acts as a consumer of input data and producer of output data. The different pipeline stages may take different execution times and execute in parallel either synchronously based on a single clock or asynchronously by communicating via request acknowledgement system. The pipeline stages are separated with buffers (E.g.: registers) to store the data temporarily.
\\
\\
To increase the speed of a pipeline, one would break down the stages into smaller and smaller units, thus lengthening the pipeline and increasing overlap in execution. Speed of a pipeline is ultimately limited by the longest processing atomic stage in the pipeline. So essentially we can picture pipelining a special case of task parallelism where there is a dependency between tasks. It is sometimes also called \emph{Fine-grained stream parallelism} because of frequent interactions between tasks which are involved in processing a data stream.
\\
\\
\textbf{Preemption} \cite{Grama.etal2003, Li.Yao2003, Buttazzo2004, Hughes.Hughes2003}\\
It is operation of suspending a running task by interrupting (using clock) its execution, saving its context in its TCB, inserting the task it into priority queue of ready tasks and then allocating processor to another task(if available and ready). Preemption is done in order to allocate processor to second ready task either because the second task has higher priority or because the time slice of the first task has expired or because of decision of OS according to its scheduling policy.
\\
\\
\textbf{Re-entrant Code or Re-entrancy} \cite{Herlihy.Shavit2008, Sutter2010,Hughes.Hughes2003}\\
A block of code is considered re-entrant if it can be safely executed concurrently by multiple threads or if the code cannot be changed while in use. By saying that "code cannot be changed" it basically means that the threads invoking the same function either do not change any shared modifiable data or, there is just no modifiable data shared between them. Re-entrant code avoids race conditions by removing references to global variables and modifiable static data. Therefore, the code can be shared by multiple concurrent threads or processes without a race condition occurring. Enabling conditions for re-entrancy of a code block or function are:
\begin{enumerate}
\item Must hold no static (or global) non-constant data.
\item Must not return the address to static (or global) non-constant data.
\item Must work only on the data provided to it by the caller.
\item Must not rely on locks to singleton resources.
\item Must not modify its own code (unless executing in its own unique thread storage)
\item Must not call non-re-entrant computer programs or routines.
\end{enumerate}

Therefore, re-entrancy is a more fundamental property than thread-safety and by definition, leads to thread-safety: \emph{Every re-entrant function is thread-safe; however, not every thread-safe function is re-entrant}
\\
\\
\textbf{Shared Memory or Shared Address Space} \cite{Grama.etal2003}\\
It is a mechanism for inter-process communication (IPC) whereby two or more address spaces that are protected from each other are allowed to intersect at points, still retaining protection over the non-intersecting regions. It opens up a pre-defined portion of a process's address space; the rest of the address space is still protected, and even the shared portion is only unprotected for those processes sharing the memory. Shared memory also reduces requirements for physical memory for example a multi-threaded program typically has number of threads which share the same address space of instructions. The key mechanism that enables shared memory is the virtual memory mechanism more precisely page table organization structure. 
\\
\\
\textbf{Strong Vs Weak Scaling} \cite{IJBush}\\
When the problem size stays fixed but the number of PE's are increased, the observed scaling is \textbf{Strong scaling}. Strong scaling of a program is said to be linear if the speedup is equal to number of PE's. Strong scaling efficiency is calculated as follows:
\\
\\
\textit{Strong scaling efficiency} = $ ( t_{n}^{s} * N )/t_{1}^{s}  * 100\% $
\\
\\
where,
\\
\\
$t_{1}^{s}$ = Time to execute a given workload with single PE\\
$N$ = number of PE's\\
$t_{n}^{s}$ = time to execute same workload with N PE's
\\
\\
On the other hand, if the workload assigned to each PE remains constant but total workload is increased in proportion with PE's than scaling observed is called \textbf{Weak Scaling}. For a program, weak scaling is said to be linear if total execution time remains constant while total workload is increased in proportion to increase in PE's.
\\
\\
\textit{Weak scaling efficiency} = $ t_{n}^{w} / t_{1}^{w}  * 100\% $
\\
\\
where,
\\
\\
$t_{1}^{w}$ = Time to execute unit workload with single PE\\
$N$ = number of PE's\\
$t_{n}^{w}$ = time to execute $N$ unit workload with $N$ PE's
\\
\\
Deviation from linear weak scaling indicates that either algorithm is not truly \BigO{N} or parallel overhead is increasing.
\\
\\
\textbf{Slowdown Theorem or Brent's Theorem}\\
When $q$ instead of $p$ processors are used, such that $q < p$, the slowdown is at most $p/q$.
\\
\\
\textbf{Speedup}\\
Speedup is a relative term expressed as ratio of time taken to execute an application for a given
workload on one processor to time taken to execute it on p-processors
\\
\\
\textit{Speedup} = $T_{1} / T_{p}$
\\
\\
Where,\\
\\
$T_{1}$ = Execution time on 1-PE
$T_{p}$ = Execution time on p-PE's
\\
\\
If the application is able to achieve a speedup

\begin{itemize}
\item Proportional to p, the speedup is called \textbf{Linear Speedup}.
\item Exactly equal to p, the speedup is called \textbf{Perfect linear speedup}
\item Greater than p, the speedup is called \textbf{Super-linear speedup}
\end{itemize}
~\\
\\
\\
\textbf{Synchronization} \cite{Grama.etal2003, Hughes.Hughes2003}\\
Synchronization is the process of coordinating execution of multiple concurrently executing threads or tasks. Synchronization is essential in order to ensure correctness of the application. Essentially, there are two variants of Synchronization

\begin{itemize}
\item \textbf{Data Access Synchronization}: Data synchronization is needed in order to control race conditions and allow concurrent threads or processes to safely access a block of memory. Data synchronization controls when a block of memory can be read or modified. We can identify four types of data access synchronization (Cartesian product of R/W access and E/C access) required by concurrently executing tasks: \emph{EREW}, \emph{ERCW}, \emph{CREW} and \emph{CRCW} (Key: C - Concurrent, E - Exclusive, R - Read, W - Write)

\item \textbf{Task Sequencing Synchronization}: It involves coordinating the order of execution of concurrent tasks. Four types of synchronization relationship exists between any two tasks whose order of execution is to be coordinated:

\begin{itemize}
\item \textbf{SS (Start-to-Start)}
\item \textbf{FF (Finish-to-Finish)}
\item \textbf{FS (Finish-to-Start)}
\item \textbf{SF (Start-to-Finish)}
\end{itemize}
\end{itemize}
~\\
\\
\\
\textbf{Task or Thread} \cite{Li.Yao2003}\\
The term \emph{task} is used in the theory of concurrency, while the term thread is commonly used in programming languages. The terms can thus be used interchangeably.
\\
\\
A \textbf{Task or Thread} is a block of code or instructions that form part of a process such that, it can be scheduled for execution by the operating system scheduler and can compete with other threads of same or different process for processor execution time. Each thread represents a flow of control or a unit of computation.
\\
\\
A process can be broken up into \emph{multiple threads} to make it a \emph{Multi-threaded Process}. The kernel stores and maintains dynamic information or state for each thread of a process called \emph{Thread Context} in a data structure called \emph{Task Control or Information Block (TCB or TIB)}. Threads of a process called as \emph{Peers} normally have following common characteristics:

\begin{itemize}
\item Have a TCB that consists of a program pointer, register set, state, priority and stack.
\item Are part of address space of the process
\item Are independent from main process control flow
\item Share data segment of the process
\item Each thread stack is part of process Stack Segment
\item Can be controlled by parent process
\item Can access and modify resources of the parent process only
\item Can acquire or create new resources
\item Can communicate and control other peers i.e. create, suspend, resume, terminate peers
\item Have a scope: Process or System level
\end{itemize}
~\\
\\
\\
\textbf{Task or Thread Scheduler} \cite{Li.Yao2003, Buttazzo2004, Ben2006, Hughes.Hughes2003}\\
A Task scheduler forms part of operating system kernel which provides the scheduling algorithms needed to determine which task or thread executes when. For a given set of tasks, a scheduling algorithm determines order of assignment of task to the processor so that each task is executed until completion. Most non-real time operating systems schedulers these days support at least the following types of scheduling:

\begin{itemize}
\item Pre-emptive priority based
\item Round-robin
\item FIFO
\end{itemize}
~\\
\\
\\
\textbf{Task Level Parallelism} \cite{Reinders2007}\\
For a given data set, we perform different and in-dependent tasks on the all elements in the data set in parallel. These separate tasks thus share the same data set. So they perform different operations on elements of same data set.
\\
\\
\textbf{Thread Safe} \cite{Herlihy.Shavit2008, Sutter2010, Hughes.Hughes2003}\\
\begin{itemize}

\item When code is written in such a way that it cannot be run in parallel without the con-currency causing problems, it is said not to be thread-safe.

\item According to Klieman, Shah, and Smaalders (1996): "\emph{A function or set of functions is said to be thread safe or re-entrant when the functions may be called by more than one thread at a time without requiring any other action on the caller's part}"

\item A thread-safe function is one that can be safely (i.e., it will deliver the same results regardless of whether it is) called from multiple threads at the same time.

\item If the functions are not thread safe, then this means the functions contain one or more of the following: static variables, accesses global data, and/or is not re-entrant.

\item Essentially means that each function can be invoked by more than one thread at the same time i.e. re-entrant code are thread safe. Be sure to use thread-safe libraries.

\item Thread safety only concerns the implementation of the function and not its external interface.

\item Any function that maintains a persistent state between invocations requires careful writing to ensure it is thread-safe. In general, functions should be written to have no side effects so that concurrent use is not an is-sue. In cases where global side effects — which might range from setting a single variable to creating or deleting a file - do need to occur, the programmer must be careful to call for mutual exclusion, ensuring that only one thread at a time can execute the code that has the side effect, and that other threads are excluded from that section of code.

\item If it is not known which functions from a library are thread safe and which are not, a programmer has three choices:

\begin{itemize}
\item Restrict use of all thread-unsafe functions to a single thread.
\item Do not use any of the thread-unsafe functions.
\item Wrap all potential thread-unsafe functions within a single set of synchronization mechanisms.
\end{itemize}

\item An additional approach is to create interface classes for all thread-unsafe functions that will be used in a multithreaded application. The unsafe functions are encapsulated within an interface class. The interface class can be combined with the appropriate synchronization objects through inheritance or composition. The interface class can be used by the host class through inheritance or composition. The approach eliminates the possibility of race conditions.
\end{itemize}
~\\
\\
\\
\textbf{Throughput}\\
It is defined as number of operations completed per unit time. It is one of the metric of system performance. It can be expressed as number of operations performed per unit time for in terms of floating point operations per second(FLOPS): \emph{GFLOPS}(Giga-), \emph{TFLOPS}(Tera-), \emph{PFLOPS}(Peta-)
\chapter{Processor Architectures and Micro-architectures}
\label{app:procarchmicro}

\textbf{Asymmetric Multiprocessors}\\
See \textit{Homogeneous VS Heterogeneous CMP}
\\
\\
\textbf{Cache Coherency} \cite{Hennessy.Patterson2006, Baer2010}\\
In shared memory or shared address space CMP architectures, each PE may locally cache data which is shared with other PE's. As each PE can independently update its own cache(or shared memory depending upon cache-write policy), other PE's may be using older cached copies of the same memory location. This problem of maintaining consistent view of the shared memory across PE's of a shared address space CMP is called \textbf{Cache coherency}. Depending upon the logical and physical structure of CMP as well as the cache protocol, there are different choices of cache coherence protocols. \textbf{Snoopy cache coherence protocol} is implemented in shared bus CMP's where cache controllers communicate with each other through shared bus to maintain coherency of their caches. \textbf{Directory Protocol} is suitable for architectures where broadcasts or multi-cast between cache controllers of respective PE is not straight-forward. Instead, a directory maintains state information of each shared memory block which may be cached.
\\
\\
\textbf{Chip Multiprocessors (CMP)}\\
It is a kind of Multiprocessor in which multiple processors are packed on a single chip. For example \textit{Multicore} and \textit{Manycore} processors. Also see \textit{Multicore Vs Manycore Processors}
\\
\\
\textbf{Chip Multithreading(CMT)} \cite{CMT2008}\\
Chip multithreading is basically terminology used by SUN to describe architecture of its UltraSPARC T1/T2 processors. According to SUN, CMT combines CMP with \textit{fine-grained multi-threading}. For example UltraSPARC T2 processor has eight processor cores each supporting two instruction pipelines, with four hardware threads per pipeline. In effect up to 64 hardware threads can be executed on the processor. The SUN CMT's are focused towards high-end server market where high throughput is more essential than latency and it calls the overall concept as "\textit{Throughput computing}".
\\
\\
\textbf{Die} \cite{Kaeslin2008, Wolf2008}\\
Integrated circuits are manufactured jointly on single piece of thin semiconductor wafer (typically silicon) of diameter of 200mm to 300mm. Later this wafer is cut apart into multiple pieces each with identical integrated circuits. This piece is what is known as a \textbf{Die}.
\\
\\
Later, one or more of these dies can be combined with similar or different dies on a hermetic package to form a chip or microchip. The I/O circuit terminals of the on-die integrated circuits are brought out from package through pins. Size of a die can range from a pinhead to a postage stamp. Many sub-functions maybe integrated on a single chip: processor cores, SRAMs, FIFOs, phase-locked loops (PLLs), RF and analog devices, standard interfaces (PCI, USB, Ethernet, WLAN, etc.), FPGA's, custom accelerators, etc.
\\
\\
\textbf{Flynn's Taxonomy} \cite{Grama.etal2003, Hennessy.Patterson2006, Baer2010, Mathew.Franklin2008, Smotherman2008}\\
As early as 1966, Michael J. Flynn classified computer organizations into four categories according to the uniqueness or multiplicity of instruction and data streams. By instruction stream is meant the sequence of instructions as performed by a single processor. The sequence of data manipulated by the instruction stream(s) forms the data stream. Accordingly, it classifies multiprocessor architectures in four categories: \textbf{SISD}, \textbf{SIMD}, \textbf{MISD} and \textbf{MIMD}
\\
\\
Key\\
S - Single, M - Multiple\\
D - Data Stream, I - Instruction Stream\\
\\
\\
\textbf{GPGPU} \cite{Nickolls.Dally2010}\\
General purpose GPU exemplifies massively parallel multithreaded GPU processor used for General purpose computing. It is essentially meant for data and control intensive applications unlike the traditional graphics applications which are mostly data parallel or task parallel.
\\
\\
\textbf{Hardware Multithreading} \cite{Hennessy.Patterson2006, Baer2010}\\
Processors with hardware multithreading basically support execution of multiple hardware threads that can share functional units. Each of these hardware threads has associated architectural state: instructions, data, registers, PC, page table etc. Such hardware threads arise from a software thread either explicitly created by the user or by the compiler. Basic motives behind hardware multi-threading is to exploit thread level parallelism to hide memory latency and increase throughput through concurrent execution of many hardware threads, improve utilization of hardware resources ( caches, tlb's, alu's etc.) of a PE by sharing them among HW threads.
\\
\\
Hardware multi-threading requires that every hardware thread has its own thread context or state stored in its private registers, PC's, etc requiring some extra hardware for this and also hardware required for choosing next thread and switching between threads.
\\
\\
Many variations of hardware multi-threading exists namely \textbf{Temporal Multi-threading}( Fine-grained and Coarse-grained ), \textbf{Simultaneous Multithreading}, \textbf{Chip multithreading}(CMT), \textbf{Speculative multithreading}.
\\
\\
\textbf{Homogeneous VS Heterogeneous CMP} \cite{Levy.Conte2009, Liu2008}\\
\textbf{Homogeneous CMP's} have several identical processor cores integrated on the same chip. For example Intel Multicores (Dual/Quad/Oct Core), AMD Opteron Multicores, NVIDIA Fermi GPU Architecture ( 512 Streaming Multiprocessor Cores). Homogeneous architectures are suitable where flexibility of programmability, portability and average performance/power for multiple domains is essential.
\\
\\
In \textbf{Heterogeneous CMP} architectures, different types of cores are integrated on the same chip. In essence, each class of core has its own instruction set and is meant to serve a given functionality say SIMD cores for multimedia (audio/video/image) vector processing, Superscalar cores for control-intensive algorithms. Such architectures have been popular in embedded domain as set of applications to be executed are fixed and high performance with low power consumption for each computation domain is an essential requirement.

Easy programmability and portability is still an unsolved problem for such architectures. However some trends indicate that desktop and server processor architectures may move towards heterogeneous approach. E.g. Notable example is IBM's Cell BE which has one Multithreaded Power Architecture based PE and multiple RISC based co-processors called SPE (Synergistic Processing Element). Another example is Imagine processor from Stanford. NVIDIA's planned future GPU architecture named Echelon project is a Manycore CMP with heterogeneous architecture. It will consist of 1024 Streaming Multiprocessor Cores and 8 latency optimized cores similar to cores of an SMP.
\\
\\
\textbf{Hyperthreading }(HT)\\
Also see \textit{Simultaneous Multithreading(SMT)}
\\
\\
\textbf{ILP (Instruction Level Parallelism)} \cite{Hennessy.Patterson2006, Mathew.Franklin2008}\\
At this level of parallelism, multiple independent instructions can be executed in parallel using pipelined and replicated hardware resources. There are two basic approaches to exploit ILP: static-compiler based (Intel's EPIC and VLIW) and dynamic-hardware based (Superscalar).
\\
\\
For Superscalar processors with multi-instruction issue per clock-cycle, ILP is limited by true data dependencies (RAW or Read After Write) causing data hazards, resource or structural dependencies (limited number of functional units) causing structural hazards, anti-data dependency (WAR or Write After Read) and output dependency ( WAW - Write After Write), control or branch dependencies.
\\
\\
\begin{itemize}
\item True data dependencies (RAW) are eliminated or reduced by register forwarding and bypassing, static and dynamic instruction scheduling (score boarding), out-of-order issue and by interleaving multiple independent threads of instructions on a single processor
\item Structural or Resource dependencies are eliminated or reduced by duplicating functional units like multiple ALU's, multiple memory ports for each cache, separate data/instruction cache, multiple pipelines each mapped to different functional units(integer vs. floating point)
\item Control or Branch dependencies are eliminated or reduced by code optimization's like loop transformations like unrolling, predicated instructions and branch predictors for speculative execution (branch prediction, value prediction, pre-fetch data)
\item False (Anti - WAR and Output - WAW) dependencies create problems with out-of-order scheduling and are eliminated by techniques like register and memory renaming hardware
\end{itemize}
~\\
\\
\\
\textbf{Massive Parallel Processors (MPP)}\\
See \textit{Chip Multi-processors} and \textit{GPGPU}
\\
\\
\textbf{MIMD} \cite{Grama.etal2003, Hennessy.Patterson2006, Baer2010}\\
It is a generic class of multi-processor architecture in Flynn's Taxonomy. In a MIMD type multiprocessor, each PE can execute a different instruction stream independently of other PE's. In practice, this means that a multiple instruction multiple data (MIMD) system could be running different programs on each processor or different parts of the same program.
\\
\\
MIMD computers exploit thread (or task) level parallelism, since multiple threads or tasks operate in parallel. MIMD are basically classified as \textit{Shared Memory} and \textit{Distributed Memory}

\begin{itemize}
\item \textbf{Shared Address Space MIMD}: In a shared memory system, all processes share a single address space and communicate with each other by writing and reading shared variables. UMA (Uniform Memory Access) and NUMA (Non-Uniform Memory Access) are two major classes of shared memory systems classified according to whether access to shared memory by PE's is uniform or non-uniform.
\item \textbf{Distributed Address Space MIMD}: In a distributed address space MIMD systems, each process has its own address space and communicates with other processes by message passing (sending and receiving messages).
\end{itemize}
~\\
\\
\\
\textbf{MISD} \cite{Grama.etal2003, Hennessy.Patterson2006, Baer2010, Mathew.Franklin2008,Smotherman2008}\\
It is a class of multi-processor architecture in Flynn's Taxonomy. As per there is no exact realization of a MISD processor. Only Streaming Processors and to some extent VLIW processor comes can be considered close to this category.
\\
\\
\textbf{Multicore}\\
A Multicore processor typically has multiple dies on the same chip package. Each die contains one or more processor cores. These multiple dies are encapsulated on a chip package and interconnected through an on-chip interconnection network or shared memory interface.
\\
\\
\textbf{Multicore Vs Manycore Processors} \cite{Nickolls.Dally2010, Kirk.Hwu2010, Levy.Conte2009}\\
There are some differences between these two kinds of processors although both are essentially based on SIMD/SPMD paradigm. A \textit{Multicore Processor} (e.g.: Intel 2/4/6[i7] Core, AMD Opteron 4/8/12 core), has small number of \textit{big processor cores} on a single die. Also there maybe multiple dies on the same chip. Each processor core on the die is a microprocessor in itself but cores on the same die can share some architectural components with lower level memory (L2/L3 cache), memory management etc. It is meant mainly for general purpose computing which are data and control intensive in nature.
\\
\\
On the other hand \textit{Manycore Processors} (e.g.: GPGPU from NVIDIA, AMD-ATI etc.) have large number of \textit{small processor cores} on the chip die. They are meant mainly for Data intensive applications close to SIMD or SPMD paradigm.
\\
\\
\textbf{Multiprocessors}\\
It is a very broad term which encapsulates computing systems that leverage multiple processors to perform computations. In such a system, multiple processors maybe organised with different physical configurations. For example Chip Multiprocessors, Multicore Processor, Manycore Processor, Clusters. \textit{Chip Multiprocessors} have all processor cores on the same chip. \textit{Multicore Processors} (a kind of CMP) on the other hand have small number of big processor cores on a single die.
\\
\\
\textbf{Multi-threaded processors}\\
See \textit{Hardware Multi-threading}
\\
\\
\textbf{NUMA Processors}\\
A NUMA CMP is a shared memory or shared address space architecture. Each PE has its own local cache hierarchy, local main memory and a globally shared address space. The global shared memory Local memory access is thus faster than non-local/shared memory. The memory latency in NUMA architecture thus depends upon the relative position of memory with respect to PE.
\\
\\
One of the advantages of NUMA is that the architecture scales in terms of memory bandwidth and local memory latency as number of PE's scale. Disadvantage is that it requires resource (interconnect, memory) replication as number of PE's is increased and so its cost is higher (area, power). Another disadvantage is that they cannot be programmed easily unless problem of \emph{cache coherency} is addressed (through inter-processor communication either programmatically in software or through hardware as in cc-NUMA architectures)
\\
\\
\emph{A NUMA is thus a shared-address-space (logically) CMP but with distributed-memory (physically)}
\\
\\
\textbf{Out-of-Order Issue/Execution} \cite{Hennessy.Patterson2006, Mathew.Franklin2008, Smotherman2008}\\
If a processor supports out of order issue/execution, then it has the ability to dynamically perform dependency analysis on a given window of instructions in the instruction stream and then issue/execute later independent instruction(s) ahead of an earlier instruction which is waiting for operands and functional units. 

Out of order issue/execution is also sometimes referred to as \textit{dynamic instruction issue/execution}. Generally long latency memory operations are launched earlier and instead of waiting for these memory operations to complete, other instructions not dependent on the previously launched memory operations are launched.
\\
\\
\textbf{Processor Architecture}\\
Processor Architecture refers to the instruction set, registers, and memory data-resident data structures that are public to a programmer. Processor architecture maintains instruction set compatibility so processors will run code written for processor generations, past, present, and future.
\\
\\
\textbf{Processor Core or Processing Element (PE)}\\
A processor core or PE is a monolithic processor that may implement architectures like Superscalar, VLIW, SIMD or Multi-threaded.
\\
\\
\textbf{Processor Coupling} \cite{Grama.etal2003}\\
The processor cores on a Multiprocessor may be tightly or loosely coupled depending upon the way the Inter-Processor communication (IPC) takes place. In \textbf{Tightly Coupled Multiprocessors}, IPC takes place on same die or chip through shared memory at some level. Example SMP, UMA, NUMA, Multicores etc. 
\\
\\
In case of \textbf{Loosely Coupled Multiprocessors}, multiple standalone processors are connected together either through an on-board interconnection network (buses etc.) or through a high speed communication network (Example Gigabit Ethernet on clusters)
\\
\\
\textbf{Processor Micro-architecture}\\
Micro-architecture refers to the implementation (Register-Transfer/Gate/Transistor/Circuit level) of processor architecture in silicon. Within a family of processors, the micro-architecture is often enhanced over time to deliver improvements in performance and capability, while maintaining compatibility to the architecture. A given Processor architecture can have many different micro-architecture implementations.
\\
\\
\textbf{Scalar Processor} \cite{Hennessy.Patterson2006, Mathew.Franklin2008,Smotherman2008}\\
A processor that fetches and executes one instruction at a time is called a \textbf{Scalar Processor}. Performance (CPI) of scalar processors has been improved through pipelining, out-of-order issue/execution.
\\
\\
\textbf{SIMD} \cite{Grama.etal2003, Hennessy.Patterson2006, Baer2010, Mathew.Franklin2008, Smotherman2008}\\
It is a class of multi-processor architecture in Flynn's Taxonomy. In a single instruction stream, multiple data stream (SIMD) processor, several identical processors perform the same set of operations on different sets of data. Each processor has its own data memory (hence Multiple Data), but there is a single instruction memory and central control processor, which fetches and dispatches instructions. They basically realize \textbf{Data Level Parallelism}. The SIMD category includes \textit{array processors, vector processors, and systolic arrays}.
\\
\\
\textbf{Simultaneous Multithreading}(SMT) \cite{Hennessy.Patterson2006, Baer2010, Kozyrakis2010, Koufaty2003}\\
This is a kind of hardware multi-threading in which more than one hardware thread can execute its given instruction in any pipeline stage in a given cycle. Such architectures basically combine Superscalar execution which exploits ILP in an instruction stream with hardware multi-threading which exploits TLP. Main idea behind this is to use multithreading to hide pipeline stalls(vertical wastage) due to high latency operations and reduce under-utilization of multiple functional units per pipeline stage ( also called horizontal waste ) due to limited ILP in a single instruction stream.
\\
\\
Such architectures are able to simultaneously issue or execute or keep in flight instructions from multiple instruction streams (or threads) in a given pipeline stage thereby filling most of the pipeline slots with work and thereby avoiding pipeline stalls(\textit{vertical waste}) and effectively increasing utilization ( \textit{reduced horizontal waste} ) of replicated hardware units per stage by sharing them between instructions from different threads ( overlapping ILP of different instruction streams in same pipeline stage ). For example if there are 3 integer and 2 floating point units in execute stage, with SMT all these units can be used together by integer and floating point instructions from different threads. Some Examples of such architectures: Compaq alpha EV8 which was a 4 thread SMT. Intel calls SMT as Hyperthreading(HT) and many of its single and multicore processors use HT for its PE's notably atom, core i3/i5/i7, P4, Itanium and Xeon(2-way HT). Another notable example is SUN UltraSPARC T2 plus (Victoria Falls) which is an 8 core single chip with 8 threads per core.
\\
\\
\textbf{SIMT}\\
See \textit{Section-6.4 CUDA Programming Model}
\\
\\
\textbf{SISD} \cite{Grama.etal2003, Hennessy.Patterson2006, Baer2010, Mathew.Franklin2008, Smotherman2008}\\
It is a class of multi-processor architecture in Flynn's Taxonomy. A conventional sequential processor fits into the single-instruction stream, single data stream (SISD) category. Only a single instruction is executed at a time in a SISD processor, although pipelining may allow several instructions to be in different phases of execution at any given time. Nevertheless, it still processes one instruction stream.
\\
\\
\textbf{SPMD} \cite{Grama.etal2003, Hennessy.Patterson2006, Baer2010, Mathew.Franklin2008, Smotherman2008}\\
Shared Program Multiple Data or SPMD is a coarser version of SIMD architecture paradigm. It basically means that a number of different processors execute the same program or set of instructions but each processor executes it on a different set of data. Each PE in an SPMD processor is assigned a sort of identifier and based on this identifier it can conditionally execute parts of the program. So although different PE's have the same program but they might be executing different instructions.
\\
\\
\textbf{Speculative Execution}\\
It involves execution of set of instructions ahead of an earlier instruction, but unlike out-of-order execution, here instruction look-head based dependency analysis is not performed and so the usefulness of the work done from the speculative execution depends upon whether speculation was correct or not. \emph{Correctness of speculation} is determined later on when results of earlier instructions are actually resolved. 
\\
\\
Basic motive behind speculative execution is to extract available ILP in instruction stream by utilising the hardware resources which would otherwise go idle even when the work obtained from the speculative execution can later on become useless because of miss-speculations. In case of miss-speculation, some extra work is done in order to undo the effects of speculative execution i.e. to rollback or restore system state and then to start execution of the correct instructions.
\\
\begin{itemize}
\item \textit{Data speculation} is done in form of value prediction and advanced Loads (or Prefetching).
\item \textit{Control Speculation} in the form of Branch prediction is used to lower control stalls of pipeline by predicting a branch as taken(or not taken) and then speculatively fetching and executing set of instructions from the predicted path(or sometimes both the paths).
\end{itemize}

Speculative execution requires support of processor hardware and also software (optimizing compiler). Normally extra hardware is required to buffer the results of speculative execution, to buffer the system state for rollback on miss-speculation and to handle precise exceptions.
\\
\\
\textbf{Speculative Simultaneous Multithreading(SST)}\\
Extends SMT architecture with speculative execution at thread level also called \textbf{Thread-level speculation}.
\\
\\
\textbf{Streaming Processor}\\
A Streaming Processor is composed of several processors each of which executes its own dedicated instruction stream but all of them process a single data stream. To some extent they can be considered as pipelining of multiple processors which process an incoming data stream and pass it on to the next processor in the pipeline for further processing.
\\
\\
\textbf{Superscalar Processor} \cite{Hennessy.Patterson2006, Baer2010, Mathew.Franklin2008}\\
The ability of a processor to issue and execute multiple instructions in each clock cycle and to do it dynamically at runtime is referred to as \textbf{Superscalar Execution}. Multiple instructions can be issued or/and executed either in-order or out-of-order(O-o-O). Superscalar processors replicate execution units, such as integer ALUs, FPUs, load/store units, etc. into one CPU so that it can fetch and execute more than one instruction at a time in same stage of pipeline. The goal of Superscalar processors is to attain fractional CPI (Cycles per Instruction) by exploiting instruction level parallelism.
\\
\\
\textbf{Symmetric Multiprocessors} (SMP) \cite{Grama.etal2003, Hennessy.Patterson2006, Baer2010, Mathew.Franklin2008}\\
A Shared Memory or Symmetric Multiprocessor is a UMA CMP where each individual PE has a local cache hierarchy and shares a centralized global memory hierarchy symmetrically with other PE's. By symmetrically it means that for each PE, the latency to access data in shared global memory is same. SMP style architectures don't scale linearly in performance with scaling of PE's and usually some sort of bottleneck emerges if we try to do so (say saturation of shared memory controller due to increased memory contention). On the other hand, NUMA style architectures normally do scale as number of processors increase.
\\
\\
\emph{An SMP is thus a CMP with shared-address-space (logically) but with a centralized-memory (physically)}
\\
\\
\textbf{Transaction Look Aside Buffer (TLB)}\\
see \textit{Virtual Memory}
\\
\\
\textbf{Temporal Multithreading} \cite{Hennessy.Patterson2006, Baer2010, Kozyrakis2010, Ienne2008}\\
This is a kind of hardware multi-threading in which only one hardware thread can execute a given instruction in any pipeline stage in a given cycle. Temporal multithreading has mainly two variations: \textbf{fine-grained multi-threading} and \textbf{coarse grained multi-threading.}
\\
\\
In \textbf{fine-grained or co-operative multi-threading}, the processor pipeline interleaves execution of multiple active threads by switching among these hardware threads on each instruction every few clock cycles(as low as 1 cc). The concept is similar in essence time-slice based pre-emptive round-robin scheduling of threads. This type of multi-threading is effective when there are sufficient number of threads with non-blocking instructions that can hide long operation latency. It is a good architecture for highly data parallel and data intensive applications where lot of time is spent waiting for data. Main advantages are overall higher throughput, fills well short stalls, less complex hardware: no or smaller caches, no register forwarding, no O-o-O execution hardware. Primary disadvantage is that for hardware threads with few or no stalls get delayed unnecessarily(poor single thread performance). Hardware overheads of this kind of multi-threaded architectures involves thread tracking in every stage, bigger shared resources to avoid contention among large number of small hardware threads. Example of such architecture: MTA ( Multi-threaded architecture ) is a CMP supported 128 threads per PE. Another example is SUN UltraSparc T1 (Niagara) which additionally is a multi-core processor with 4,6 or 8 PE's with each PE can handle up to 4 hardware threads scheduled in round-robin fashion.
\\
\\
In \textbf{coarse grained or block multi-threading}, thread context switch is done by hardware only when an existing thread blocks or stalls the pipeline due to certain events like long latency cache misses on loads, I/O, synchronization with another hardware thread etc. These sort of multi-threaded architectures are much less likely to slowdown non-blocking threads unlike fine-grained architectures. Other advantage is that fills well long stalls. However, one key drawback of this approach is that it requires freezing or flushing the instruction pipeline when a thread stalls and loading instructions from a waiting thread. This whole procedure takes some time and can degrade performance if there are large number of threads with short stalls. It is much more useful if most of the blocking threads result in long stalls.
\\
\\
\textbf{UMA}\\
Uniform Memory Access CMP architectures are essentially \textit{shared memory} or \textit{shared address space }architectures where all PE's have same latency to access shared main memory. The shared main memory could be a single memory or organized as multiple memory banks. Example of UMA CMP is an SMP (Symmetric or Shared Memory Multiprocessor)
\\
\\
\textbf{Vectorization or Vector Processors}\\
Vectorization is essentially 'SIMDization' or in essence a \textit{data parallel} memory access and execution where multiple data items stored in a vector or array are accessed and operated in parallel by multiple functional units of the processor. For example Multiplication of two 1-D integer vectors.
\\
\\
Vector processors also \textit{pipeline} the operations on the individual elements of a vector. The pipeline includes not only the arithmetic operations (multiplication, addition, and so on), but also memory accesses and effective address calculations. Net effect is amortized main memory access, lesser data/control hazards due to reduced instruction bandwidth, more parallelism due to replication and pipelining of functional units. Ideas of vector processors which are ubiquitous in all super-computing systems were slowly introduced in the ISA of Superscalar processors in the form Instruction set extensions (Streaming SIMD extensions like MMX, SSE/SSE2/SSE in Intel Processors)
\\
\\
\textbf{VLIW Processor} \cite{Hennessy.Patterson2006, Mathew.Franklin2008}\\
A VLIW (Very Long Instruction Word) processor belongs to another category of ILP processors which also issue and execute multiple instructions in each clock cycle. It distinguishes itself from a Superscalar processor in terms of simpler hardware architecture (control complexity) but more complex compiler in order to extract ILP statically in instruction stream.
\\
\\
It was Joseph Fischer and Bob Rau who devised the idea of VLIW, and characterised it as an architecture that can issue one Very Long Instruction per cycle, where each Very Long Instruction encapsulates many tightly coupled independent operations each of which execute in a small and statically predictable number of cycles. In such a system, the task of grouping independent operations into a very long instruction word is done by a compiler. The processor freed from the cumbersome task of dependence analysis has to merely execute in parallel the operations contained within a very long instruction. This leads to simpler and faster processor implementations. Later, Intel incorporated ideas of VLIW architecture in its EPIC (Explicit Parallel Instruction Set Computing) styled Itanium IA-64 ISA.
\chapter{Graph theory - Small world networks}
\label{app:algograph}
~\\
\\
\\
\textbf{Power Law}\\
The power law states that the probability that any vertex of a graph is connected to $k$ other vertices is proportional to $1/k^{n}$ where $n > 1$.In simpler terms, when an independent variable $x$ increases, the dependent variable $f(x)$ decreases and vice versa. Power law distributions are similar to heavy-tailed, pareto or zipf distributions.
\\
\\
\textbf{Scale free networks} \cite{Barabasi2003}\\
Networks with power law degree distribution are called as \textbf{Scale-free networks}. Such networks have some set of vertices which have relatively high degree compared to the average degree of the network. Such vertices are called \textbf{hubs}. Scale free networks are distinguished from random networks in the sense that in random networks such hubs are relatively negligible and most vertex degrees do not diverge too much from average network degree.
\\
\\
High degree of hubs in scale-free net-works is correlated to networks fault tolerance. Even if a hub fails, the network maintains its connectedness due to remaining hubs. However if most hubs fail the network can become disconnected. So hubs are both strength and weakness of the scale-free networks. Examples of scale free networks(vertices and arcs) is for example world-wide web(web-pages and links) \cite{Barabasi2000}, internet(routers and physical/optical connections) \cite{Faloutsos1999}, citations network connecting scientific papers \cite{Redner1998}, alliance networks in business(companies and collaborations/alliances), protein regulatory networks(protein and interaction)\cite{Seshasayee2006, Barrat2004}.
\\
\\
\textbf{Small world networks} \cite{Watts2006}\\
Scale free-networks in which most actors are not neighbours of one another, but most actors are reachable from other actors of the network are known as \textbf{Small world networks}. If a network has a degree-distribution which can be fit with a power law distribution, it is taken as a sign that the network is small-world. For example network of airline flights, network of connected proteins, many large-scale neural networks in the brain \cite{Eguiluz2005}, such as the visual system and brain stem, exhibit small-world properties.
\chapter{CUDA Programming Model}
\label{app:cudamodel}
\textbf{Compute Unified Device Architecture or CUDA}\cite{Tesla2008, Sanders.Kandrot2010, Kirk.Hwu2010} is NVIDIA's hardware software co-processing architecture cum programming model that allows programmer to write applications for its GPGPU's. CUDA programming model requires programmer to specify structured parallelism in form of shared-memory data parallel programs(SPMD style). Unlike general purpose processors, most GPGPU's do not support dynamic parallelism\footnote{In Dynamic parallelism, one parallel thread can dynamically spawn new parallel threads at runtime. Dynamical parallelism is a usual feature of any General purpose or Multicore processors but is not supported on Tesla or Fermi GPGPU's. Recently, NVIDIA also added support for dynamic parallelism on its Kepler GPGPU's}
\\
\\
NVIDIA calls the main system processor plus its DRAM as \textbf{Host} and the GPGPU plus its GDDR DRAM as \textbf{device}. Host acts as a \textbf{\emph{Master}} which fully manages the GPGPU \textbf{\emph{Slave or co-processor}}. In other words, the GPGPU cannot access CPU memory, cannot allocate local memory or cannot start computing unless host CPU tells it to.
\\
\\
Using minimal set of CUDA abstractions(\emph{Kernel and device functions}) for expressing parallelism, a programmer can write application for CUDA based GPGPU's in languages such as C, C++, OpenCL, etc. A CUDA program is a unified source code containing interleaving of ANSI C code that executes on host called \textbf{Host Code} and CUDA C (minimal extensions to ANSI C) code that runs on GPGPU called \textbf{device code or kernel}. Programmer needs to carefully divide and map the application parts depending upon capability of host CPU and GPGPU device. Usually, highly data parallel part of application is accelerated on GPGPU and inherently sequential or less data parallel part is executed on host CPU. For executing such CUDA program, this unified source code needs to be compiled using NVCC compiler driver. 
\\
\\
In addition, a CUDA program generally involves transferring data to be processed on GPGPU from CPU memory to GPGPU GDDR RAM. Once the GPGPU has finished execution of CUDA kernel, the results are subsequently copied back to CPU memory. Following is a typical CUDA program execution flow managed by the host CPU:

\begin{itemize}
\item Allocate host data [\emph{malloc()} function ]
\item Allocate device memory where host data will be copied [\emph{cudaMalloc()} function]
\item Copy data from host memory to device memory [\emph{cudaMemcpy()} synchronous function]
\item Invoke kernel on device by calling it with following parameters: thread hierarchy configuration and memory pointers to allocated device memory
\begin{verbatim}
// asynchronous function call
kernel_function<<<thread_hierarchy_config>>>(device_memory_ptrs,...); 
\end{verbatim}
\item Do some execution on host ( happens in parallel with kernel execution )
\item Optionally, host waits for previous calls to complete [\emph{cudaThreadSynchronize()} function]
\item Copy result data from device to host[\emph{cudaMemcpy()} synchronous function]
\item Deallocate host and device memory [\emph{free()} and \emph{cudaFree()}]
\end{itemize}

\subsection{Thread Hierarchy}
\label{subsec:cudathreads}
\begin{figure}[ht!]
	\centering
	\includegraphics[scale=0.8]{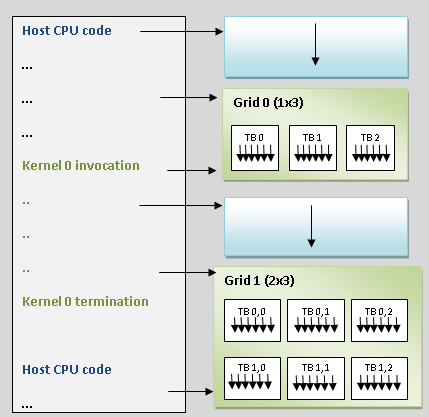}
	\caption{CUDA Program - interleaving of host code and kernel invocations}\label{fig:gpucudaprog}
\end{figure}

\begin{figure}[hb!]
    \centering
	\includegraphics[scale=0.3]{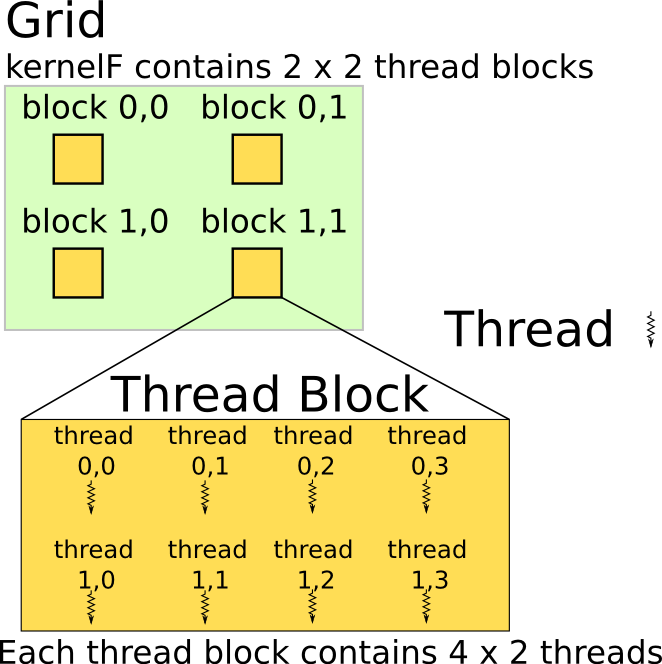}
	\caption{Multidimensional grid and thread block}\label{fig:gpucudathreadhierch}
\end{figure}

A CUDA \textbf{\emph{kernel}} is a sequential piece of C/C++ code that can be executed by one or more CUDA hardware threads. Any C/C++ function can be made a kernel function by putting \emph{\_\_global\_\_} prefix to its function definition. Such a kernel can be called from host and will be executed on device. Similarly any function that is to be called and executed on the device should be prefixed with \emph{\_\_device\_\_} keyword. Functions without any function qualifiers are by default host functions(called and executed on host). A function can be made both host and device function by putting both \emph{\_\_host\_\_} and \emph{\_\_device\_\_} qualifier to the function definition. 
\\
\\
CUDA requires programmer to specify a hierarchy of threads through \emph{thread\_hierarchy\_config} parameter passed to the device kernel during invocation. This parameter basically specifies organization of CUDA threads in terms of \textbf{thread blocks} and number of \textbf{threads per block} (\emph{threads\_per\_block}). This means a total number of (\emph{thread blocks x threads\_per\_block}) CUDA threads will be created on kernel invocation.
\\
\\
Further, all the thread blocks that are generated during a kernel invocation are collectively called as \textbf{grid}. Each thread block executes independently of other thread blocks. When all threads of a kernel complete execution, the corresponding grid terminates. Every time device kernel is invoked and terminated, a corresponding grid is created and terminated. Any CUDA program is thus an interleaving of CPU code and CUDA grids or kernel invocations (Figure~\ref{fig:gpucudaprog}).
\\
\\
Within a grid, each thread block has a \emph{thread block id} which can be accessed within kernel function by threads of the block through variables \textbf{blockIdx.x} and \textbf{blockIdx.y} (2-dimensions for compute capability 1.3). Similarly each thread within a block has an associated id which can be referred in kernel function through variables \textbf{threadIdx.x, threadIdx.y and threadIdx.z}. Additional built-in variables, \textbf{gridDim} and \textbf{blockDim}, provide the dimension of the grid and the dimension of each thread block respectively. During execution of kernel by CUDA threads, each thread will see different value for its blockIdx and threadIdx variables. These four variables: blockIdx, threadIdx, gridDim and blockDim can be accessed within a kernel function and are basically used for calculating index values which can be used by thread executing the kernel to read or write data to appropriate memory locations. Accordingly we can specify \emph{thread\_hierarchy\_config} parameter as follows:

\begin{verbatim}
    thread_hierarchy_config = grid_config, threadblock_config}
    grid_config             = (threadBlock.X, threadBlock.Y)}
    threadblock_config      = (threadIdx.X, threadIdx.Y, threadIdx.Z)}
\end{verbatim}

CUDA provides a C struct type named \textbf{dim3} to create \emph{grid\_config} and \emph{threadblock\_config} parameters. For example, invocation of kernel function $kernelF \lll dim3(2,2,1),dim3(4,2,1) \ggg$ will generate grid and thread blocks as shown in Figure~\ref{fig:gpucudathreadhierch}.

\subsection{Thread Scheduling}
\label{subsec:cudathreadschedul}
\begin{figure}[ht!]
	\centering
	\includegraphics[scale=0.5]{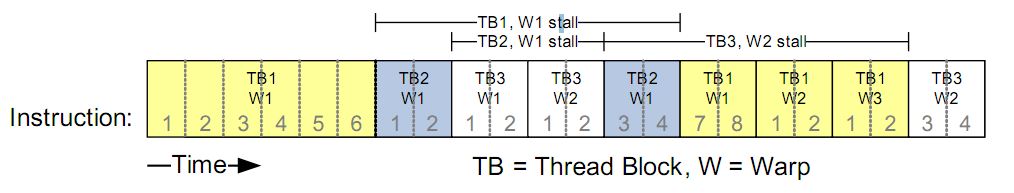}
	\caption{Scheduling of warps and thread blocks. \textit{Source}: Mark Silberstein, Technion, "Lectues notes - Programming massively parallel processors"}\label{fig:gpucudawarpsschedul}
\end{figure}

The CUDA thread execution is somewhat analogous to SIMD/SPMD model of execution but NVIDIA calls this as Single Instruction Multiple Thread(SIMT) architecture \cite{Tesla2008}. Just as a SIMD processor controls a vector of multiple data lanes, a SIMT processor controls bunch of threads(\textit{warp}). The SIMT architecture manages CUDA threads in bunch of 32 parallel threads of the same type(vertex, geometry, pixel, or compute) called as \textbf{Warp} (Figure~\ref{fig:gpucudaprog}). 
\\
\\
With \textbf{SIMT execution}, the hardware fetches and issue same instruction for each thread of a warp independently, before moving to the next instruction of the warp. If threads within a warp stall( Figure~\ref{fig:gpucudawarpsschedul} ) due to an instruction causing long latency operation like global memory loads, floating point arithmetic, etc., the Multithreaded issue/warp scheduler stops execution of the warp and swaps it with another waiting warp ready to be executed. If there are one or more ready warps waiting to be executed, some priority scheme is used to identify which warp is to be scheduled next. NVIDIA says the overhead of scheduling a new warp is almost zero if there are one or more warps ready for execution. In other words, CUDA fine grained Multithreading basically interleaves execution of warps. Therefore, on a CUDA device, a warp is the smallest schedulable executable unit of parallelism and is thus the granularity of hardware multithreading.
\\
\\
According to NVIDIA recommendations, an SM realizes full efficiency and performance when all threads of a warp take the same execution path. If some threads within a warp \textbf{\emph{diverge}} due to say a data-dependent conditional branch, or other programming constructs (loop iterations etc.), the warp execution serializes leading to performance  deterioration equal to number of divergent paths.

\subsection{Thread communication and synchronization}
\label{subsec:cudathreadsynch}
CUDA provides various mechanism for thread communication and synchronization primitives \cite{WongPapadopoulou2010} depending upon level in thread hierarchy (see Table~\ref{tab:cudathreadcommsynch})
\begin{table}
  \centering
  \caption{CUDA thread communication and synchronization}
  \begin{tabular}{ p{5cm} | l | p{4cm} }
    \hline
    \textbf{Thread hierarchy}                     & \textbf{Communication}  & \textbf{Synchronization} \\ \hline \hline
    Threads within a warp                         & shared/global memory    & implicit synchronization \\ \hline
    \multirow{2}{*}{Warps within a thread block}  & shared memory           & barrier synchronization (\_\_syncthreads()) or Atomics \\ \cline{2-3}
                                                  & global memory           & CUDA Atomic operations \\ \hline
    Warps of different thread blocks              & global memory           & No mechanism \\ \hline
    Thread blocks within a given grid or kernel   & global memory           & atomic memory operations \\ \hline
    Thread blocks from different grids or kernels & global memory           & implicit synchronization \\
    \hline
  \end{tabular}  
  \label{tab:cudathreadcommsynch}
\end{table}

\subsection{Memory Model}
\label{subsec:cudamemorymodel}
\begin{figure}[ht!]
	\centering
	\includegraphics[scale=0.7]{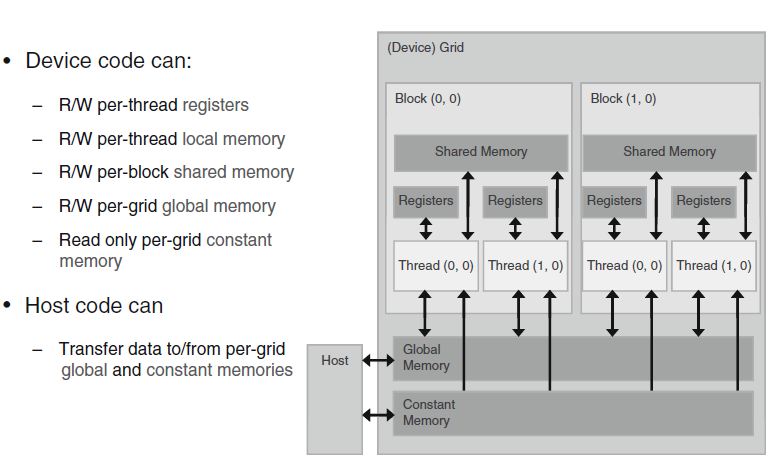}
	\caption{CUDA memory model \cite{Kirk.Hwu2010}}\label{fig:gpucudamemmodel}
\end{figure}

In CUDA memory model(see Figure~\ref{fig:gpucudamemmodel}), there are multiple levels of memory: \textbf{per thread registers} and \textbf{per thread local memory} (not shown in figure), \textbf{per thread block shared memory}, \textbf{per grid global and constant memory} 

\begin{table}[hb!]
  \centering
  \caption{CUDA Variable Qualifiers}
  \begin{tabular}{ l | p{5cm} }
    \hline
    \textbf{Variable Qualifier} & \textbf{Variable placed in memory ?}\\ \hline \hline
    \_\_device\_\_     & device global memory(GDDR)\\ \hline
    \_\_constant\_\_   & Constant Memory \\ \hline
    \_\_shared\_\_     & Shared Memory(On-chip) \\ \hline
    No qualifier       & Registers(On-chip). Register spills go into local memory(GDDR) \\
    \hline
  \end{tabular}  
  \label{tab:cudavar}
\end{table}

\begin{sidewaystable}
  \centering        
  \caption{Memory organization of Tesla C1060 T20/GT200 with CUDA compute capability 1.3}
  \begin{tabular}{ | p{3cm} | p{3cm} | p{3cm} | p{3cm} | p{3cm} | p{3cm} |}
    \hline
	& \textbf{Registers} & \textbf{Local memory} & \textbf{Shared memory} & \textbf{Global memory} & \textbf{Constant Memory}\\ \hline
	\textbf{On-chip/Off-chip} & On-chip & Off-chip GDDR & On-chip & Off-chip GDDR & Off-chip\\ \hline
	\textbf{Scope} & Private per thread & Private per thread & Threads of the thread block & All threads of a grid & All threads of a grid\\ \hline
	\textbf{Lifetime} & Thread & Thread & Thread block & Application & Application\\ \hline
	\textbf{Read(R)/Write(W) by device?} & R/W & R/W & R/W & R/W & R\\ \hline
	\textbf{R/W access by host?} & No & No & No & R/W & R/W\\ \hline
	\textbf{Size} & 64KB per SM or TB (32bit x 16K) (64 logical banks/SM) & $<$ 4 GB MB & 16KB per SM or TB (1KB x 16 banks) & 4 GB & 64 KB total (8 KB/SM)\\ \hline
	\textbf{Is Cached?} & NA & No &No &No &Yes, Constant cache\\ \hline
	\textbf{Latency \cite{WongPapadopoulou2010} (clock cycles)} &  \textasciitilde 1 cc & \textasciitilde 400-450 cc & \textasciitilde 30-40 cc & \textasciitilde 400-450 cc &  \textasciitilde 1-3 cc cache hit.  \textasciitilde 100 cc cache miss\\ \hline
	\textbf{Bandwidth} & \textasciitilde 8-10 TB/s & 170 GB/s &  \textasciitilde 1000-1400 GB/s aggregate for 30 SM &  \textasciitilde 100-150 GB/s &  \textasciitilde 1000 GB/s\\
    \hline
  \end{tabular}  
  \label{tab:cudamemspecs}
\end{sidewaystable}

CUDA API provides appropriate qualifiers (see Table~\ref{tab:cudavar}) that can be used to place variables in appropriate CUDA memory. In case of absence of variable qualifier(automatic variables), the compiler allocates register memory to the variable(thread local). In case there are not sufficient registers, the variable is placed in local memory(GDDR). Thread local Arrays can also be allocated in registers provided there are enough registers otherwise they also spill to local memory. In addition, shared memory can be statically allocated at compile time and also dynamically allocated at runtime.
\\
\\
Table~\ref{tab:cudamemspecs} shows specifications of different kind of memories in CUDA memory model for GPGPU with GT200 CUDA architecture \cite{Kirk.Hwu2010, Silberstein2011, Siegel2009, Yang, Farber2008} and compute capability 1.3. All memories in CUDA memory model belong to two types of GPU physical memories: \textbf{on-chip} and \textbf{off-chip}. All on-chip CUDA memories provide low latency and high bandwidth access to the data. To get good performance for memory bound algorithms on such architectures, it is important to utilize the on-chip CUDA memories effectively and minimize interaction with off-chip memories.

\chapter{Counters for computeprof profiler}
\label{app:comprof}
Table~\ref{tab:computeprofcounters} shows description of the profile counters for the computeprofile command-line tool (\textit{Source}: NVIDIA Compute Visual Profiler).
\tablecaption{Profile counters for computeprof tool}\label{tab:computeprofcounters}
\tablehead{\hline \textbf{Counter} & \textbf{Description}\\ \hline}
\tabletail{\hline}
\begin{xtabular}{ | p{5cm} | p{10cm} |}
\textbf{Counter}	& \textbf{Description}\\ \hline
warp\_serialize	& This counter gives the number of thread warps that serialize on address conflicts to either shared or constant memory.\\ \hline
instructions	&Number of Instructions executed\\ \hline
instruction throughput (derived counter)	&This is the ratio of achieved instruction rate to peak single issue instruction rate. The achieved instruction rate is calculated using the profiler counter "instructions". This is calculated as: (instructions) / (gpu\_time * clock\_frequency)\\ \hline
cta\_launched & Number of threads blocks launched on a TPC\\ \hline
sm\_cta\_launched & Number of threads blocks launched on a SM\\ \hline
branch & Number of unique branch instructions in program ( includes barrier instructions )\\ \hline
divergent\_branch & Number of unique branches that diverge\\ \hline
Divergent branches (\%) (derived counter) & The percentage of branches that are causing divergence within a warp amongst all the branches present in the kernel. Divergence within a warp causes serialization in execution. This is calculated as: (100*divergent branch)/(divergent branch + branch)\\ \hline
dynsmemperblock	& Size of dynamically allocated shared memory per block in bytes for a kernel launch.\\ \hline
stasmemperblock	& Size of statically allocated shared memory per block in bytes for a kernel launch\\ \hline
regperthread	 & Number of registers used per thread for a kernel launch.\\ \hline
gld\_request	 & Number of executed load instructions per warp in a SM\\ \hline
gst\_request	 & Number of executed global store instructions per warp in a SM\\ \hline
gld\_64b	 & 64-byte global memory load transactions\\ \hline
gst\_64b	 & 64-byte global memory store transactions\\ \hline
glob mem read throughput (derived counter) & Global memory read throughput in gigabytes per second. For compute capability $<$ 2.0 this is calculated as: (((gld\_32*32) + (gld\_64*64) + (gld\_128*128)) * TPC) / (gputime * 1000)\\ \hline
glob mem write throughput (derived counter) & Global memory write throughput in gigabytes per second. For compute capability $<$ 2.0 this is calculated as: (((gst\_32*32) + (gst\_64*64) + (gst\_128*128)) * TPC) / (gputime * 1000)\\ \hline
glob mem overall throughput (derived counter)	&Global memory overall throughput in gigabytes per second. This is calculated as: Global memory read throughput + Global memory write throughput\\ \hline
gld\_incoherent	& Non-coalesced (incoherent) global memory loads\\ \hline
gld\_coherent	& Number of Coalesced (coherent) global memory loads\\ \hline
gst\_incoherent	& Number of Non-coalesced (incoherent) global memory stores\\ \hline
gst\_coherent	& Number of Coalesced (coherent) global memory stores\\ \hline
\end{xtabular}
\chapter{Business Plan}
A business plan describes an idea and analysis on how to start a new business of any topic of interest to the student and is a mandatory part of the master thesis for ALaRI Business Track students. The business plan that has been developed in scope of my studies in group with Darayus Patel. This business plan proposes a business in the Embedded system domain in India but is not related to topic of this Master thesis. Therefore, the details of this business plan has been left out of this thesis report and business plan report is instead made available on the ALaRI platform(privately - As report is confidential can be accessed only with permission).
\backmatter
\chapter{Glossary}

\begin{verbatim}
    API     Application Programming Interface
    ALU     Arithmetic and Logical Unit
    CMOS    Complementary Metal Oxide Semiconductor
    CMP     Chip Multiprocessor
    CPI     Cycles per Instruction
    CUDA    Compute Unified Device Architecture
    DAG     Directed Acyclic Graph
    ESL     Electronic System Level
    GPP     General Purpose Processor
    GPU     Graphics Processing Unit
    GPGPU   General Purpose Graphics Processing Unit
    IC      Integrated Circuit
    IP      Intellectual Property
    IPC     Instructions per Cycle or Inter Processor Communication
    ILP     Instruction Level Parallelism
    ISA     Instruction Set Architecture
    ITRS    International Technology Roadmap for Semiconductors
    MIPS    Million Instructions per Second
    MPSoC   Multiprocessor System on Chip
    NoC     Network on chip
    NUMA    Non-Uniform Memory Access
    PE      Processing Element
    PCI-E   Peripheral Component Interface - Express
    SIMD    Single Instruction Stream Multiple Data Stream
    SFU     Special Function Unit
    SM      Streaming Multiprocessor
    SP      Scalar Processor or CUDA Core
    SMP     Symmetric Multiprocessors or Shared Memory Processors
    TCB     Task Control Block
    TLP     Thread Level Parallelism
    TBB     Threading Building Blocks
    UMA     Unified Memory Access
    VLIW    Very Long Instruction Word
\end{verbatim}
\bibliographystyle{IEEEtran}
\bibliography{thesis}
\cleardoublepage
\end{document}